\documentclass[twocolumn]{aastex63}

\usepackage{amsmath}
\usepackage{natbib}
\usepackage{multirow} 
\usepackage{subfigure} 
\usepackage{placeins}

\usepackage{anyfontsize}
\usepackage{indentfirst}
\usepackage{comment}
\usepackage{booktabs}
\usepackage{bigints}
\usepackage{mathrsfs,amsmath} 
\usepackage{makecell}
\usepackage{graphicx}

\usepackage[normalem]{ulem}
\useunder{\uline}{\ul}{}
\usepackage{hyperref}
\usepackage{lineno}
\hypersetup{linkcolor=magenta,citecolor=cyan,filecolor=yellow,urlcolor=blue}

\received{ddmmyyyy}
\revised{\today}
\accepted{ddmmyyyy}
\published{ddmmyyyy}

\shorttitle{A Novel Method of Estimating GRB Peak Energies Beyond the \emph{Swift}/BAT Limit}

\shortauthors{Li \& Wang}

\begin{document}

\title{A Novel Method of Estimating GRB Peak Energies Beyond the \emph{Swift}/BAT Limit}


\author[0000-0002-1343-3089]{Liang Li}

\affiliation{Institute of Fundamental Physics and Quantum Technology, Ningbo University, Ningbo, Zhejiang 315211, People's Republic of China}

\affiliation{School of Physical Science and Technology, Ningbo University, Ningbo, Zhejiang 315211, People's Republic of China}

\affiliation{INAF -- Osservatorio Astronomico d'Abruzzo, Via M. Maggini snc, I-64100, Teramo, Italy}

\author{Yu Wang}

\affiliation{ICRA and Dipartimento di Fisica, Universit\`a  di Roma ``La Sapienza'', Piazzale Aldo Moro 5, I-00185 Roma, Italy}

\affiliation{ICRANet, Piazza della Repubblica 10, I-65122 Pescara, Italy}

\affiliation{INAF -- Osservatorio Astronomico d'Abruzzo, Via M. Maggini snc, I-64100, Teramo, Italy}

\correspondingauthor{Liang Li, Yu Wang}
\email{liliang@nbu.edu.cn; yu.wang@icranet.org}
 
\begin{abstract}

The \emph{Swift} Burst Alert Telescope (BAT), operating in the 15--150 keV energy band, struggles to detect the peak energy ($E_{\rm p}$) of gamma-ray bursts (GRBs), as most GRBs have $E_{\rm p}$ values typically distributed between 200-300 keV, exceeding BAT's upper limit. To address this, we develop an innovative method to robustly estimate the lower limit of $E_{\rm p}$ for GRBs with $E_{\rm p}>150$ keV. This approach relies on the intrinsic curvature of GRB spectra, which is already evident within the BAT energy range for such GRBs. By fitting BAT spectra with a cutoff power-law model and extrapolating the spectral curvature beyond BAT's range, we, therefore, can estimate the cutoff energy ($E^{'}_{\rm c}$) beyond 150 keV and the corresponding peak energy ($E^{'}_{\rm p}$). We applied this method to 17 GRBs, categorizing them into two main groups. Group I (10 bursts) maintains $\alpha$ within a typical range (from $\sim$ -0.8 to $\sim$ -1.20) with increasing $E_{\rm c}$; Group II (2 bursts) maintains $E_{\rm c}$ within a typical range (300-500 keV) but with varying $\alpha$. Our results show that for $E_{\rm c}\lesssim $1000 keV, the estimated $E^{'}_{\rm c}$ aligns well with observed values. Moreover, the reliability of $E^{'}_{\rm c}$ also depends on $\alpha$: bursts with harder $\alpha$ (e.g., $\alpha \gtrsim -2/3$) show reduced accuracy, while bursts with softer $\alpha$ (e.g., $\alpha \lesssim -2/3$) yield more precise estimates. In conclusion, this method is well-suited for GRB spectra with moderately observed $E_{\rm c}$ ($E_{\rm p}$) values and $\alpha$ indices that are not too hard.

\end{abstract}


\section{Introduction} \label{sec:intro}

Gamma-ray bursts (GRBs) are among the most energetic events in the Universe, characterized by brief yet intense emissions of high-energy emission (mostly in $\gamma$-ray). Understanding the spectral properties of GRBs is essential for uncovering the physical mechanisms driving these events\citep[e.g.,][and references therein]{Preece2000, Kaneko2006, Goldstein2012, Goldstein2013, Gruber2014, Yu2016, Acuner2018, Li2019a, Li2019c, Li2019b, Li2021a, Li2023c}. The photon spectrum of a GRB provides critical insights into the energy distribution of the radiation, with several models proposed to characterize the observed spectra\citep[e.g.,][]{Preece1998, Lloyd2000a, Geng2018, Ryde2019, Acuner2020}. Based on observational data, one of the most frequently used spectral models in the GRB spectral study is the Band function \citep{Band1993}, an empirical model effectively fits the photon spectra across a wide energy range. Phenomenologically, the photon number spectrum of the Band function is defined as
\begin{eqnarray}
N_{\rm Band}(E)=A \times \left\{ \begin{array}{ll}
(\frac{E}{E_{\rm piv}})^{\alpha} \rm exp (-\frac{{\it E}}{{\it E_{\rm 0}}}), & E < (\alpha-\beta)E_{\rm 0}  \\
\lbrack\frac{(\alpha-\beta)E_{\rm 0}}{E_{\rm piv}}\rbrack^{(\alpha-\beta)} \rm exp(\beta-\alpha)(\frac{{\it E}}{{\it E_{\rm piv}}})^{\beta}, & E\ge (\alpha-\beta)E_{0}\\
\end{array} \right.
\label{eq:Band} 
\end{eqnarray}
where $A$ is the normalization factor in units of ph cm$^{-2}$keV$^{-1}$s$^{-1}$, $E_{\rm piv}$ is the pivot energy always fixed at 100 keV, $\alpha$ and $\beta$ are the low- and high-energy asymptotic power-law photon indices, respectively, $E_{0}$ is the characteristic break energy correlated with the peak energy of $\nu F_{\nu}$ spectrum (assuming $\beta<-2$) by $E_{\rm p}=(2+\alpha)E_{\rm 0}$, representing the turning point at which the spectral shape changes from one form to another. 

According to the definition, the Band model includes a four-free model parameter ($A$, $\alpha$, $E_{\rm p}$, and $\beta$). Observationally, the spectral indices ($\alpha$ and $\beta$) and the peak energy ($E_{\rm p}$) are typically distributed around $\alpha \simeq$-0.8 (below the break energy), $\beta \simeq$-2.2 (above the break energy), and $E_{\rm p} \simeq$220 keV, respectively \cite[e.g.,][]{Preece2000,Goldstein2013,Gruber2014,Yu2019,Li2021b,Li2022,Li2023a}. 
It provides a phenomenological fit to GRBs' photon spectra, capturing the emission's complex behavior across different energy ranges and characterizes two distinct spectral regimes: a low-energy power-law component with a photon index $\alpha$ and a high-energy power-law tail with a photon index $\beta$, which are smoothly connected at a characteristic break energy, ($\alpha$-$\beta$)$E_{\rm 0}$. This smooth transition ensures continuity in the spectrum while fitting the photon spectra across a wide energy range. The model has proven highly effective in describing the emission spectra of the majority of observed GRBs, capturing essential features without assuming specific physical mechanisms \cite[e.g.,][]{Yu2019,Li2022}. 

In some cases, the Band function can be replaced by a simpler cutoff power-law function (CPL), which describes a power law with an exponential cutoff. This is particularly useful when the high-energy power-law index ($\beta$) is poorly constrained. The CPL function (the first portion of the Band function) is given by 
\begin{equation}
N_{\rm CPL}(E) =A \left(\frac{E}{E_{\rm piv}}\right)^{\alpha}\rm exp(-\frac{\it E}{\it E_{\rm c}}),
\label{CPL}
\end{equation}
where $E_{\rm c}$ is the cut-off energies, which defines the energy at which the photon spectrum transitions from the low-energy power law to the high-energy exponential cutoff, and correlated with the peak energy $E_{\rm p}$ of the $\nu F_{\nu}$ spectrum through $E_{\rm p}=(2+\alpha)E_{\rm c}$. 

Observationally, GRB prompt emission is well known to have strong spectral evolution \citep[e.g.,][and references therein]{Yu2019, Li2019b, Li2021b, Li2023a}. The Band model is often preferred for time-integrated spectral analysis, as it accumulates more high-energy photons, ensuring that the high-energy segment can be better constrained. Conversely, the CPL model is frequently used for time-resolved spectral analysis, as short timescales often lack sufficient high-energy photons for robust fitting. Both models are standard tools for characterizing the observed spectra, providing nearly equivalent fits to observational data and being widely adopted in nearly all GRB studies \citep[e.g.,][and references therein]{Kaneko2006, Goldstein2012, Gruber2014, Li2019a, Li2022, Li2023c}. 

The peak energy, $E_{\rm p}$, refers to the energy at which the photon spectrum reaches its maximum in the $\nu F_{\nu}$ representation, represents the energy at which most of the energy of the selected spectrum (time-resolved analysis) or the entire burst (time-integrated analysis) is released. Therefore, $E_{\rm p}$ may holds important information and clues for understanding the radiation mechanism of GRBs. Here, we discuss the potential physical meaning of $E_{\rm p}$ and its relevance to GRB studies.

\begin{itemize}

\item Distinguishing radiation mechanisms. For synchrotron radiation, $E_{\rm p}$ is typically in the range of a few hundred keV, emitted by relativistic electrons in a magnetic field. In contrast, inverse Compton scattering, where these relativistic electrons upscatter lower-energy seed photons, shifts $E_{\rm p}$ to tens to hundreds of MeV range \citep{Piran2004}.

\item Dynamical properties of the relativistic jet. The observed value of $E_{\rm p}$ is affected Doppler boosting depending on the Lorentz factor of the jet. A larger Lorentz factor results in a higher observed $E_{\rm p}$, making it a key diagnostic of the jet's velocity \citep{Lithwick2001}.

\item Differentiating thermal and non-thermal components. Thermal emission, typically associated with the GRB photosphere, produces a well-defined Planck-like peak near $E_{\rm p}$, reflecting the characteristic temperature of the emitting region. In contrast, non-thermal processes, such as synchrotron radiation from relativistic particles, produce a broader and more complex spectral shapes around $E_{\rm p}$ \citep{Peer2011}.

\item GRB classification. SGRBs, typically resulting from compact object mergers \citep{Eichler1989,Berger2014,Abbott2017}, exhibit higher $E_{\rm p}$ values compared to LGRBs, which are associated with the collapse of massive stars \citep{Woosley1993,MacFadyen1999,Hjorth2003}. This difference is attributed to the jet properties: SGRB jets are generally less baryon-rich and are launched with higher Lorentz factors, leading to more energetic photons and a higher $E_{\rm p}$. By contrast, LGRBs, with slower jets and more massive outflows, tend to produce lower $E_{\rm p}$ values \citep{Kouveliotou1993}.

\item Redshift and cosmological implications (standard candle). The empirical correlation between $E_{\rm p}$ and the isotropic energy $E_{\gamma,\rm iso}$, known as the Amati relation \citep{Amati2002}, along with similar correlations such as the Yonetoku relation between $E_{\rm p}$ and the peak luminosity  \citep{Yonetoku2004}, establishes GRBs as tool for measuring the redshfits. These relations highlight the potential of GRBs as standard candles, similar to Type Ia supernovae, for probing the universe's expansion history by the observational quantities \citep{Nava2011a}.

\end{itemize}

Since 2004, NASA's \emph{Swift} satellite has revolutionized GRB studies \citep{Gehrels2004,Burrows2005,Berger2005,Campana2006}. It enabled rapid localization and multi-wavelength observations of these events. Major breakthroughs include the discovery of GRB afterglows and confirming distinct origins for long and short GRBs. \emph{Swift} also advanced research on supernovae, black hole formation, and high-redshift cosmology \citep{Gehrels2009}. The satellite carries three instruments: the Burst Alert Telescope \citep[BAT;][]{Barthelmy2005}, the X-Ray Telescope \citep[XAT;][]{Burrows2005a}, and the UV-Optical Telescope \citep[UVOT;][]{Roming2005}. The \emph{Swift}/BAT (15-350 keV) instrument\footnote{The effective energy is 15-150 keV.}, while highly effective in detecting and localizing GRBs, has limitations in its energy range, which is restricted to 15-150 keV. This constraint makes it less sensitive to higher-energy emissions, necessitating complementary data from other instruments like Fermi/GBM \citep{Meegan2009} for broader spectral coverage. 

Observational data from multiple satellites with a wider spectral window (e.g., Fermi/GBM) have revealed that the $E_{\rm p}$ distribution of GRBs covers at least 3 orders of magnitude, with peak values typically falling within the range of 200-300 keV. Consequently, for most GRBs, the typical $E_{\rm p}$ exceeds the upper limit of the effective energy range (15-150 keV) observed by the \emph{Swift}/BAT. This situation implies that many GRBs detected by \emph{Swift} do not have their $E_{\rm p}$ measured within the BAT energy band, leading to challenges such as uncertainties in spectral $k$-corrections (1-10$^{4}$ keV) and accurate calculations of isotropic total release energy $E_{\gamma, \rm iso}$. To estimate a $E_{\rm p}$ for \emph{Swift}/BAT GRBs, several attempt has been made by many authors. For instance, \cite{Sakamoto2009} has employed a simple empirical correlation ($\Gamma_{\rm BAT}-E_{\rm p}$) between the photon index $\Gamma_{\rm BAT}$ and the peak energy ($E_{\rm p}$) to provide rough estimates of $E_{\rm p}$, where $\Gamma_{\rm BAT}$ is the photon index derived from the power-law model fit. However, this approach suffers from significant random uncertainties for individual GRBs, which can result in inaccurate estimates. Moreover, some recent studies \citep{Yu2019,Li2021b} have suggested that the $\alpha-E_{\rm p}$ correlation may not always exhibit strong consistency if we consider that $\Gamma_{\rm BAT}$ measured from \emph{Swift}/BAT is the same as $\alpha$ obtained from \emph{Fermi}/GBM; variations in correlation have been observed across different samples and even among distinct bursts.

In practice, when analyzing a GRB spectrum within a relatively narrow energy band (e.g., \emph{Swift}/BAT GRBs)-especially when the upper limit of the observed energy range is much lower than $E_{\rm p}$-the spectrum may only be adequately fit by a single power-law function, which can be defined as $N(E)=A(E/E_{\rm piv})^{-\hat{\Gamma}}$. However, the intrinsic GRB spectrum is likely curved across a broader energy range. Due to the limitations imposed by the narrow observational band, $E_{\rm p}$ and significant non-linear spectral features may go undetected. Consequently, the relationship $\Gamma\neq \alpha$ may arise. This discrepancy may helps to explain the inconsistencies observation between the typical $\Gamma$ values obtained from \emph{Swift}/BAT and the typical  $\alpha$ values from \emph{Fermi}/GBM, with the former typically yielding softer spectral indices compared to the latter.

Following these lines of argument, well-determined $E_{\rm p}$ of \emph{Swift}/BAT GRBs is important for both the GRB physics and observation. Together with XRT and UVOT instruments, one can provide a full data analysis of multi-wavelength observation covering both prompt emission and afterglow emission from the same satellite and therefore ensure a consistency data analysis. On the other hand, we cannot rule out the possibility that, although some statistical criteria to assess the goodness-of-fit, such as reduced chi-squared $\chi^{2}_{\rm r}$ (GBM Band) is better than $\chi^{2}_{\rm r}$ (BAT PL), when we perform model fitting, the intrinsic spectral data may contribute more weight to the fitting in the high-energy range ($>$150 keV).

In this paper, we develop a novel method (spectral shape curvature extrapolation method) to search GRB peak energies $E_{\rm p}$ beyond the \emph{Swift}/BAT limit. Further, we also compare several other possible approaches (e.g., empirical correlation method). The paper is organized as follows. The methodology is presented in Sections \ref{sec:Methodology}. The results are presented in Section \ref{sec:results}. The discussions are presented in Section \ref{sec:Dis} and our conclusions are summarized in Section \ref{sec:Con}. Throughout the paper, the standard $\Lambda$-CDM cosmology with the parameters $H_{0}= 67.4$ ${\rm km s^{-1}}$ ${\rm Mpc^{-1}}$, $\Omega_{M}=0.315$, and $\Omega_{\Lambda}=0.685$ are adopted \citep{PlanckCollaboration2018}.

\section{Methodology}\label{sec:Methodology}

The \emph{Swift}/BAT instrument, with its narrow energy range (15-150 keV), struggles to detect the peak energy ($E_{\rm p}$) of GRBs. Some new methods for estimating $E_{\rm p}$ beyond the \emph{Swift}/BAT limit, such as the curvature extrapolation method based on spectral shape, can be attempted. The underlying principle is that GRB spectra are intrinsically curved. Therefore, even when the intrinsical $E_{\rm p}$ of a GRB spectrum is significantly higher than the observational limit of the \emph{Swift}/BAT instrument, a well-defined spectral shape within the BAT energy range may still provide valuable information, such as curvature. By using the properties of this curvature, it is possible to estimate a lower bound for $E_{\rm p}$ above 150 keV with reasonable accuracy.

\subsection{Spectral Model Selection: Band versus Cutoff Power Law}\label{sec:Model}

The purpose of this study is to estimate the peak energy ($E_{\rm p}$) of GRB spectra for those bursts whose detected by the \emph{Swift}/BAT instrument, where $E_{\rm p}$ lies beyond the BAT limit, based on the standard GRB spectral model (either Band or CPL model) using an extrapolation method. For the Band model, $E_{\rm p}$ can be directly measured from the observed spectral data. However, the CPL model provides the cutoff energy ($E_{\rm c}$) as the measured parameter, with $E_{\rm p}$ being closely related to both the low-energy spectral index ($\alpha$) and the cut-off energies ($E_{\rm c}$).

By definition, the Band function consists of two components: the low-energy CPL-function component and the high-energy power-law $\beta$-function component. These components are mathematically independent, with no intrinsic connection between their power-law indices. The high-energy $\beta$ function has minimal influence on the curvature $\kappa$ and other properties of the low-energy spectrum. Consequently, $\beta$ contributes little to estimates based on the low-energy component. Observationally, $E_{\rm p}$ is already estimated from the low-energy component, and the lack of direct observations in the high-energy $\beta$ regime ($E > E_{\rm p}$) makes $\beta$ values even more uncertain. In contrast, the cutoff energy $E_{\rm c}$ from the CPL model directly reflects the low-energy component and is closely related to parameters like the spectral index $\alpha$ and curvature $\kappa$. This connection makes $E_{\rm c}$ crucial in shaping the low-energy spectrum. In addition, we also test the Band function for estimating $E_{\rm p}^{'}$, both with a fixed $\beta$ and with $\beta$ as a free parameter, yielded suboptimal results for GRBs 140426A and 090510 (see Appendix). Practically, whether the intrinsic GRB spectrum is CPL-like or Band-like, $E_{\rm c}$ can be estimated from the spectral curvature of the low-energy component. The peak energy $E_{\rm p}$ can then be calculated using $E_{\rm p} = (2+\alpha)E_{\rm c}$. Therefore, our primary method relies on the CPL function.

\subsection{Sample Selection}\label{sec:Sample}

Our sample selection primarily relies on the Fermi/GBM burst catalog \citep{Li2023c} for its comprehensive coverage of GRBs with known redshifts and detailed spectral parameters. The detailed selection considerations:

1. All the bursts in our target sample are selected from the Fermi/GBM observation. The Fermi/GBM observational bandwidth encompasses the energy range of \emph{Swift}-BAT, allowing us to select the lower-energy portion corresponding to the BAT energy range to mimic \emph{Swift}-BAT fitting. The higher-energy data are then used to test our method for estimating $E_{\rm p}$ beyond the BAT energy limit.

2. Since the effective energy range of \emph{Swift}-BAT spans 15-150 keV, our sample excludes bursts with observed $E_{\rm p}$ values below 150 keV, as these are directly measurable within the BAT energy range.

3. To capture the intrinsic curvature of spectral data within the \emph{Swift}/BAT energy range, we prioritize bright bursts. This is because high-quality spectral data are more readily obtained from bright bursts compared to weaker ones, allowing for a more detailed investigation of the curvature features of the CPL function in the low-energy region below $E_{\rm p}$.

4. Previous studies \citep[e.g.,][]{Ryde2010} revealed that a significant fraction of GRBs may exhibit coexisting thermal (Planck function-like) and non-thermal (Band-like) spectral components, with the thermal component typically appearing as a sub-dominant feature on the left shoulder of the spectrum. A recent study \citep{Li2019c} suggest that the thermal component plays a crucial role in determining the observed Band-like $E_{\rm p}$ values and low-energy spectral index $\alpha$. To mitigate contamination from such thermal contributions, we ensure that our sample only includes cases where the CPL (or Band) model is the optimal spectral representation, excluding events like GRB 110721A \citep[e.g.,][]{Axelsson2012,Iyyani2013}, which is characterized by a prominent sub-thermal component.

5. To assess whether the effectiveness of our extrapolation method varies with increasing $E_{\rm c}$ values, we select 10 GRBs (see the upper-panel of Table \ref{tab:CPLresults}) from \cite{Li2023c} that exhibit gradually increasing observed $E_{\rm c}$ values and the optimal spectral representation is the CPL model. Moreover, the low-energy spectral index $\alpha$ could also affect $E_{\rm c}$ estimations, we therefore select $\alpha$ is approximately around -1, with minimal fluctuations, specifically within the range of -0.80 to -1.20. This is because we aimed to ensure that our selected bursts have consistent $\alpha$ values thereby isolating $E_{\rm c}$ as the primary variable in our analysis. These bursts were incorporated into the sample and are collectively referred to as Group I in our analysis. 

6. Similarly, to evaluate whether the effectiveness of our extrapolation method is influenced by varying $\alpha$ values, we select a subset of GRBs including two interesting events (see the lower-panel of Table \ref{tab:CPLresults}) from \cite{Li2023c} that exhibit a gradual change in observed $\alpha$ values and the optimal spectral representation is the CPL model. As the observed $E_{\rm p}$ might also affect the accuracy of the estimated $E^{'}_{\rm p}$, we focus on GRBs with observed $E_{\rm p}$ values approximately within 200-400 keV, ensuring minimal fluctuations. These bursts are incorporated into our sample and are collectively referred to as Group II in the subsequent analysis. 

7. A similar test is also applied to events where the optimal spectral representation is the Band model, including five interesting bursts (see Table \ref{tab:Bandresults}). These bursts are included in our sample and are collectively referred to as Group III in the subsequent analysis. 

\subsection{Modeling Peak Energies Beyond 150 keV}\label{sec:Modeling}

We develop a method that integrates several fixed-$E_{\rm p}$ spectral fittings, Markov Chain Monte Carlo (MCMC) sampling, and information criterion analysis to explore the parameter space and infer a robust lower limit on $E_{\rm p}$ when it exceeds the instrument's upper energy threshold. As detailed in Section \ref{sec:Methodology}, our approach employs the CPL function, which parametrizes the photon flux $N(E)$ as a combination of a smoothly connected power-law and an exponential decay. Main steps include:

1. To retrieve the original spectral data, we employed {\tt 3ML} ({\tt the Multi-Mission Maximum Likelihood Framework}, see \citealt{Vianello2015}), following to the standard procedures \citep{Li2019b,Li2019c,Yu2019,Burgess2019,Li2020,Li2021a,Li2021b} outlined by the Fermi Science Term. These procedures include the selection of detectors, sources, and background intervals. We rebin the spectral data for each GRB based on its observed characteristics.  For bright bursts, we set the photon count per bin to 100 for the NAI detector and 50 for the BGO detector. For relatively weaker bursts, these thresholds were adjusted downward as needed to maintain an acceptable signal to noise ratio ($>3$) of each bin, excessively large error bars were addressed by increasing photon counts per bin. This procedure ensures error bars are reasonable and spectral data points are sufficient, which are critical for capturing the intrinsic spectral curvature.

2. To estimate a lower limit on $E_{\rm c}$, we fix $E_{\rm c}$ at a series of values extending beyond the BAT's upper limit (e.g., 22 values from 150 keV to several MeV). For each fixed $E_{\rm c}$, we use MCMC sampling to explore the posterior distributions of the free parameters ($A$, $\alpha$, and $\beta$). We employ the affine-invariant ensemble sampler \citep{Goodman2010}, which is efficient for exploring complex parameter spaces. Typically, 20 walkers are used with 6000 iterations per walker. The first 3000 iterations are discarded as burn-in to ensure convergence. This process provides robust posterior estimates for the model parameters.

3. For each fixed $E_{\rm c}$, we compute several statistical criteria to assess the goodness-of-fit while accounting for model complexity: Akaike Information Criterion (AIC, \citealt{Akaike1974}):
\begin{equation}
\text{AIC} = 2k-2{\rm ln}\mathcal{L},
\end{equation}
where $k$ is the number of free parameters in the model and $\mathcal{L}$ is the maximized value of the likelihood function for the model.
Bayesian Information Criterion (BIC, \citealt{Schwarz1978}):
\begin{equation}
\text{BIC} = k{\rm ln} N-2{\rm ln}\mathcal{L},
\end{equation}
where $k$ is the number of model parameters, and $N$ is the number of data points, $\mathcal{L}$ is the maximum likelihood. Deviance Information Criterion (DIC, \citealt{Spiegelhalter2002}):
\begin{equation}
\text{DIC} =-2\text{log}[p(\text{data}\mid\hat{\theta})]+2p_{\rm DIC},
\end{equation}
where $\hat{\theta}$ is the posterior mean of the parameters over the MCMC samples, and $p_{\rm DIC}$ is a term to penalize the more complex model for overfitting \citep{Gelman2014}.
Reduced Chi-squared ($\chi^2_\nu$):
\begin{equation}
\chi^2_\nu = \frac{1}{\nu} \sum_{i=1}^{n} \left( \frac{O_i - E_i}{\sigma_i} \right)^2= \frac{\chi^2}{\text{DOF}},
\end{equation}
where $O_i$ and $E_i$ are the observed and expected counts, $\sigma_i$ are the measurement uncertainties, and $\nu = (n - k)$ is the degrees of freedom (DOF), $\chi^2$ is the sum of squared residuals normalized by the uncertainties, and $\text{DOF}$ (degrees of freedom) is the number of data points minus the number of fitted parameters.

4. We analyze how the information criteria vary with changing fixed $E_{\rm c}$. As $E_{\rm c}$ exceeds the true peak or cutoff energy, improvements in fit quality diminish, creating a ``bending point'' or plateau in the criteria curves ($E_{\rm c}$ versus Information Criterion). This bending point provides a robust lower limit for $E_{\rm c}$. To estimate the uncertainty in this lower limit, we examine the spread of bending points across different criteria. Additionally, the posterior distributions from MCMC sampling provide credible intervals for $A$ and $\alpha$ at each fixed $E_{\rm c}$.

5. To quantitatively evaluate the quality of the model comparisons, we primarily adopt DIC and also referred to Reduced Chi-squared statistic ($\chi^2_{\text{r}}$) for supplementary insight. The choice of DIC as the main criterion is motivated by the full application of Bayesian analysis and MCMC sampling in this study. The difference in DIC values between two models is defined as $\Delta \text{DIC} = \text{DIC}_1 - \text{DIC}_2$ and serves an indicator of model preference. Four levels are commonly used to interpret $\Delta \text{DIC}$. 
\begin{itemize}
\item Level 1 ($\Delta \text{DIC} < 2$): The two models are statistically indistinguishable. 
\item Level 2 ($2 \leq \Delta \text{DIC} < 5$): Model 2 is weakly preferred over Model 1.
\item Level 3 ($5 \leq \Delta \text{DIC} < 10$): Model 2 is significantly better than Model 1
\item Level 4 ($\Delta \text{DIC} \geq 10$):  Model 2 is strongly favored over Model 1. A difference of $\Delta \text{DIC} \geq 10$ is often considered robust evidence favoring one model over another. 
\end{itemize}
These thresholds are consistent with the principles of other model selection criteria, such as AIC and BIC, and are well-supported by Bayesian model selection theory \citep{Spiegelhalter2002, Burnham2004}. The advantage of DIC lies in its ability to handle hierarchical models and account for model complexity through ($p_{\rm D}$).

Additionally, the reduced chi-squared statistic ($\chi^2_{\text{r}}$) is a commonly used measure of model fit in observational sciences and the absolute value of $\chi^2_{\text{r}}$ should be considered. Ideally, $\chi^2_{\text{r}} \approx 1$, indicating that the residuals are consistent with the assumed uncertainties. However, the interpretation of $\chi^2_{\text{r}}$ depends on factors such as dataset size and the uncertainty estimation. For instance, a $\chi^2_{\text{r}}$ slightly above 1 may be acceptable for datasets with high degrees of freedom. Typical interpretation of $\chi^2_{\text{r}}$ include: $\chi^2_{\text{r}} \approx 1$: The model fit well, with residuals appropriately reflecting uncertainties. $\chi^2_{\text{r}} < 1$: The model may be overfitting the data, potentially overestimating complexity or uncertainty. $\chi^2_{\text{r}} > 1$: The model may be underfitting, failing to capture the observed variance. 

When comparing two fits, the difference in reduced chi-squared values, denoted as $|\Delta \chi^2_{\text{r}}|$, can provide insights into model preference. For statistical equivalence between two models, $|\Delta \chi^2_{\text{r}}|$ should be close to 0. The $\chi^2$ statistic follows a chi-squared distribution, allowing the calculation of a corresponding $p$-value to assess the model's compatibility with the data. A $p$-value $< 0.05$ suggests that the model does not adequately describe the observations, implying the null hypothesis (that the mode fits the data) is unlikely. This dual consideration of $\chi^2_{\text{r}}$ and $p$-value provides a comprehensive framework for evaluating model fits and comparing alternatives in a statistically rigorous manner.

\section{Results}\label{sec:results}

Existing catalogs of GRB spectral parameters from wide-bandwidth satellites, such as BATSE \citep{Preece2000, Goldstein2013} and GBM \citep{Goldstein2012, Gruber2014}, reveal that the typical peak energy $E_{\rm p}$ distribution lies around 200-300 keV. This value significantly exceeds 150 keV, indicating that for a substantial fraction of GRBs observed by \emph{Swift}, their $E_{\rm p}$ values cannot be determined using \emph{Swift} data alone.

\begin{figure*}
\centering
\includegraphics[angle=0,scale=0.7]{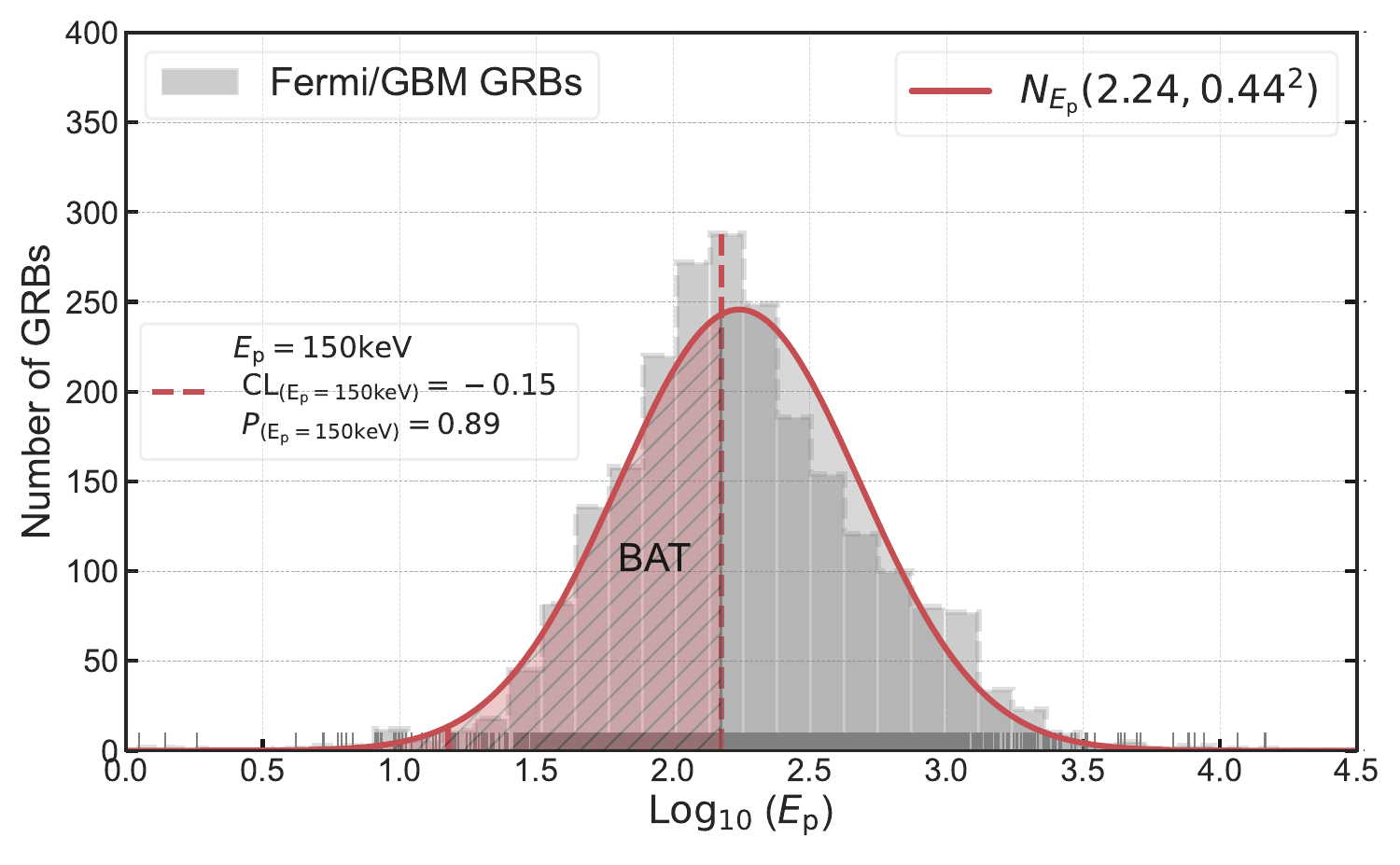}
\caption{Distribution of $E_{\rm p}$ for the complete Fermi/GBM burst catalog (grey shaded region). The red solid lines represent its best Gaussian fits. The $E_{\rm p}$ distribution covers by the BAT energy range is highlighted by the red slash-shaded region.}\label{fig:Ep_dis}
\end{figure*}

Figure \ref{fig:Ep_dis} shows the $E_{\rm p}$ distribution for the complete Fermi/GBM burst catalog, available at the NASA/HEASARC database\footnote{\url{https://heasarc.gsfc.nasa.gov/W3Browse/fermi/fermigbrst.html}}. This catalog (represented by the grey shaded area) includes a total of 3762 bursts (as of May 2024). The $E_{\rm p}$ distribution is fitted with a log-normal distribution, $N(\mu,\sigma^{2})$, where $\mu$ and $\sigma$ represent the mean and standard deviation, respectively. The best-fit parameters yield $N(2.24,0.44^{2})$, shown by the red line in Figure \ref{fig:Ep_dis}.

To estimate the fraction of bursts with $E_{\rm p}$ observed within the \emph{Swift}/BAT energy range (15-150 keV) based on the full Fermi/GBM burst sample, we define two separation lines respectively corresponding to $E_{\rm p}=15$ keV and $E_{\rm p}=150$ keV, and the area enclosed by the two lines represents the BAT energy range (highlighted by the red slash-shaded region in Figure \ref{fig:Ep_dis}). The cumulative distribution for the BAT energy range, denoted as $F(E)=P_{(15 {\rm keV}\leq E_{\rm p}\leq 150 {\rm keV})}$, is calculated to be $F(E)=0.43$. This result indicates that only approximately 43\% of the GBM-detected bursts have their $E_{\rm p}$ values falling within the BAT energy range, while the remaining 57\% have their $E_{\rm p}$ values located outside this range.

\subsection{Simulation Results}

GRB spectra often exhibit curvature, a key feature requiring quantitative investigation. To study this, we conducted a series of simulations using the CPL model. First, a synthetic CPL spectrum was generated with typical parameters: normalization $\hat{A} = 1$, photon index $\hat{\alpha} = -1$, and cutoff energy $\hat{E}_\mathrm{cut} = 300\ \mathrm{keV}$. Using these baseline parameters, we simulated 500 spectral data points within the \emph{Swift}/BAT energy range (15-150 keV), adding Gaussian noise with a standard deviation of 10\% of the intrinsic values. To assess the effect of varying $E_\mathrm{cut}$, we fixed $A$ and $\alpha$ and generated CPL spectra for $E_\mathrm{cut}$ values of 100, 120, 150, 200, 300, 500, and 1000 keV, producing seven spectra. To account for amplitude differences, $A$ was normalized for each $E_\mathrm{cut}$ to ensure comparability (details in Appendix). For each case, we calculated the spectral ratio (data-to-model ratio) and the corresponding residuals to examine deviations from the model.

\begin{figure}
\includegraphics[angle=0,scale=0.30]{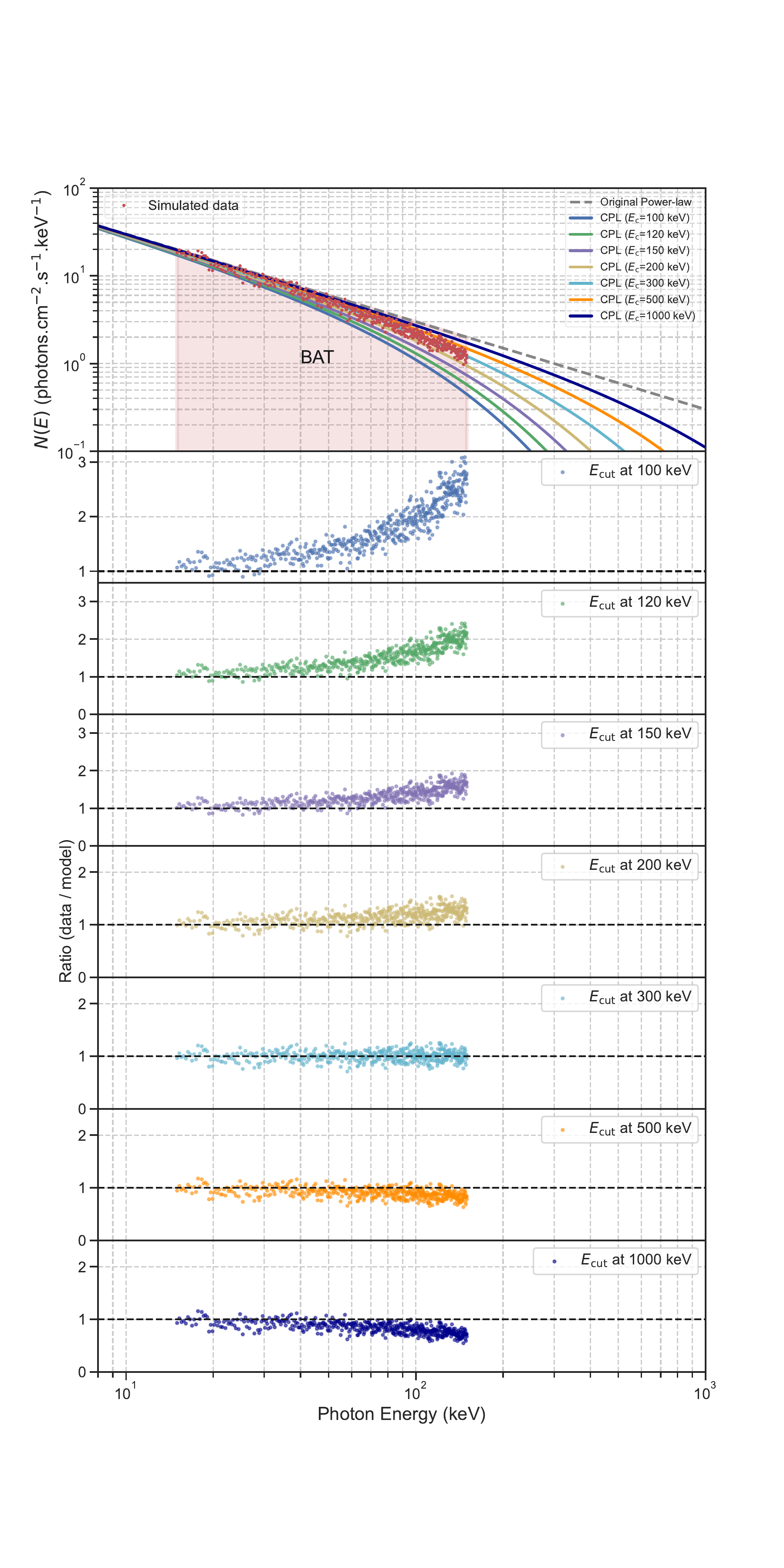}
\includegraphics[angle=0,scale=0.30]{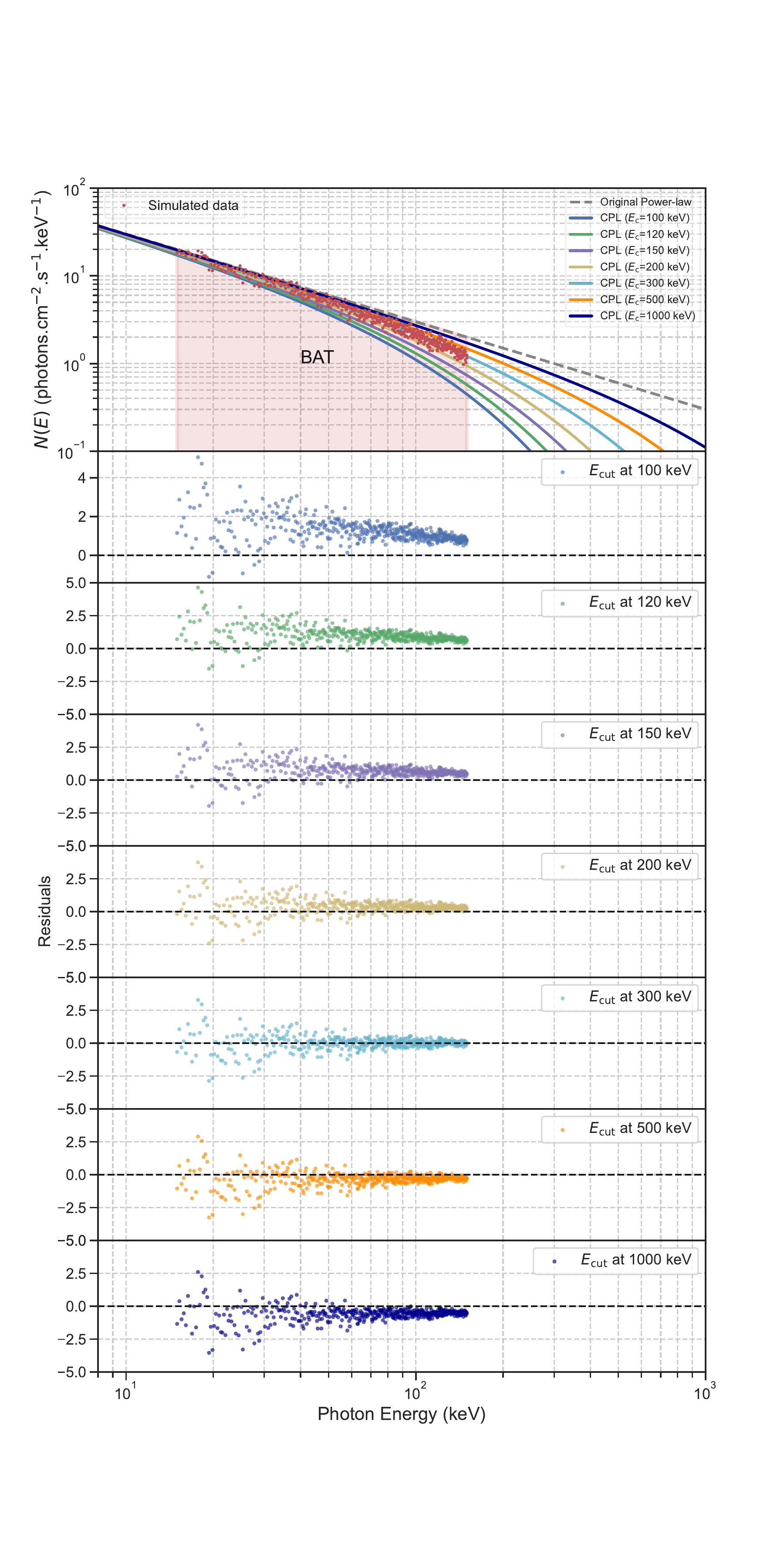}
\caption{Simulation results for GRB spectra using the CPL model. Left-panel: Top panel shows the simulated data points (red) in the 15--150 keV band, CPL model curves for various $E_\mathrm{cut}$ values (colored lines), and the power-law baseline (gray dashed line). The lower panels show the ratio evolution for $E_\mathrm{cut}$ values ranging from 100 keV to 1000 keV. Right-panel: Similar layout, but showing residual evolution for the same $E_\mathrm{cut}$ values.}\label{fig:Simulation}
\end{figure}

Figure~\ref{fig:Simulation} shows the simulation results. In the top panels of Fig.~\ref{fig:Simulation}a and Fig.~\ref{fig:Simulation}b, the red scatter points represent the 500 simulated data points in the 15--150 keV range based on the CPL model using typical model parameters, while the solid curves of various colors depict CPL spectra corresponding to different $E_\mathrm{cut}$ values across the broader 1--10,000 keV range. The gray dashed line indicates a power-law function with the same $A$ and $\alpha$ values as the CPL model, providing a comparative baseline. The BAT energy range is highlighted by the red slash-shaded region.

The lower panels of Figure~\ref{fig:Simulation}a present the ratio evolution for $E_\mathrm{cut} = 100$, 120, 150, 200, 300, 500, and 1000 keV, respectively. For each $E_\mathrm{cut}$, at low energies (near 15 keV), the ratios for all $E_\mathrm{cut}$ values converge to approximately 1, indicating a good agreement between the data and the model. However, at higher energies (close to 150 keV), significant deviations emerge. For $E_\mathrm{cut} < \hat{E}_{\mathrm{cut}}$ (e.g., $E_\mathrm{cut} = 100\ \mathrm{keV}$), the ratio increases above 1 as energy increases, reflecting a systematic overestimation of the data by the model. When $E_\mathrm{cut} = \hat{E}_{\mathrm{cut}}$, the ratio remains near 1 across the entire energy range, with minimal fluctuation. Conversely, for $E_\mathrm{cut} > \hat{E}_{\mathrm{cut}}$ (e.g., $E_\mathrm{cut} = 1000\ \mathrm{keV}$), the ratio decreases below 1 at higher energies, indicating an underestimation of the data. These trends underline the sensitivity of spectral ratios to variations in $E_\mathrm{cut}$, providing a diagnostic tool for identifying the most physically representative model parameters.  

The lower panels of Figure~\ref{fig:Simulation}(b) show the residual evolution for the same $E_\mathrm{cut}$ values. Notably, the scatter of residuals is larger at low energies (near 15 keV) compared to high energies (near 150 keV), particularly for smaller $E_\mathrm{cut}$ values (e.g., $E_\mathrm{cut} = 100\ \mathrm{keV}$). This behavior may reflect the combined effects of statistical noise and intrinsic model limitations at lower energy ranges. For $E_\mathrm{cut} < \hat{E}_{\mathrm{cut}}$, the residuals predominantly reside in the positive domain, implying systematic mismatches between the data and the model. When $E_\mathrm{cut} = \hat{E}_{\mathrm{cut}}$, the residuals fluctuate near zero, signifying excellent model agreement. For $E_\mathrm{cut} > \hat{E}_{\mathrm{cut}}$, the residuals are mostly negative, indicating systematic underfitting of the data by the model.  

This methodology allows us to constrain $E_{\rm c}$ even when it exceeds the instrument's energy range. The MCMC approach offers a full probabilistic characterization of the parameters. However, it is important to note that this method normally provides a lower limit for $E_{\rm c}$, rather than an exact value. Its precision depends on factors such as the signal-to-noise ratio of the data and how much the true $E_{\rm c}$ exceeds the instrument's capabilities.

\subsection{Application to Observational Data}\label{sec:AppObs}

When applying our method to real observational data, we address the following critical questions: How well does the $E^{'}_{\rm c}$ obtained via the extrapolation method correlated with the observed $E_{\rm c}$? Does $E^{'}_{\rm c}$ evolve with changes in $E_{\rm c}$? If so, what is the nature of this evolution? Does different $\alpha$ values affect the $E^{'}_{\rm c}$ results? To explore these questions, we conducted targeted data analysis as outlined below. In our analysis, we employ a two-step approach to determine the spectral peak energy ($E_{\rm p}$) and associated parameters: preliminary estimation and refined estimation.

\subsubsection{Preliminary Estimation of $E^{'}_{\rm c}$}\label{sec:PreEstimation}

When analyzing the spectrum of a BAT-detected GRB, the true value of $E_{\rm c}$ is unknown. Moreover, given that the distribution of $E_{\rm c}$ spans at least three orders of magnitude, preliminary estimation must be performed over a wide energy range with relatively large intervals between selected $E_{\rm c}$ values.  We therefore selected 22 $E_{\rm c}$ values ranging from 100 keV to 5000 keV for a preliminary estimation. These values are 100 keV, 120 keV, 150 keV, 180 keV, 200 keV, 220 keV, 250 keV, 300 keV, 400 keV, 500 keV, 600 keV, 700 keV, 800 keV, 900 keV, 1000 keV, 1200 keV, 1500 keV, 1800 keV, 2000 keV, 3000 keV, 4000 keV, and 5000 keV.

This preliminary step enables an efficient and straightforward determination of a rough $E_{\rm c}$ value. However, the downside is that the estimation may have a relatively large uncertainty. Once a rough $E_{\rm c}$ value is obtained, the next step involves a refined estimation by focusing on this value and exploring a narrower range with smaller intervals. This refined process allows for a more accurate determination of $E_{\rm c}$ within a smaller margin of error.

Another critical factor influencing the results is the selection criterion for the DIC. In our analysis, there exists a global minimum DIC value corresponding to the best-fit model. The difference between the DIC value of other models and this global minimum, $\Delta \text{DIC}$, provides an intuitive measure of the relative quality of model selection. If the selection criterion is too loose (e.g., $5<\Delta\text{DIC}\leq 10$), the estimated $E^{'}_{\rm c}$ derived from the selected model might be significantly lower than the true observed value $E_{\rm c}$, resulting in reduced estimation accuracy. Conversely, if the criterion is too stringent (e.g., $0<\Delta\text{DIC}\leq2$), the selected model becomes statistically indistinguishable from the best-fit model. While this improves accuracy, it undermines the conservative estimation of the lower limit of $E_{\rm c}$, as the selected $E^{'}_{\rm c}$ might even exceed the true observed $E_{\rm c}$. Therefore, we adopted an intermediate criterion, choosing the first model satisfying $2<\Delta \text{DIC}\leq 5$ to determine $E^{'}_{\rm c}$, which serves as a reliable lower limit estimation for $E_{\rm c}$.

\begin{figure*}
\centering
\includegraphics[angle=0,scale=0.40]{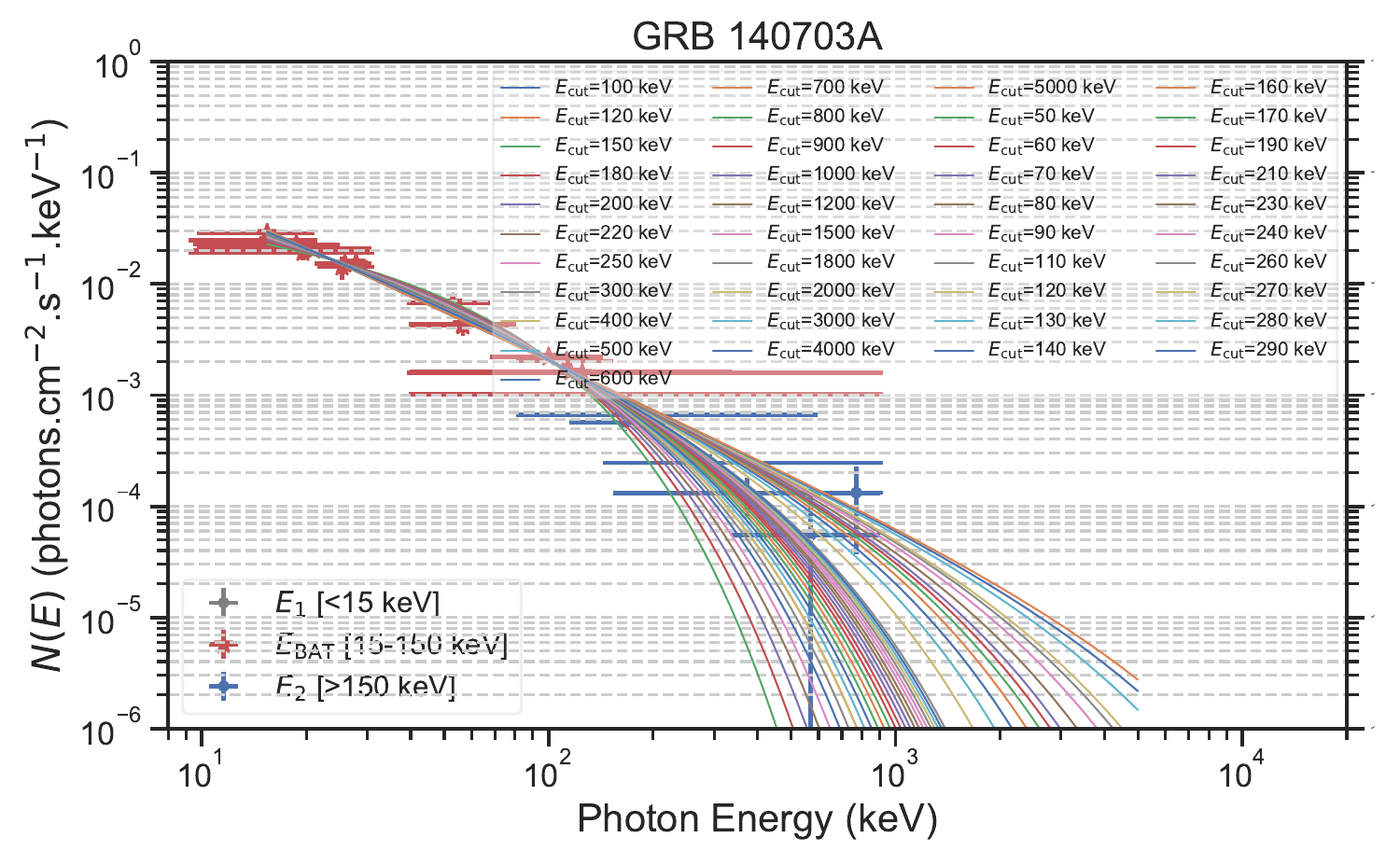}
\includegraphics[angle=0,scale=0.40]{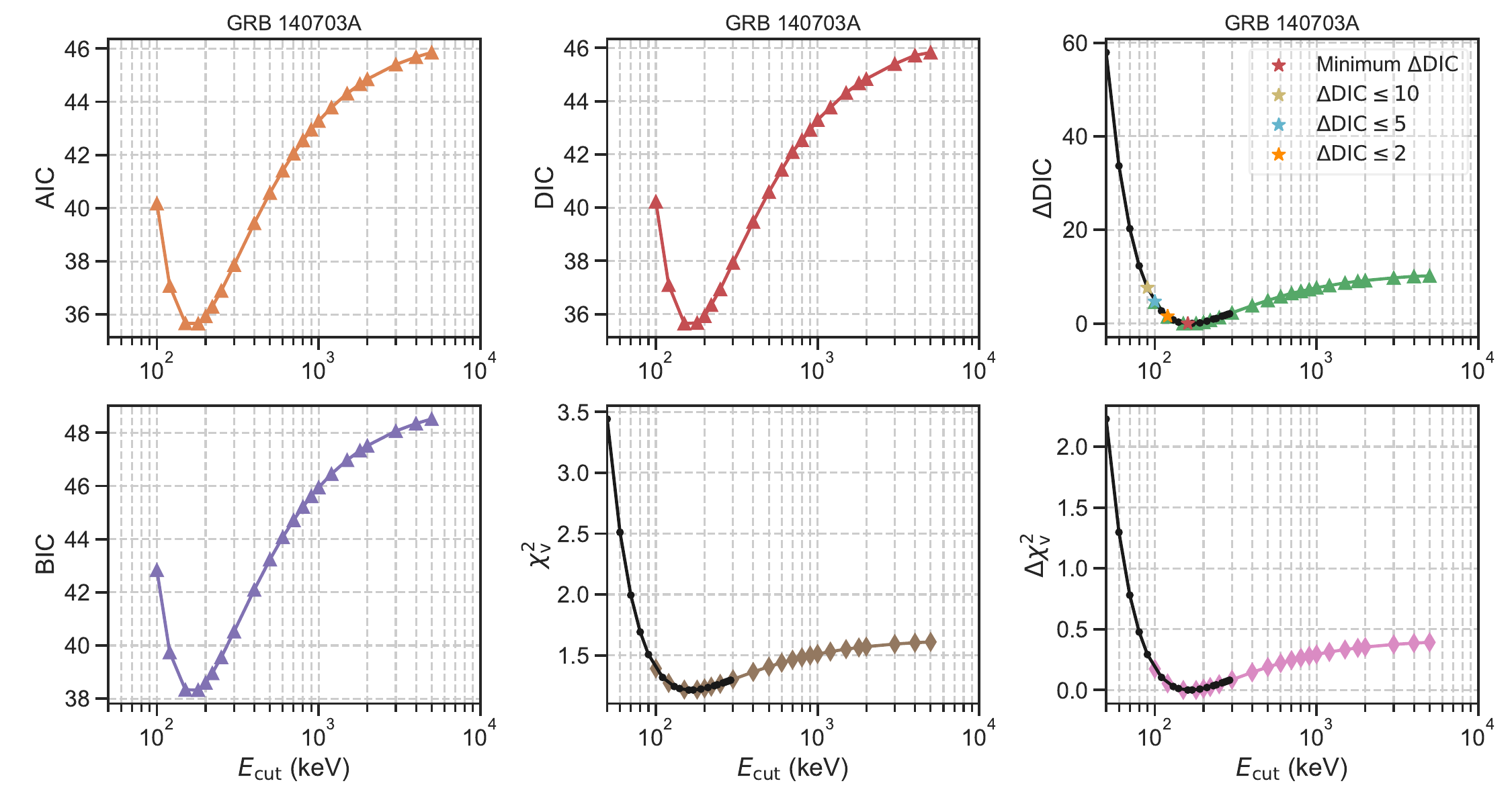}
\includegraphics[angle=0,scale=0.40]{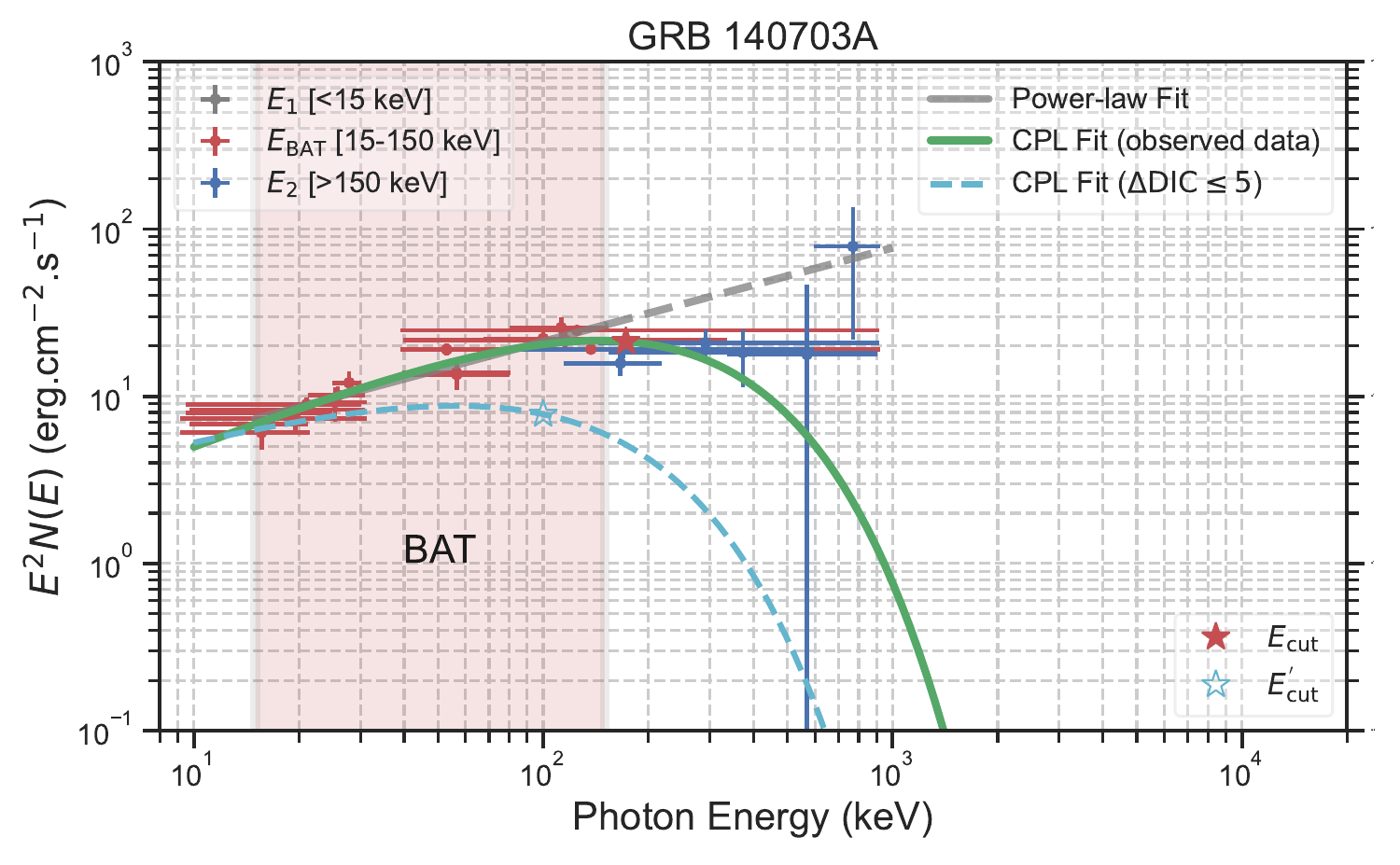}
\caption{The results of GRB 140703A. {\bf Top panel}: The spectral data obtained from the GBM observation, separated into three distinct energy ranges: $E_1$[$<$15] keV (grey), $E_{\rm BAT}$ [15-150 keV] (red), and $E_2$ [$>$150] keV (blue). Solid colored lines represent the CPL spectral model with varying $E_{\rm c}$ values (see Table \ref{tab:140703A}). {\bf Middle panel}: The statistical criteria (AIC, BIC, DIC, $\chi^{2}$) and their deviations ($\Delta$DIC, $\Delta$$\chi^{2}$) from the minimum value plotted as function as $E^{'}_{\rm c}$. Five-pointed stars in different colors denote the locations corresponding to various $\Delta$DIC thresholds used for evaluation on the $\Delta$DIC-$E_{\rm c}$ diagram. The data points indicated by black circles represent the refined estimations, while the others represent the preliminary estimations. {\bf Bottom panel}: The original spectral data from GBM observation (grey, red, and blue points) alongside the best CPL model fit (green solid line), with the fitted model parameters: $E_{\rm c}$=173$^{+51}_{-38}$ keV, $A$=(7.6$^{+3.3}_{-2.3}$)$\times$10$^{-1}$, $\alpha$=-1.16$^{+0.11}_{-0.12}$. The grey sold and dashed lines represent the best power-law fits to the spectral data within and outside the BAT energy range. Dashed lines in different colors correspond to CPL fit derived from model parameters selected using different $\Delta$DIC thresholds ($\Delta {\rm DIC} \leq 10$, $\Delta {\rm DIC} \leq 5$, $\Delta {\rm DIC} \leq 2$, and minimum $\Delta {\rm DIC}$).}\label{fig:140703A}
\end{figure*}

\begin{figure*}
\centering
\includegraphics[angle=0,scale=0.45]{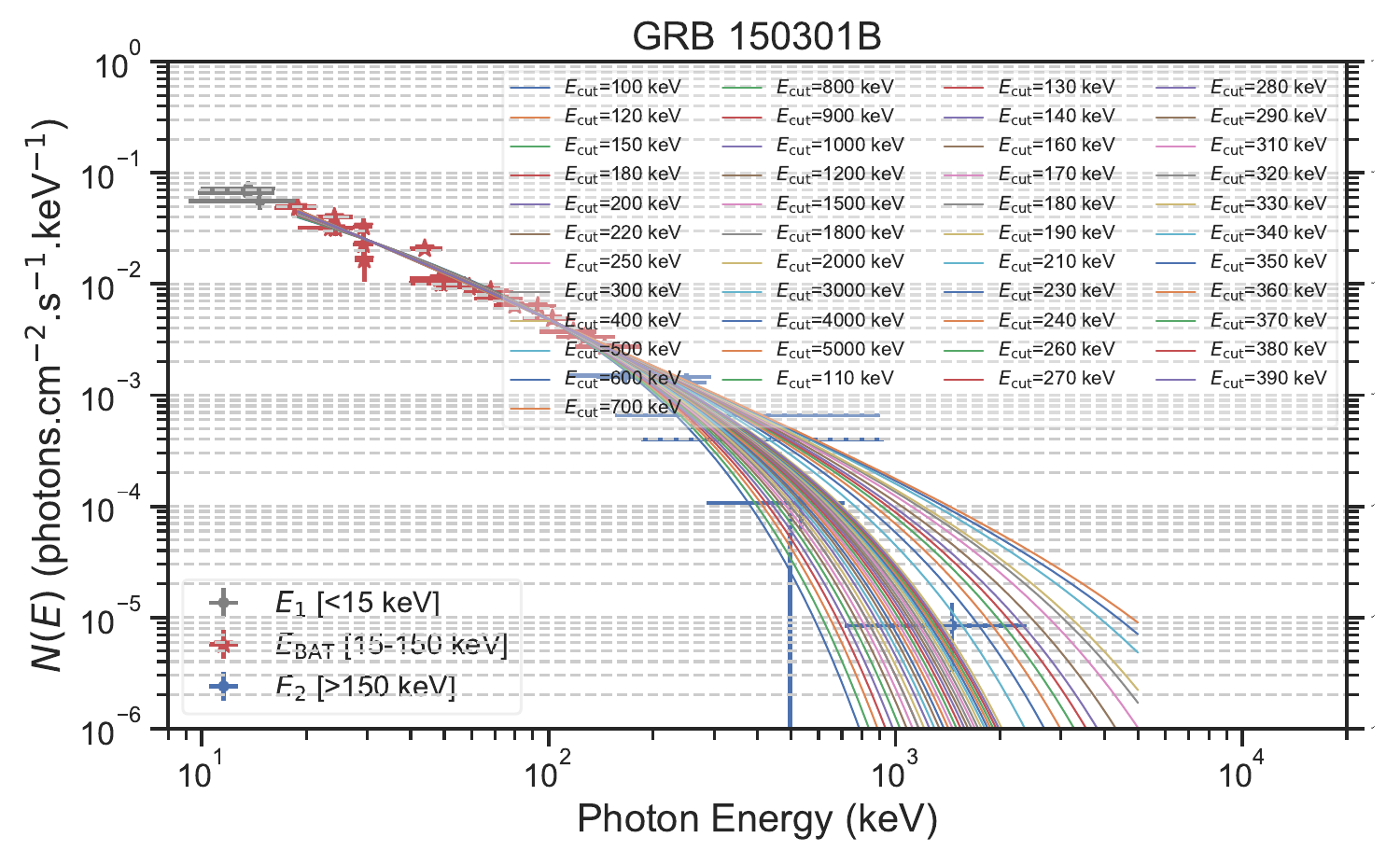}
\includegraphics[angle=0,scale=0.45]{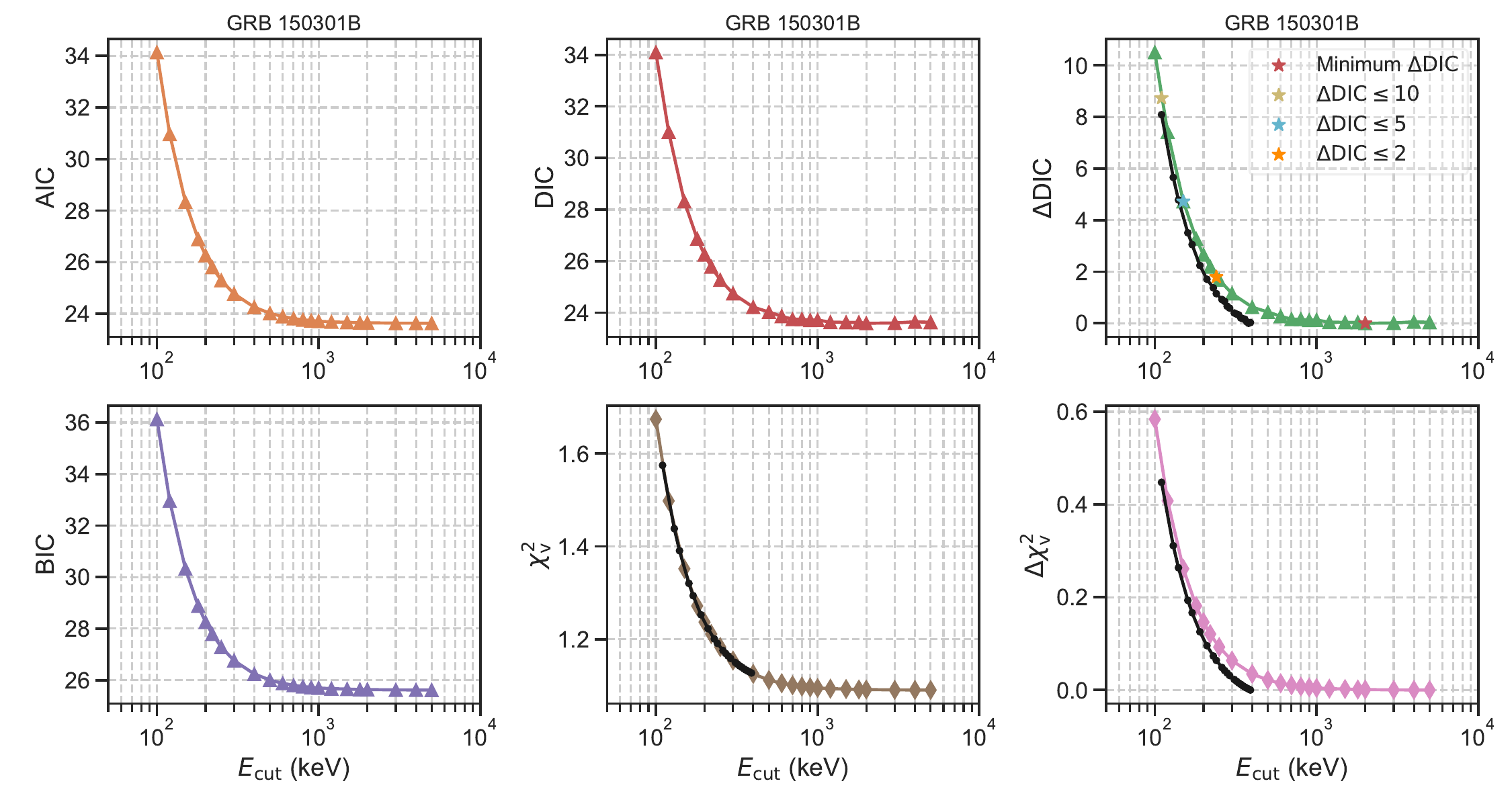}
\includegraphics[angle=0,scale=0.45]{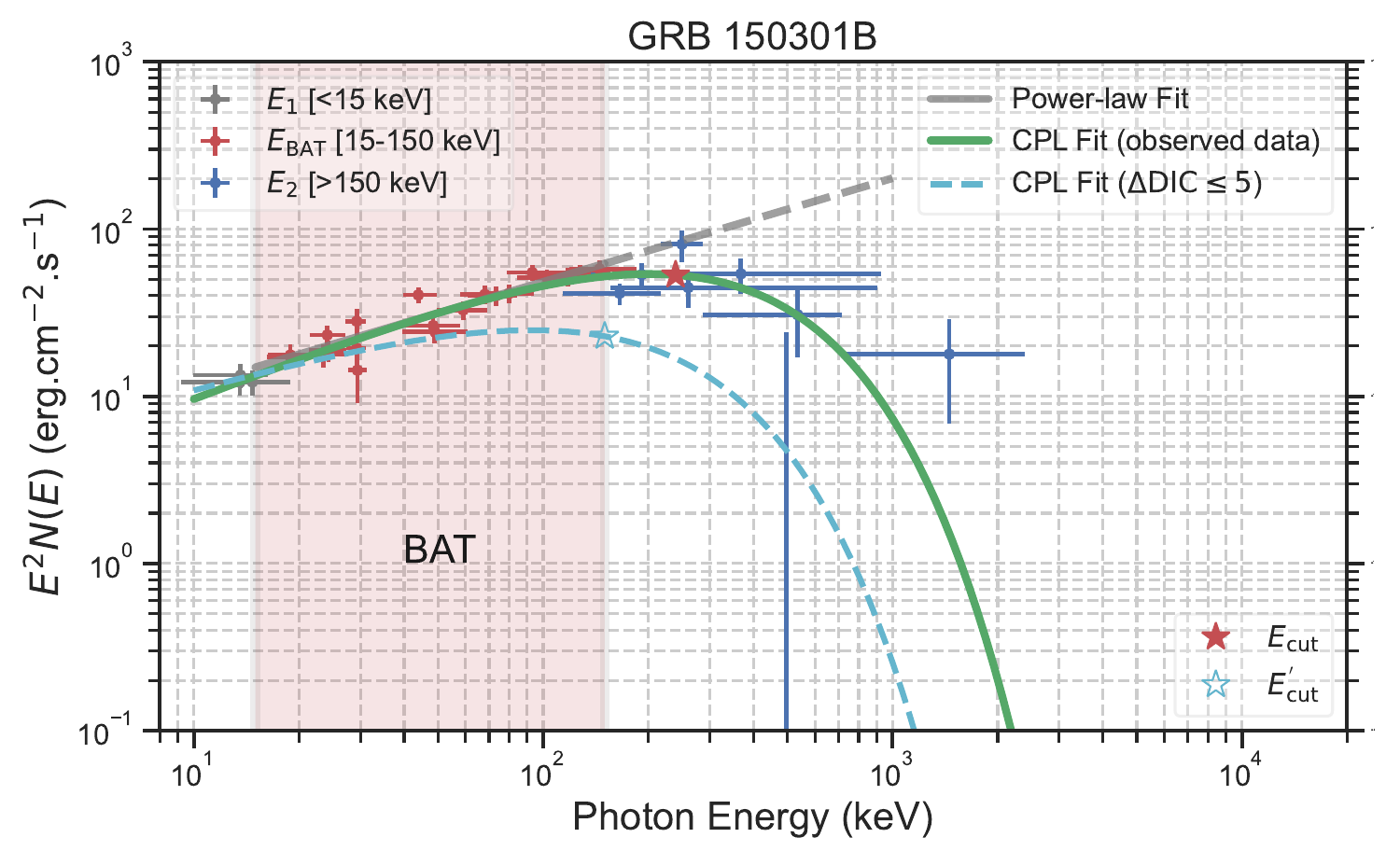}
\caption{Same as Figure \ref{fig:140703A} but for GRB 150301B. Note that the best CPL model fit to the original spectral data from GBM observation gives $E_{\rm c}$=239$^{+178}_{-100}$ keV, $A$=(1.5$^{+1.5}_{-0.7}$), $\alpha$=-1.16$^{+0.21}_{-0.22}$.}\label{fig:150301B}
\end{figure*}

\begin{figure*}
\centering
\includegraphics[angle=0,scale=0.45]{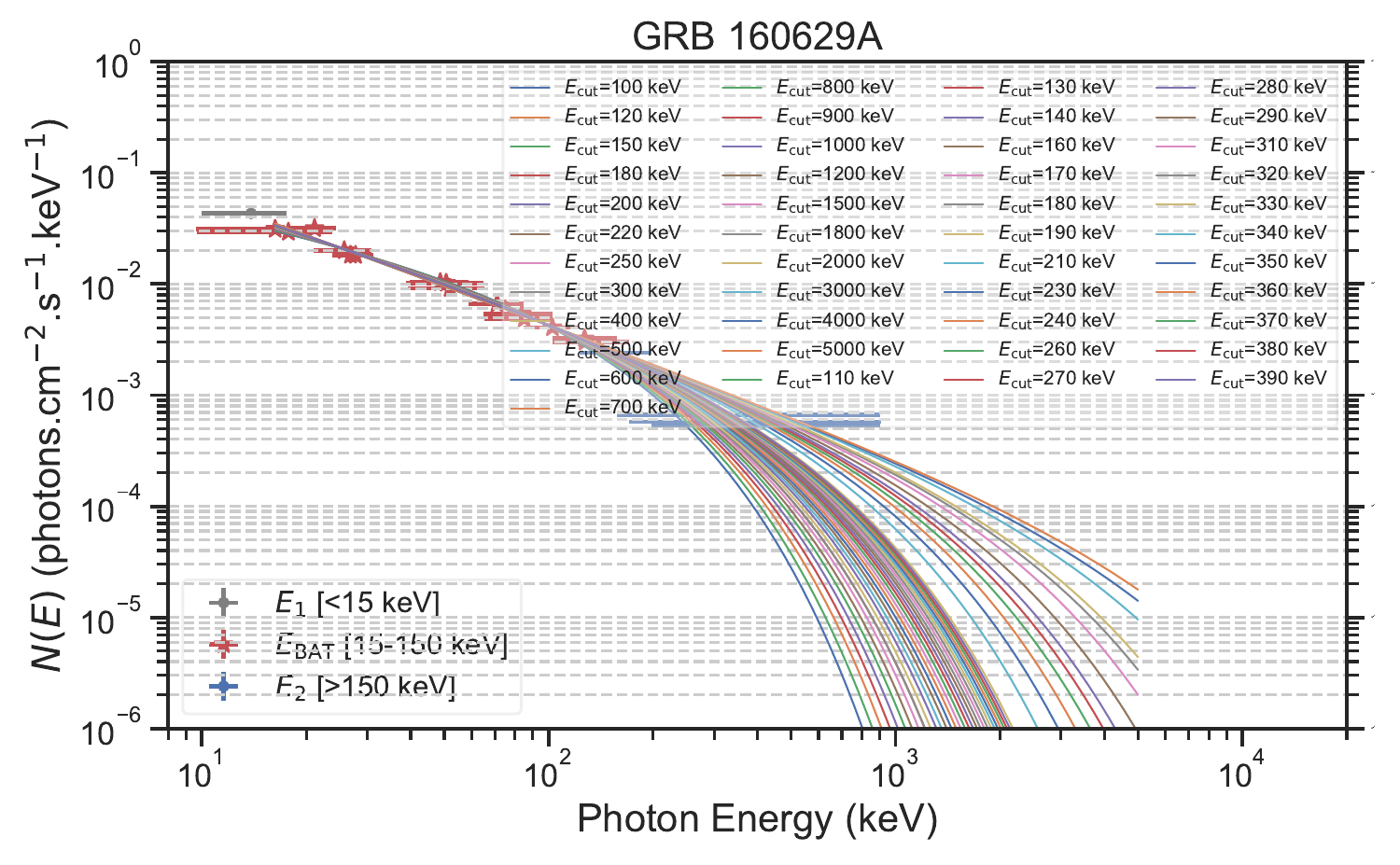}
\includegraphics[angle=0,scale=0.45]{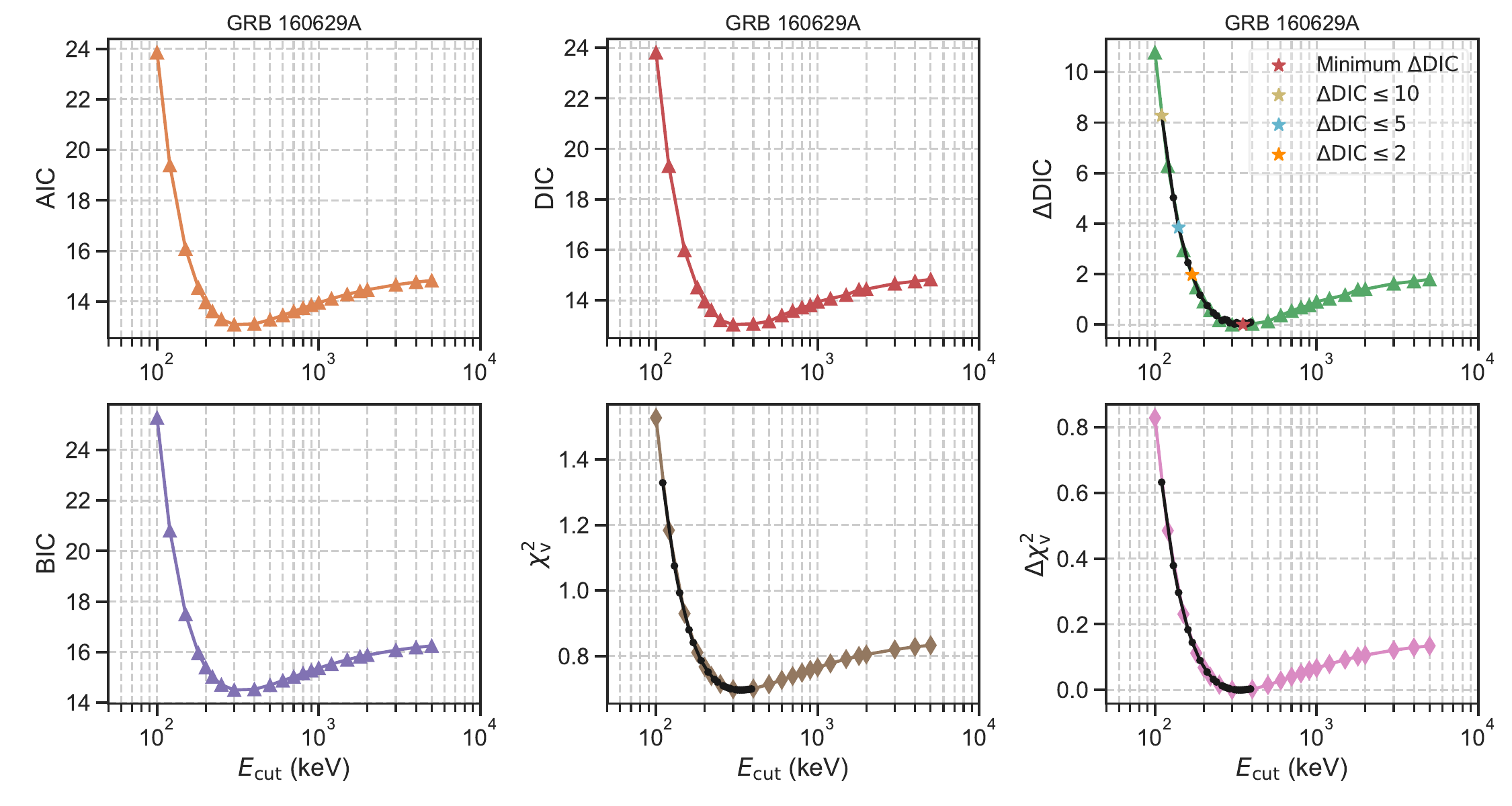}
\includegraphics[angle=0,scale=0.45]{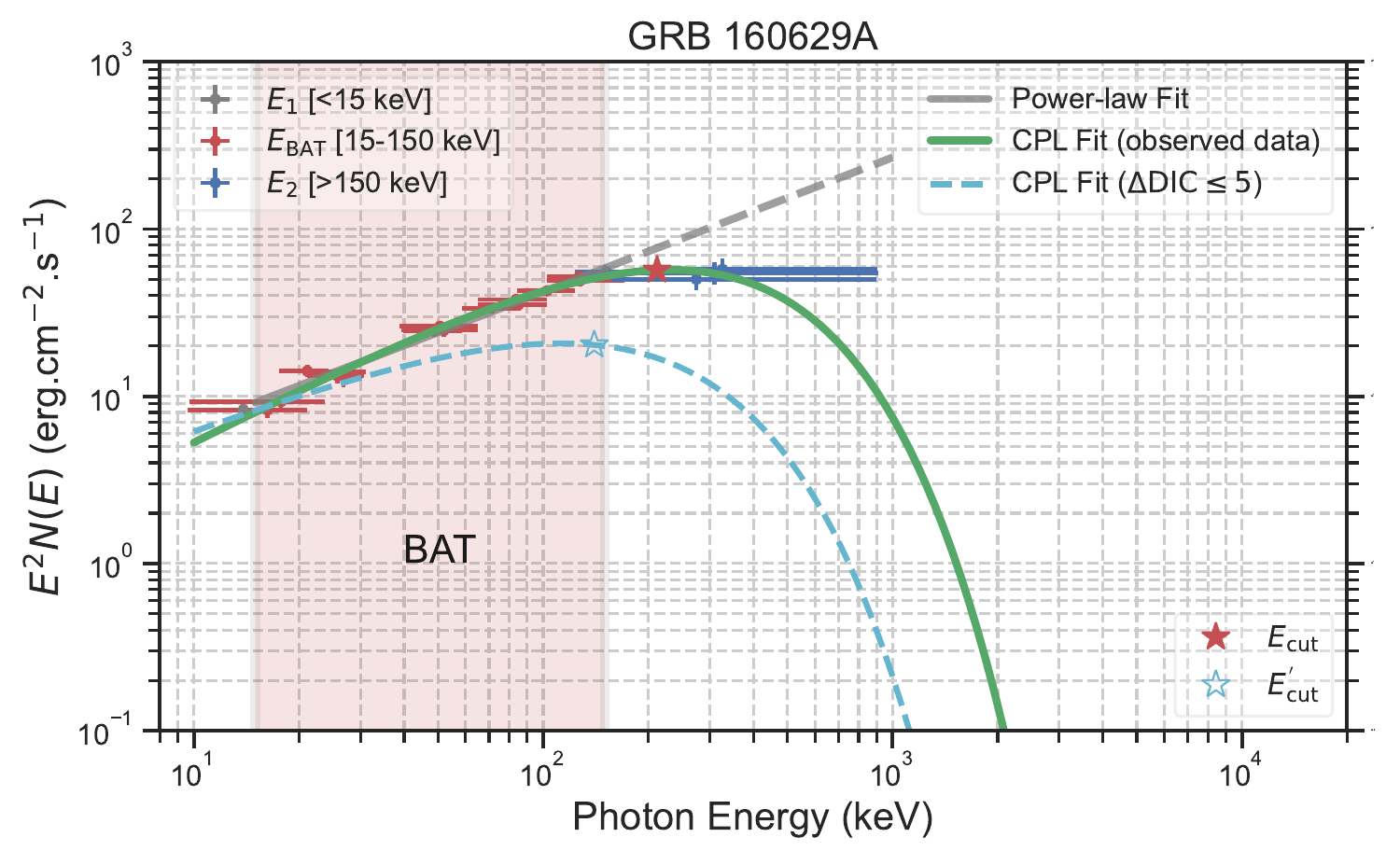}
\caption{Same as Figure \ref{fig:140703A} but for GRB 160629A. Note that the best CPL model fit to the original spectral data from GBM observation gives  $E_{\rm c}$=212$^{+64}_{-49}$ keV, $A$=(4.5$^{+2.0}_{-1.4}$)$\times$10$^{-1}$, $\alpha$=-0.91$^{+0.12}_{-0.11}$.}\label{fig:160629A}
\end{figure*}

\begin{figure*}
\centering
\includegraphics[angle=0,scale=0.45]{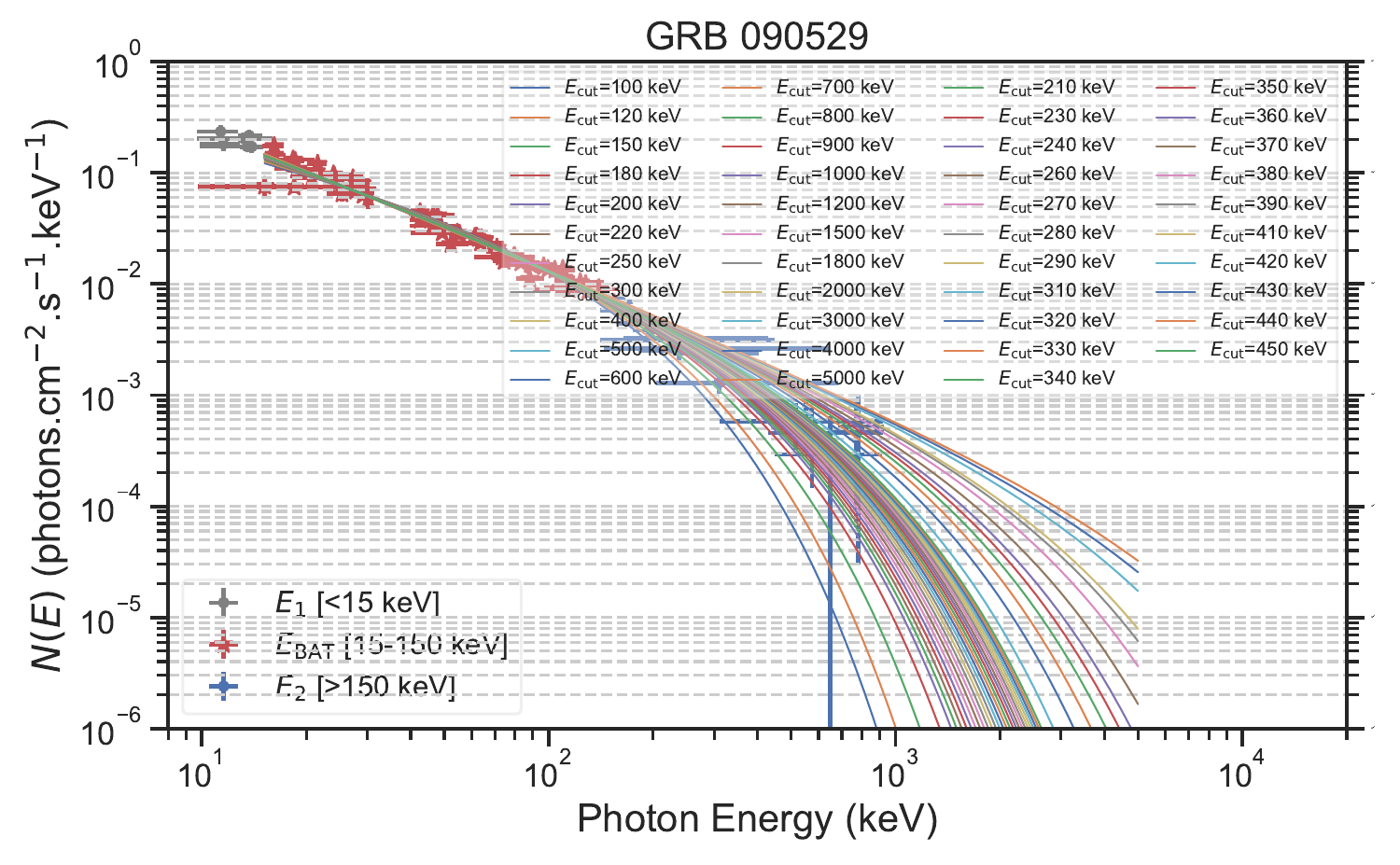}
\includegraphics[angle=0,scale=0.45]{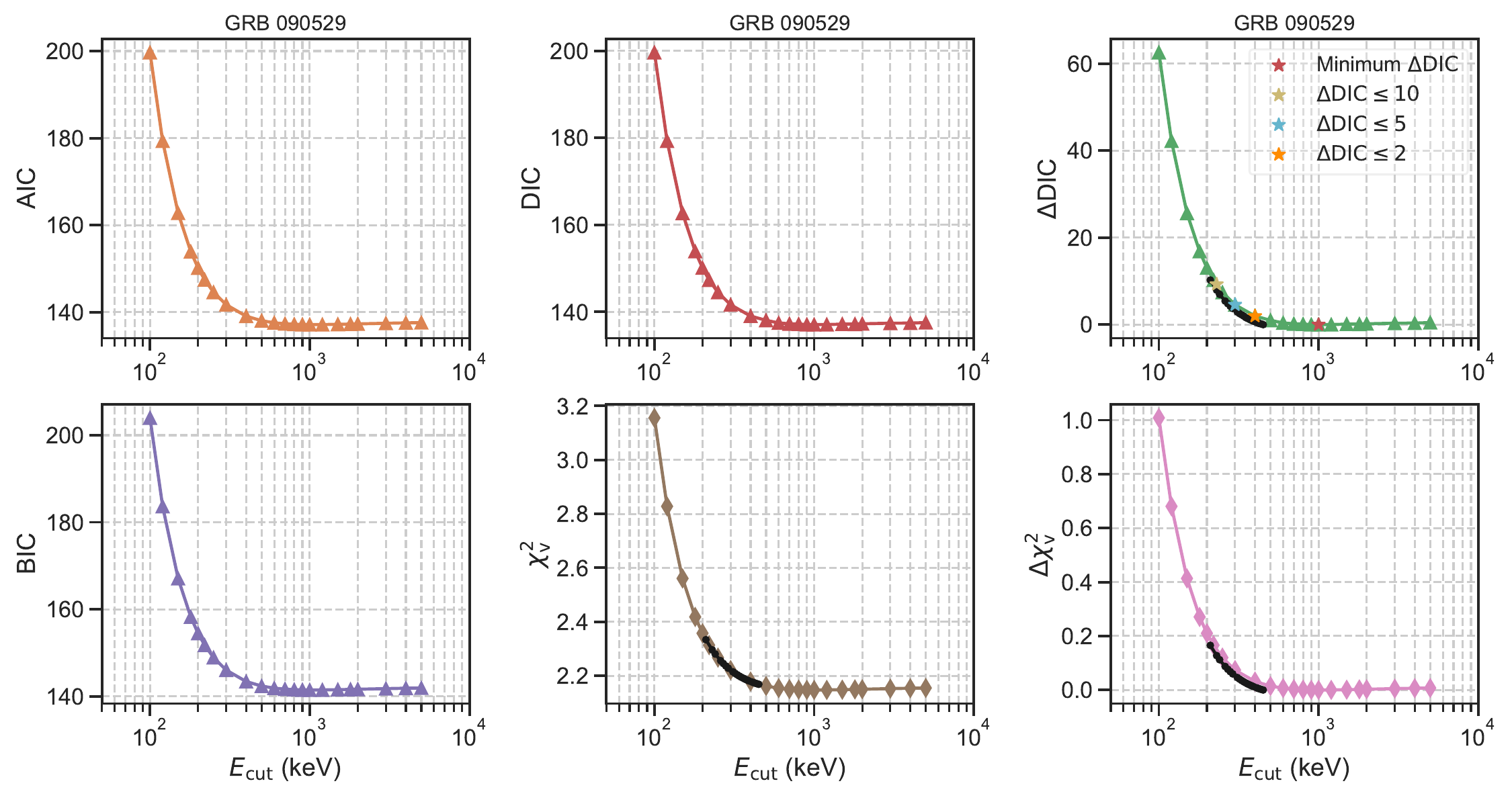}
\includegraphics[angle=0,scale=0.45]{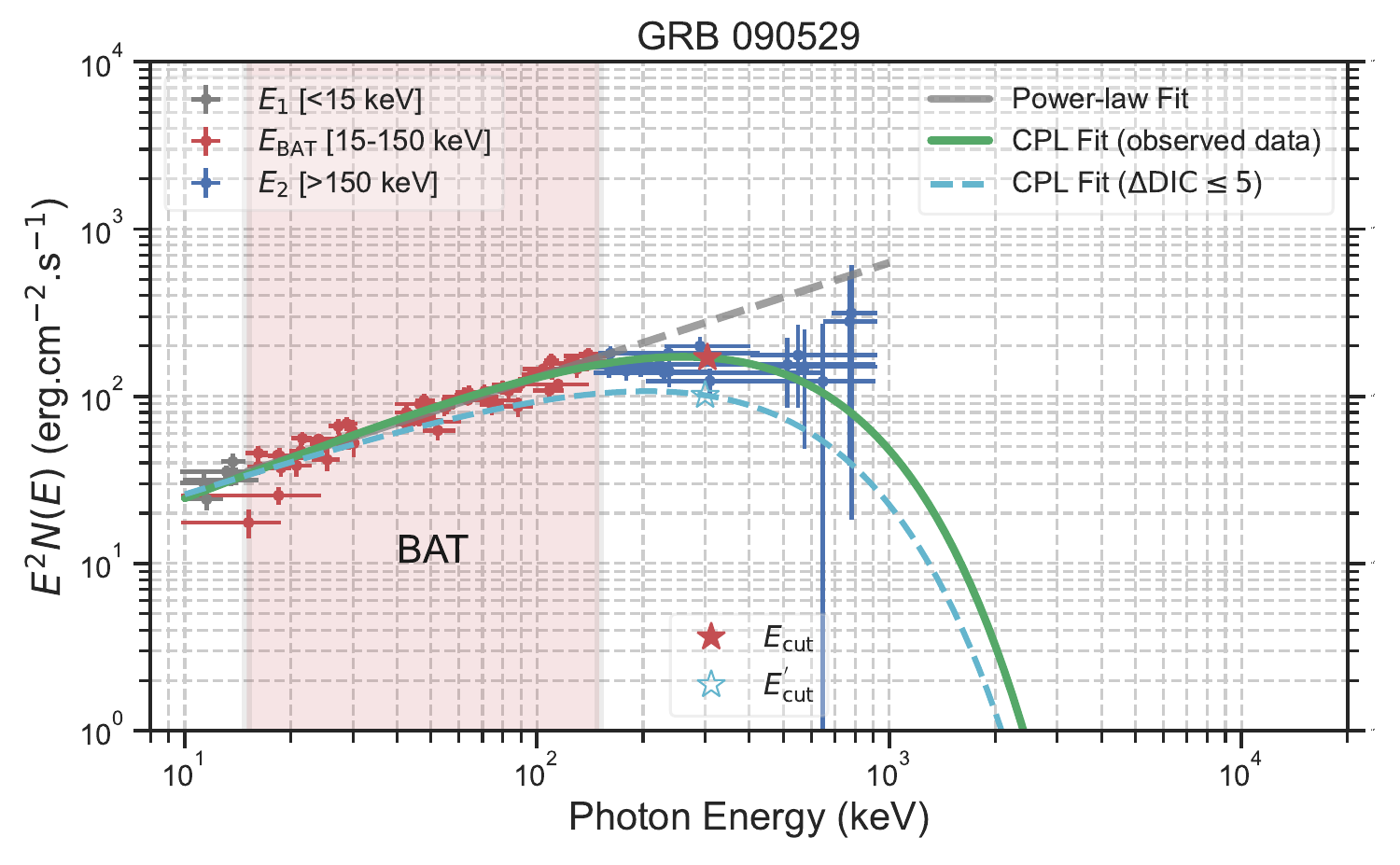}
\caption{Same as Figure \ref{fig:140703A} but for GRB 090529. Note that the best CPL model fit to the original spectral data from GBM observation gives $E_{\rm c}$=305$^{+66}_{-55}$ keV, $A$=(3.6$^{+0.8}_{-0.6}$), $\alpha$=-1.15$^{+0.06}_{-0.06}$.}\label{fig:090529}
\end{figure*}

\begin{figure*}
\centering
\includegraphics[angle=0,scale=0.45]{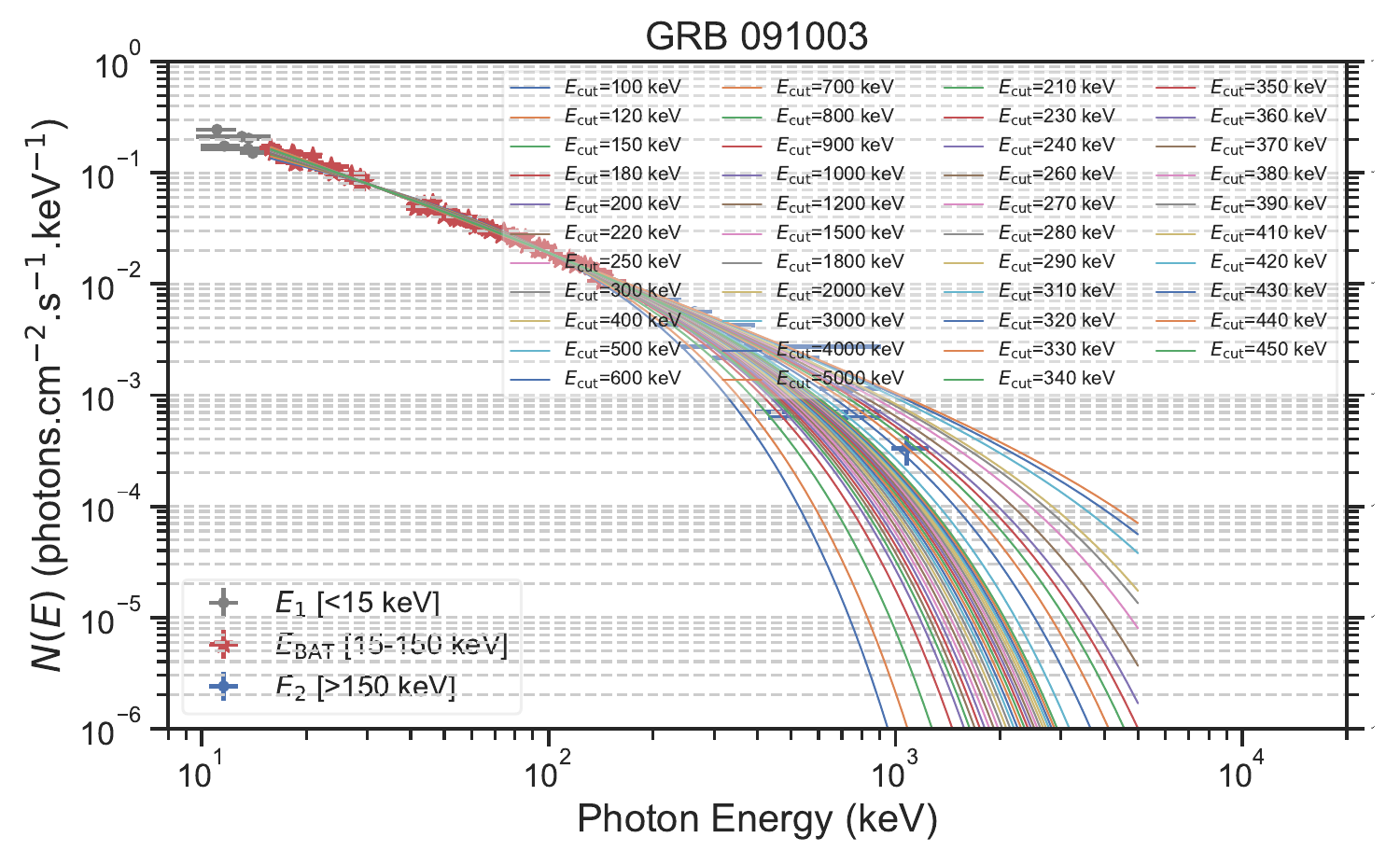}
\includegraphics[angle=0,scale=0.45]{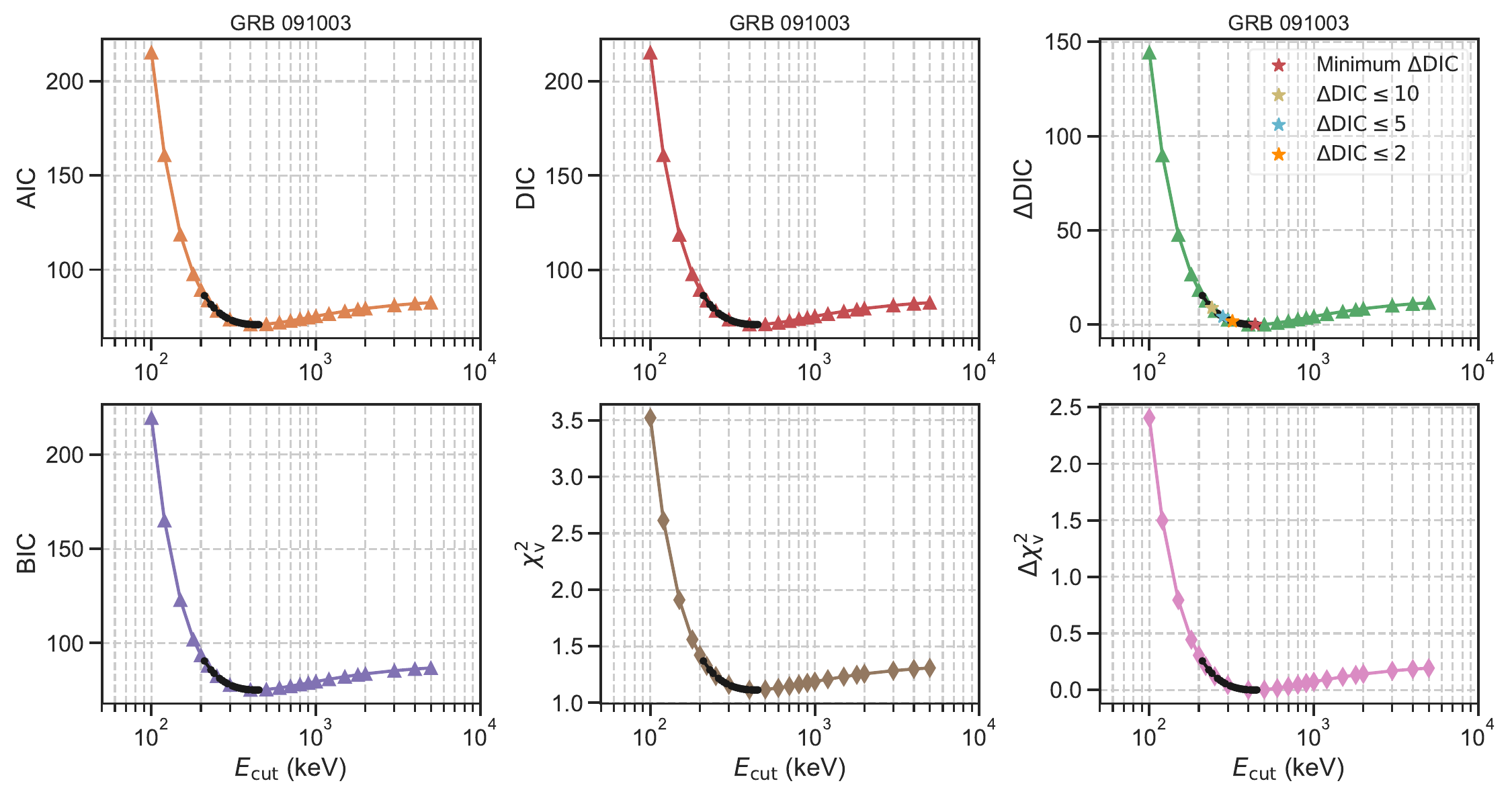}
\includegraphics[angle=0,scale=0.45]{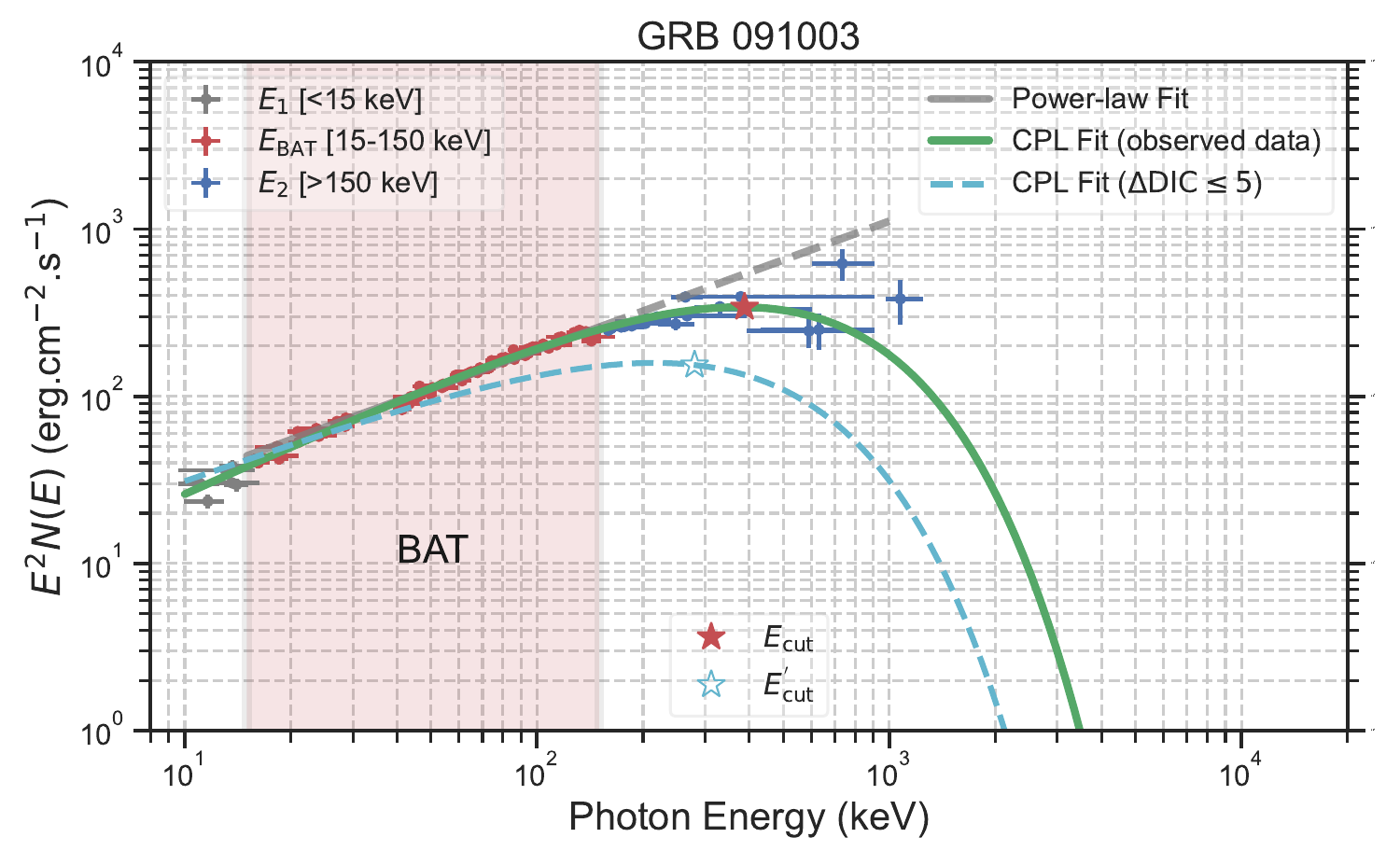}
\caption{Same as Figure \ref{fig:140703A} but for GRB 091003. Note that the best CPL model fit to the original spectral data from GBM observation gives $E_{\rm c}$=388$^{+41}_{-38}$ keV, $A$=(2.9$^{+0.3}_{-0.3}$), $\alpha$=-1.03$^{+0.03}_{-0.03}$.}\label{fig:091003}
\end{figure*}

\begin{figure*}
\centering
\includegraphics[angle=0,scale=0.45]{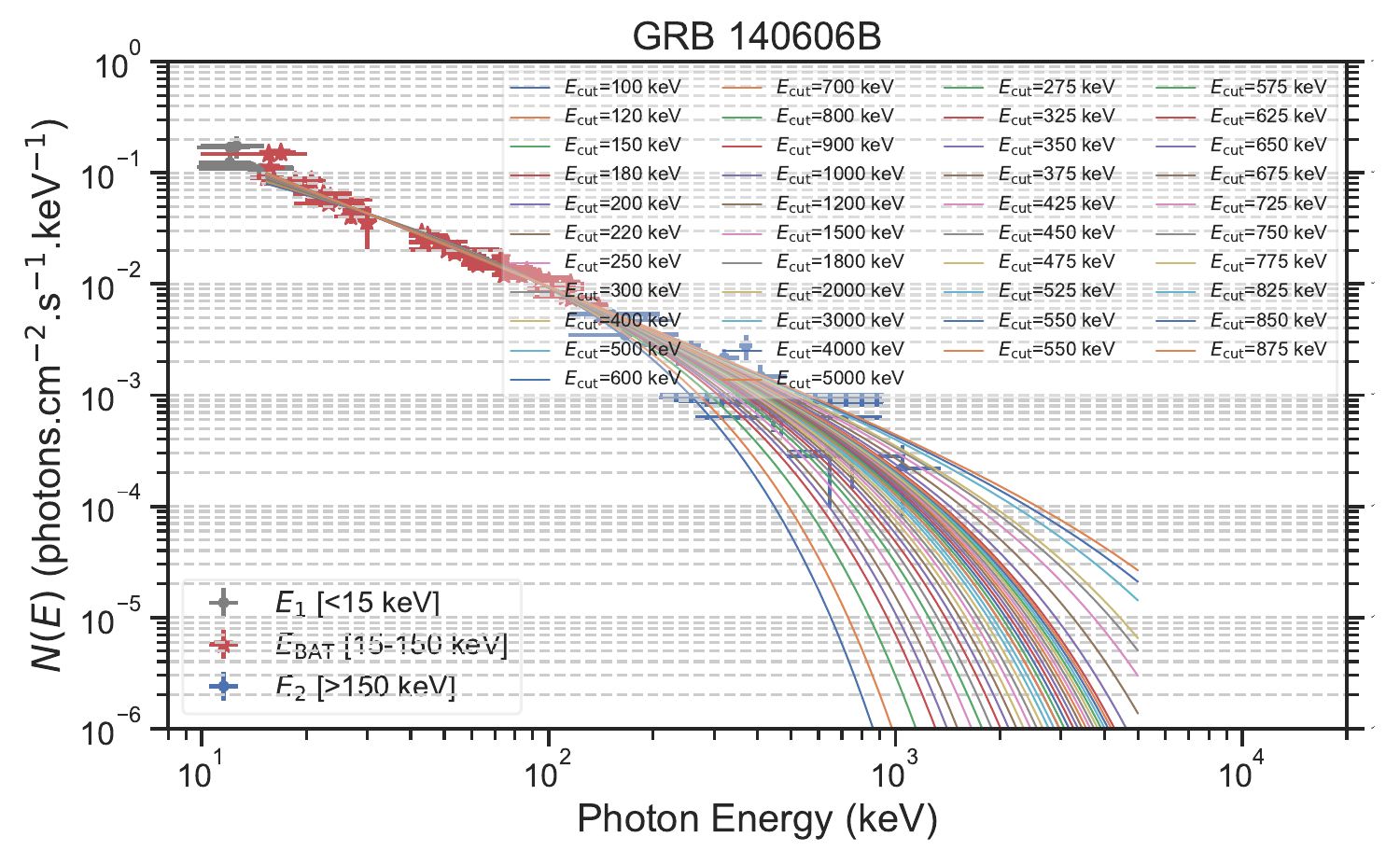}
\includegraphics[angle=0,scale=0.45]{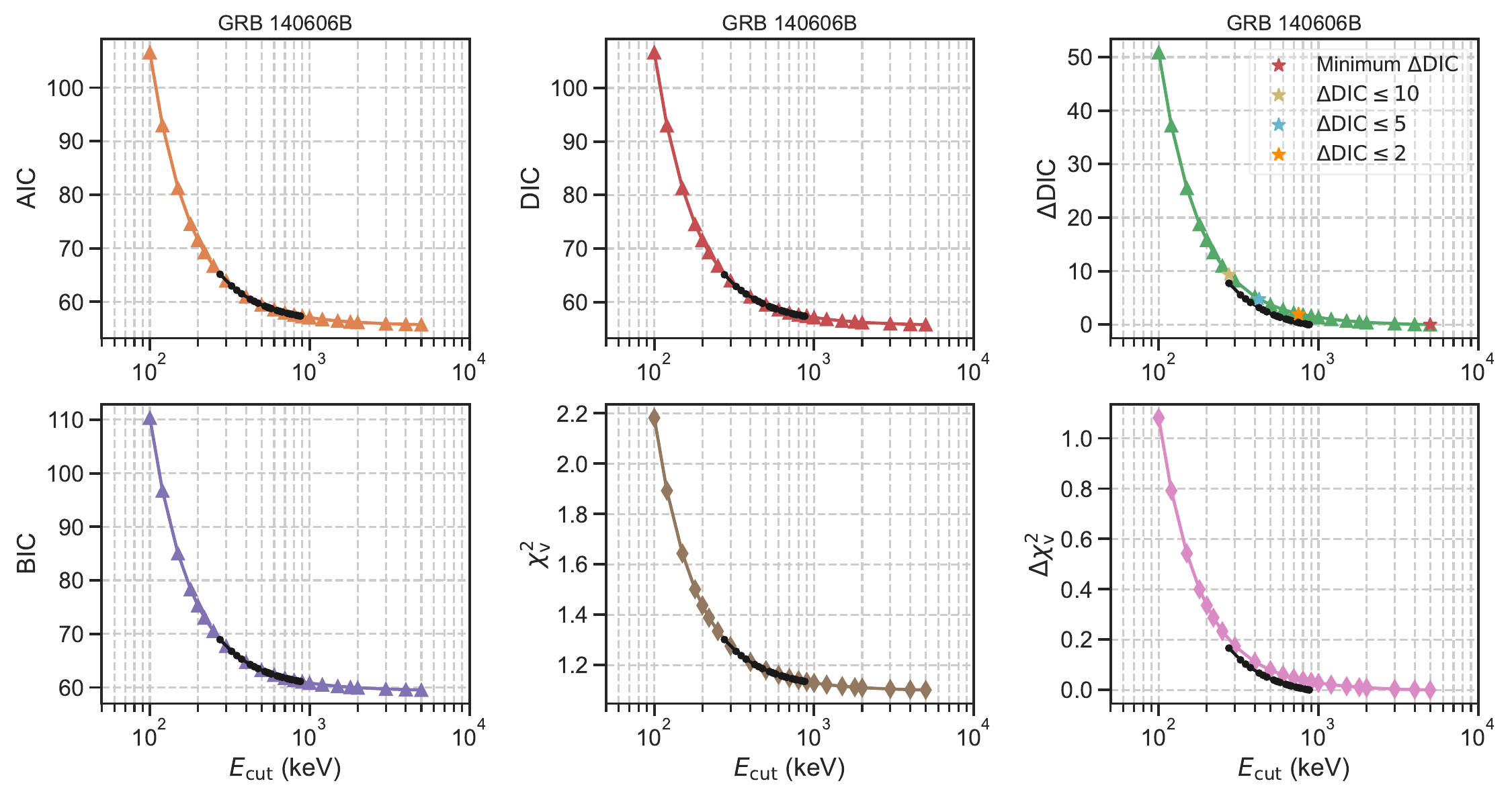}
\includegraphics[angle=0,scale=0.45]{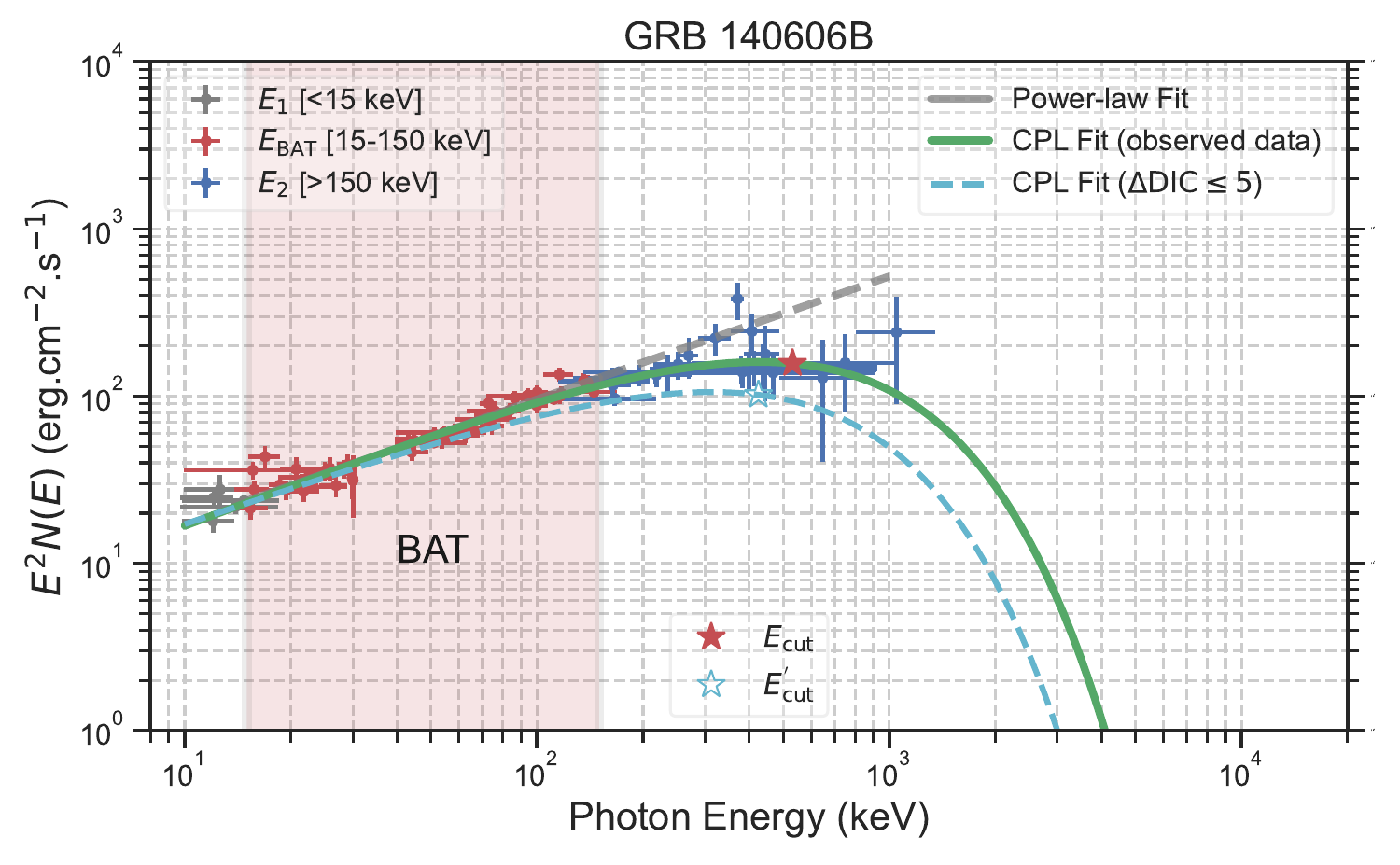}
\caption{Same as Figure \ref{fig:140703A} but for GRB 140606B. Note that the best CPL model fit to the original spectral data from GBM observation gives $E_{\rm c}$=531$^{+188}_{-136}$ keV, $A$=(2.7$^{+0.7}_{-0.5}$), $\alpha$=-1.19$^{+0.07}_{-0.07}$.}\label{fig:140606B}
\end{figure*}

\begin{figure*}
\centering
\includegraphics[angle=0,scale=0.45]{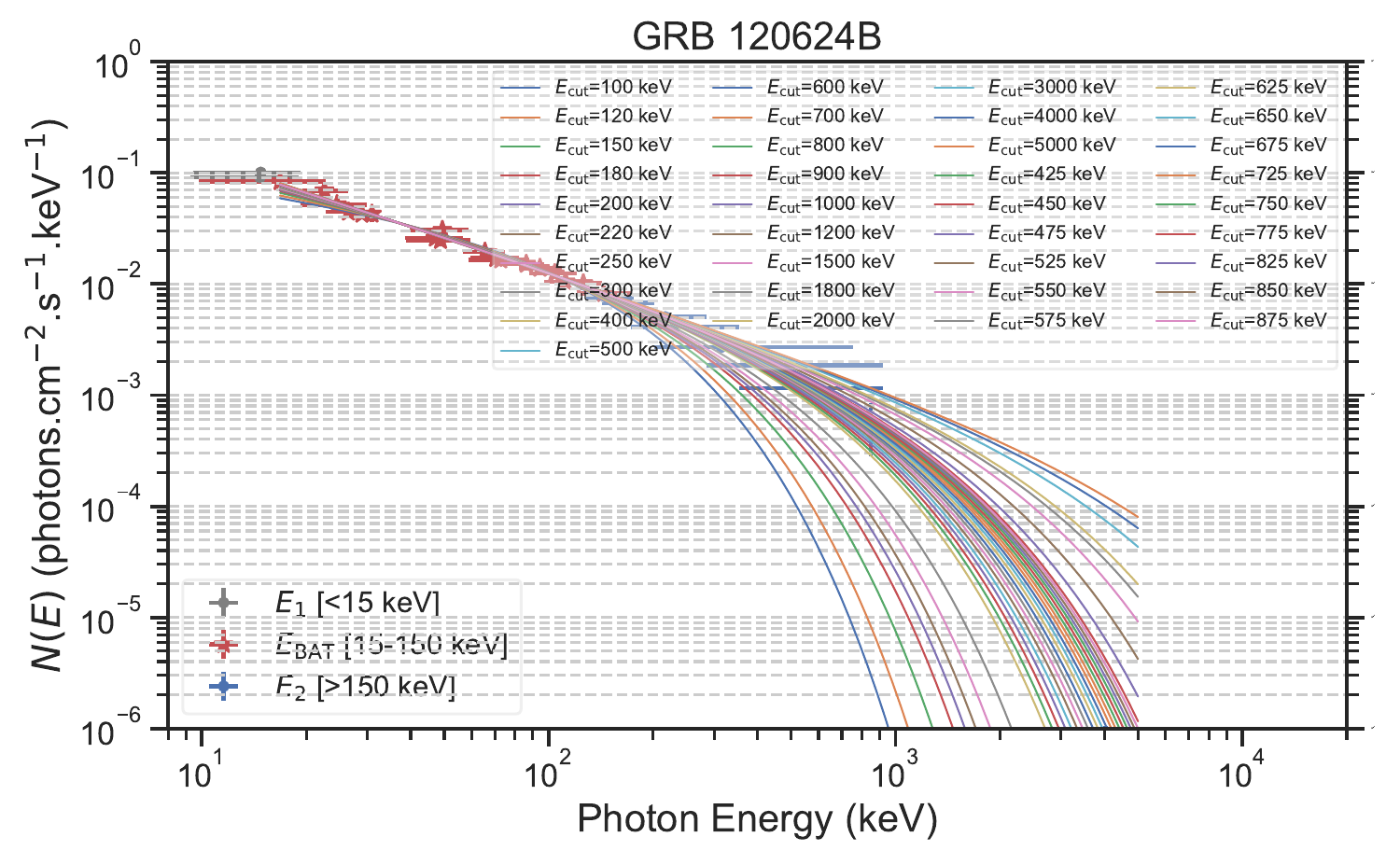}
\includegraphics[angle=0,scale=0.45]{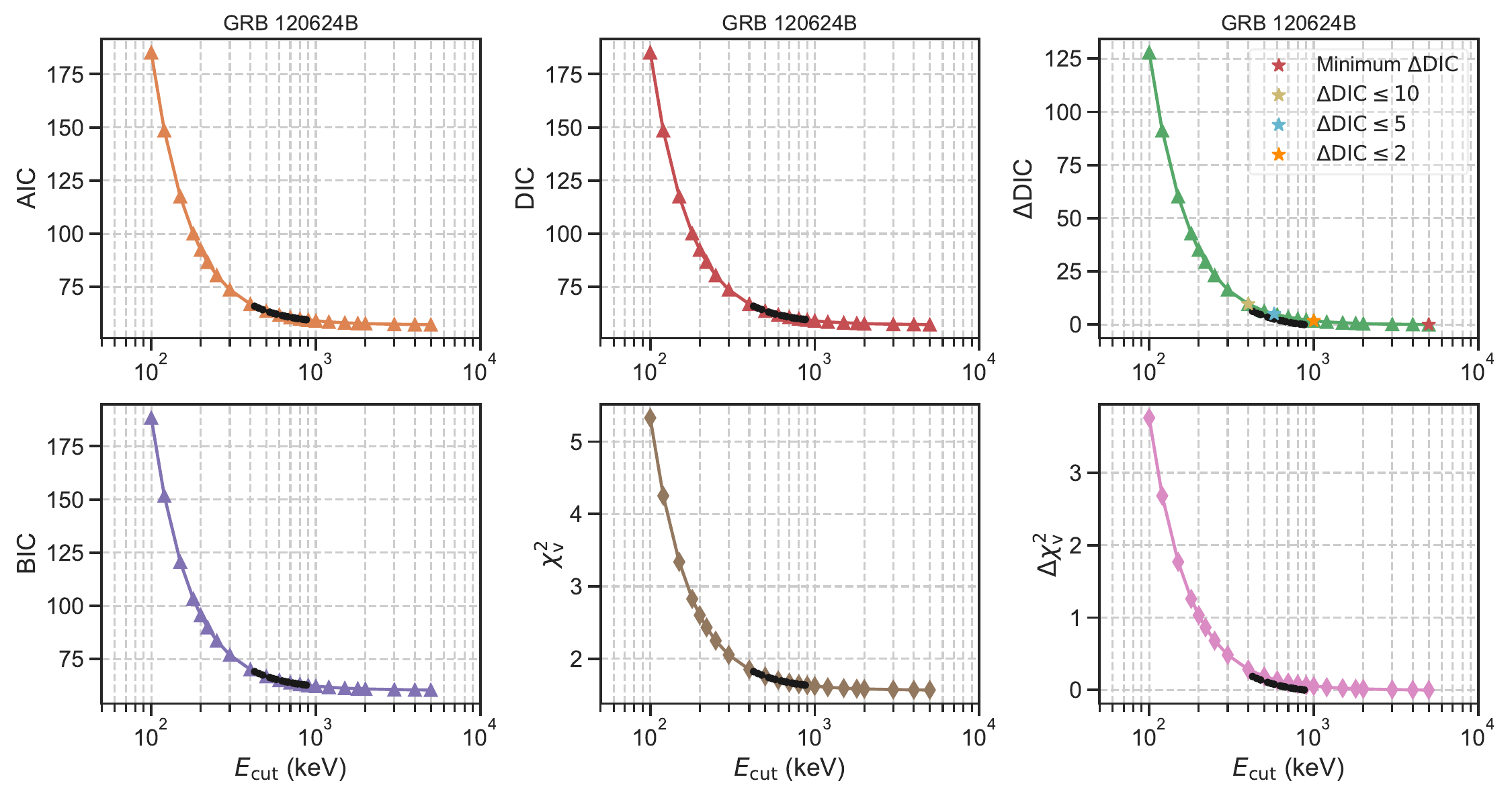}
\includegraphics[angle=0,scale=0.45]{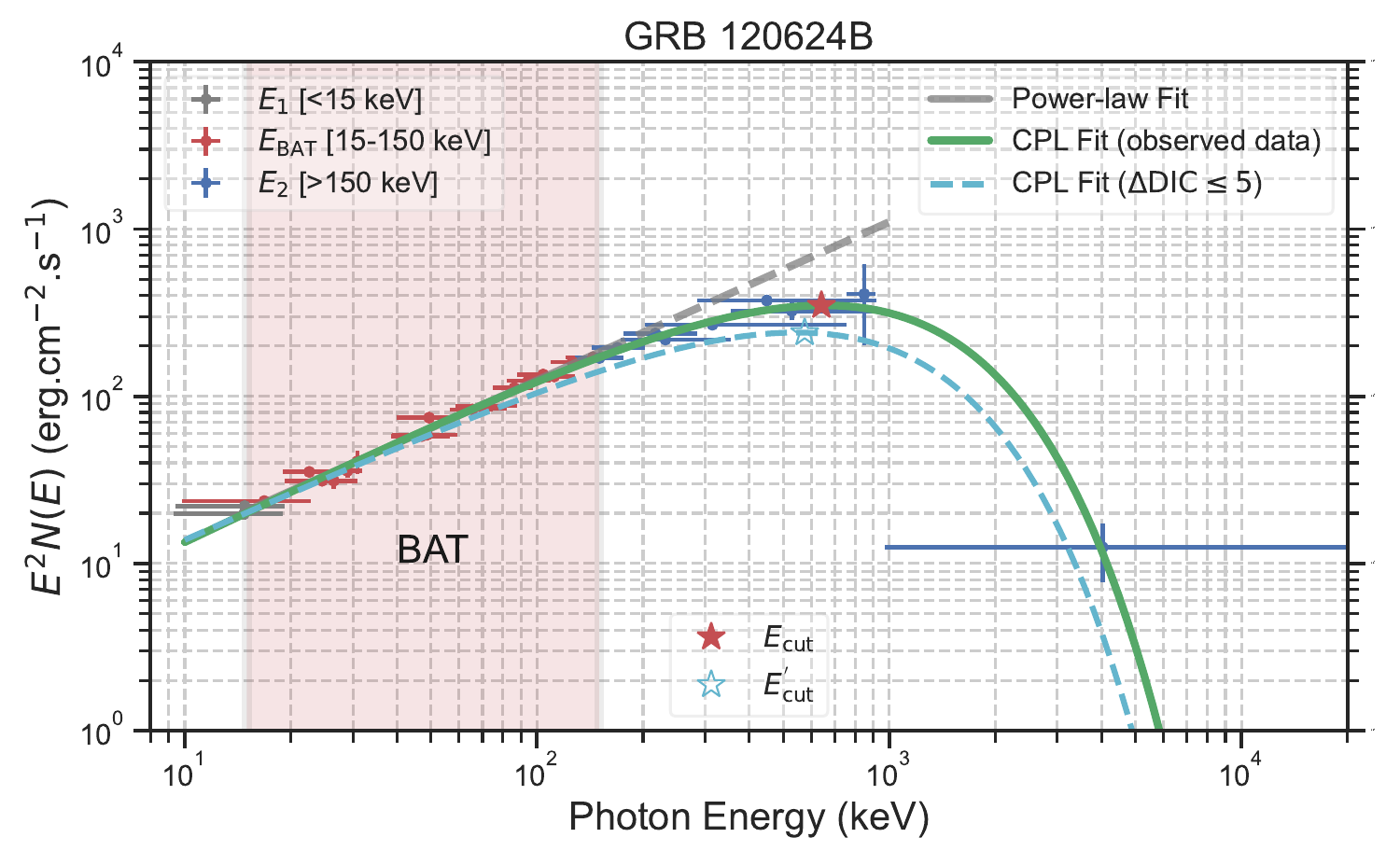}
\caption{Same as Figure \ref{fig:140703A} but for GRB 120624B. Note that the best CPL model fit to the original spectral data from GBM observation gives $E_{\rm c}$=626$^{+65}_{-60}$ keV, $A$=(1.30$^{+0.18}_{-0.16}$), $\alpha$=-0.98$^{+0.04}_{-0.04}$.}\label{fig:120624B}
\end{figure*}

\begin{figure*}
\centering
\includegraphics[angle=0,scale=0.45]{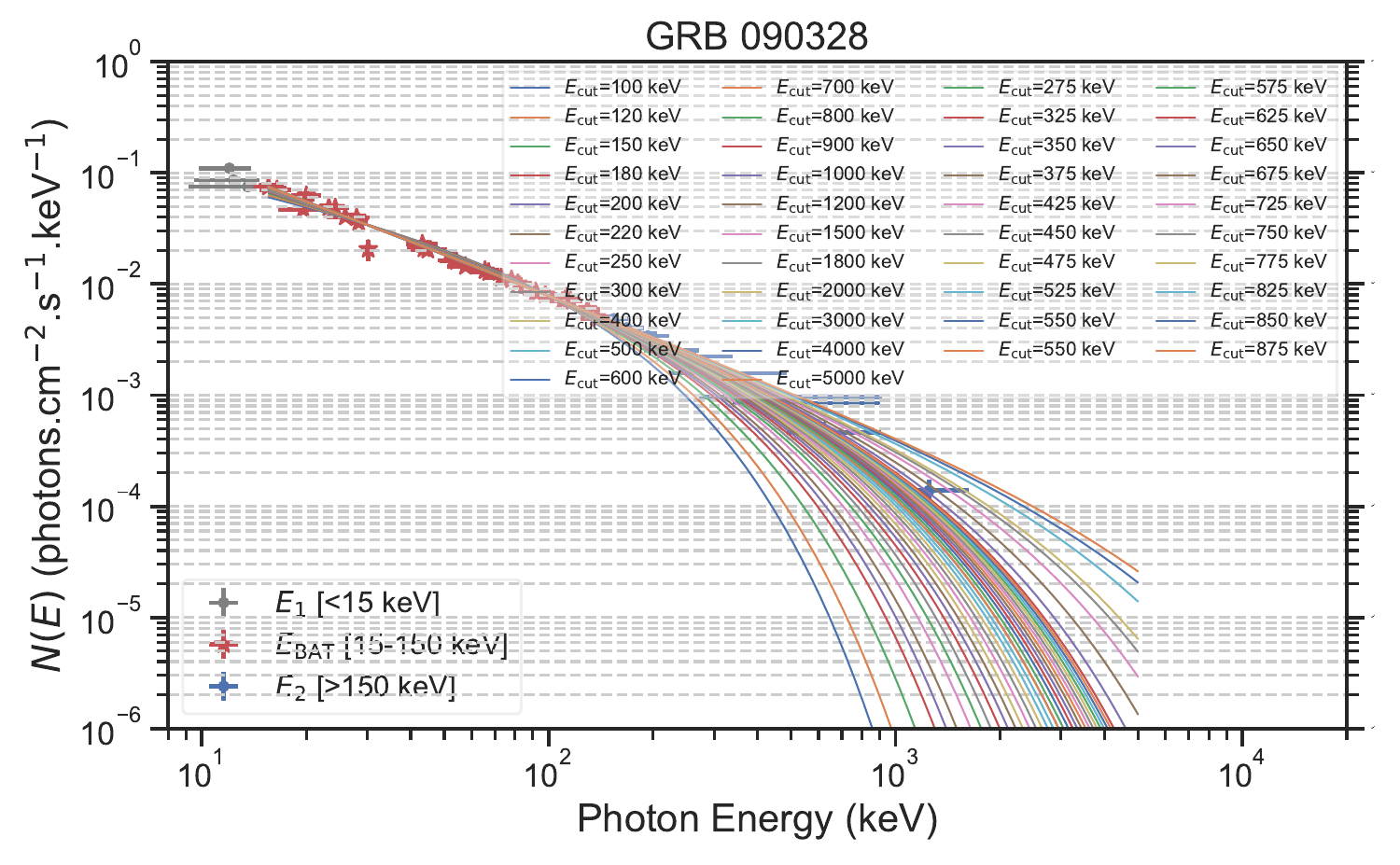}
\includegraphics[angle=0,scale=0.45]{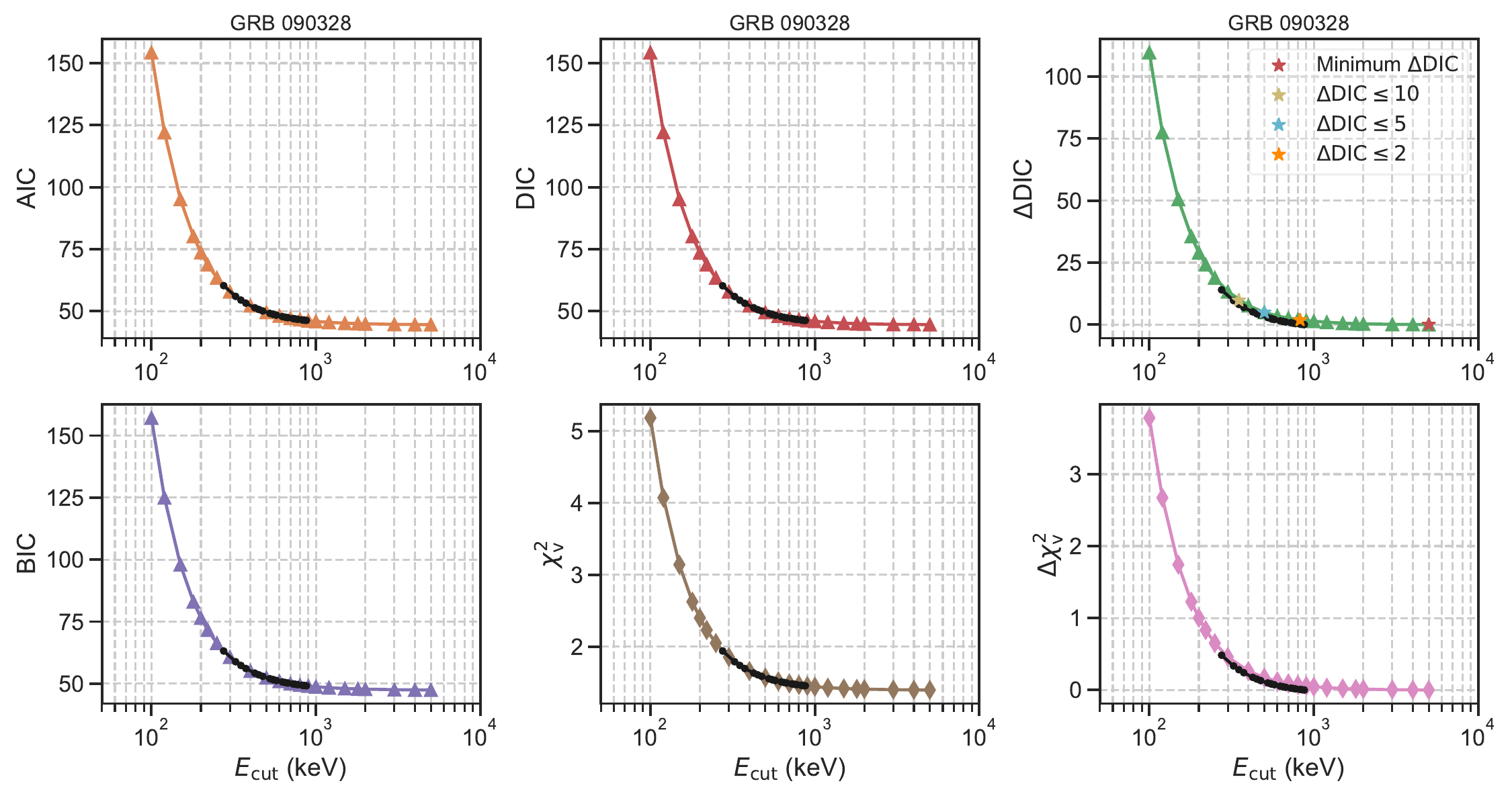}
\includegraphics[angle=0,scale=0.45]{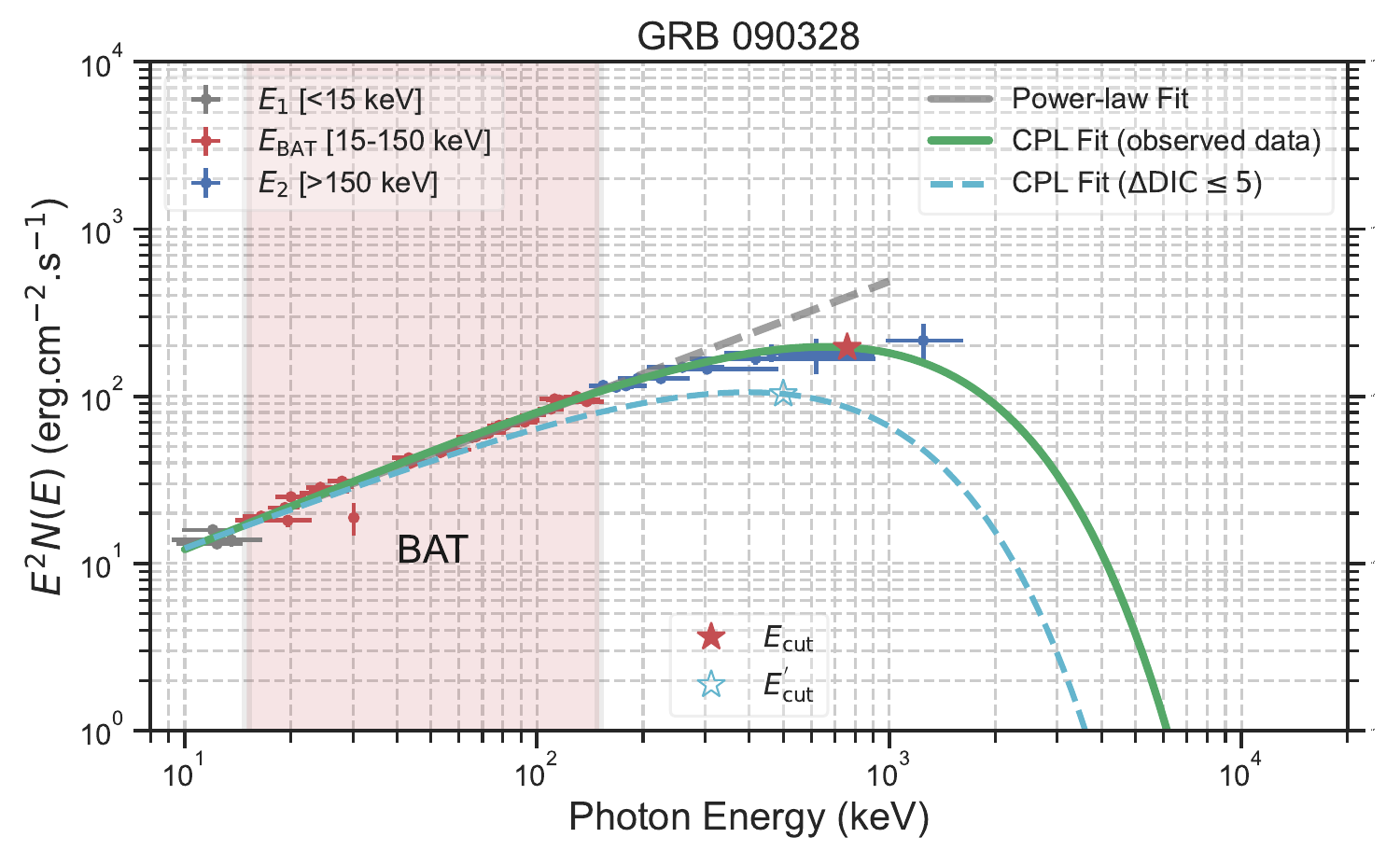}
\caption{Same as Figure \ref{fig:140703A} but for GRB 090328. Note that the best CPL model fit to the original spectral data from GBM observation gives $E_{\rm c}$=759$^{+104}_{-95}$ keV, $A$=(1.7$^{+0.1}_{-0.1}$), $\alpha$=-1.13$^{+0.03}_{-0.02}$.}\label{fig:090328}
\end{figure*}

\begin{figure*}
\centering
\includegraphics[angle=0,scale=0.45]{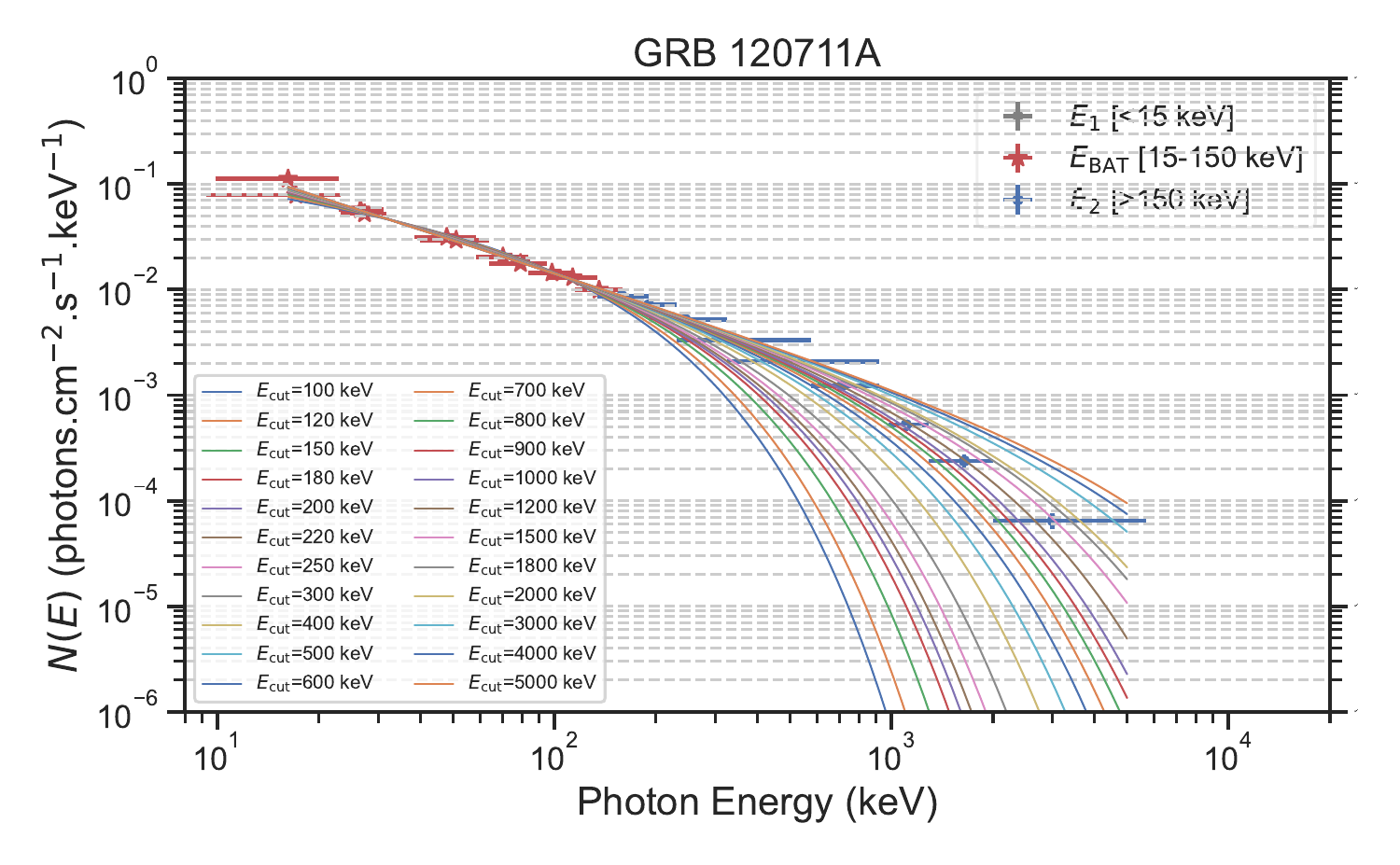}
\includegraphics[angle=0,scale=0.45]{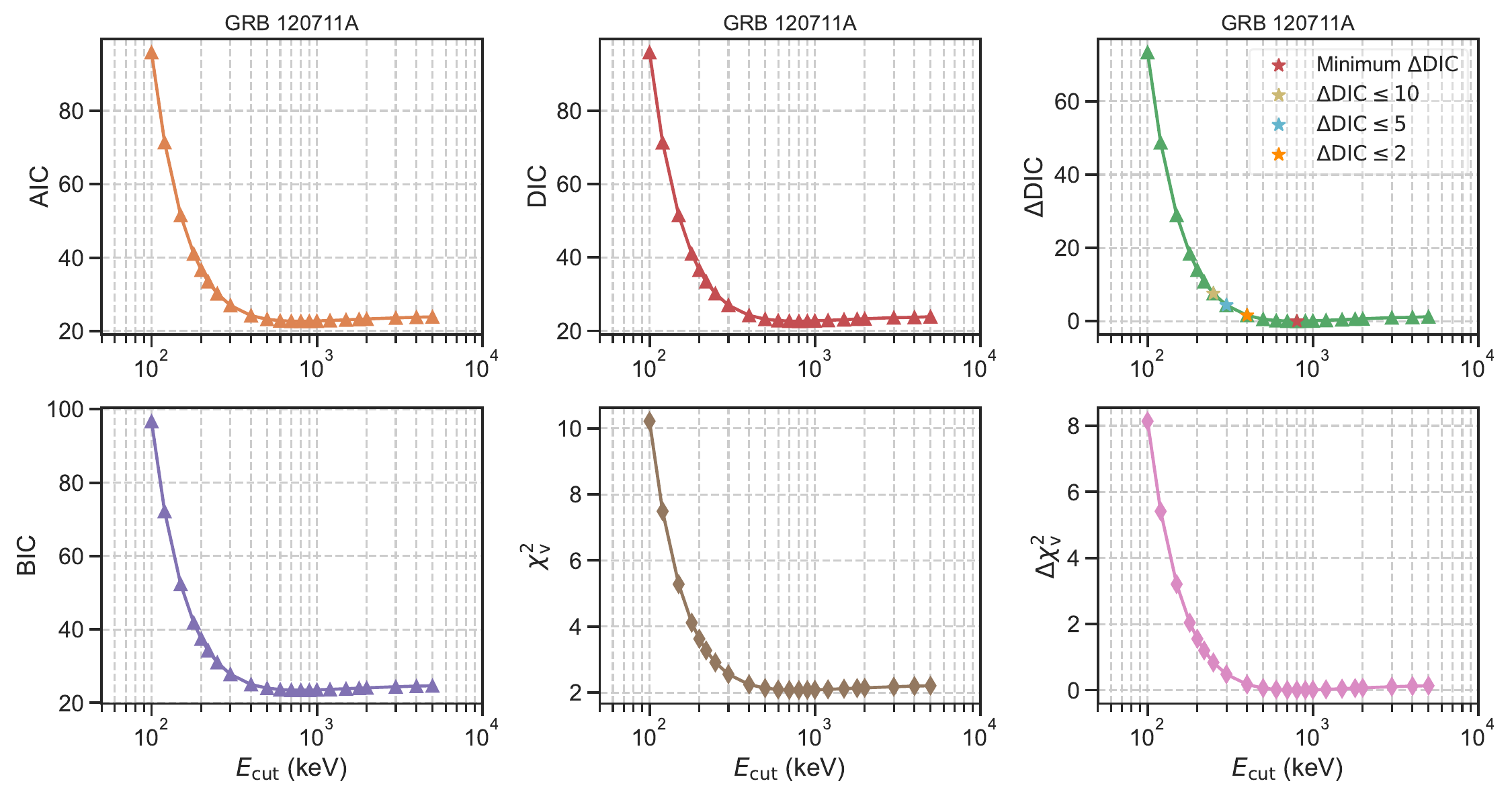}
\includegraphics[angle=0,scale=0.45]{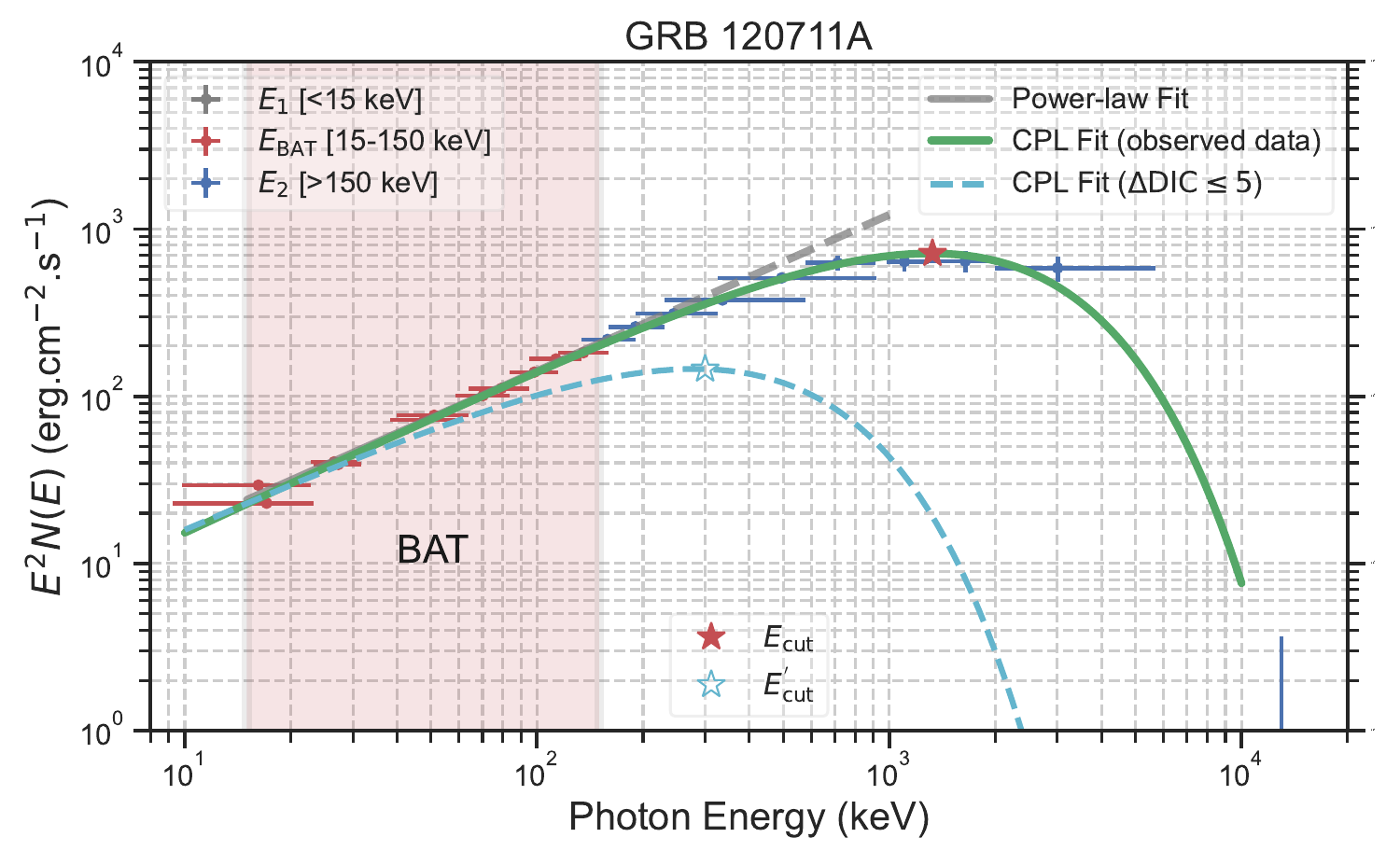}
\caption{Same as Figure \ref{fig:140703A} but for GRB 120711A. Note that the best CPL model fit to the original spectral data from GBM observation gives $E_{\rm c}$=1330$^{+140}_{-130}$ keV, $A$=(1.58$^{+0.15}_{-0.14}$), $\alpha$=-1.01$^{+0.02}_{-0.02}$).}\label{fig:120711A}
\end{figure*}

\begin{figure*}
\centering
\includegraphics[angle=0,scale=0.45]{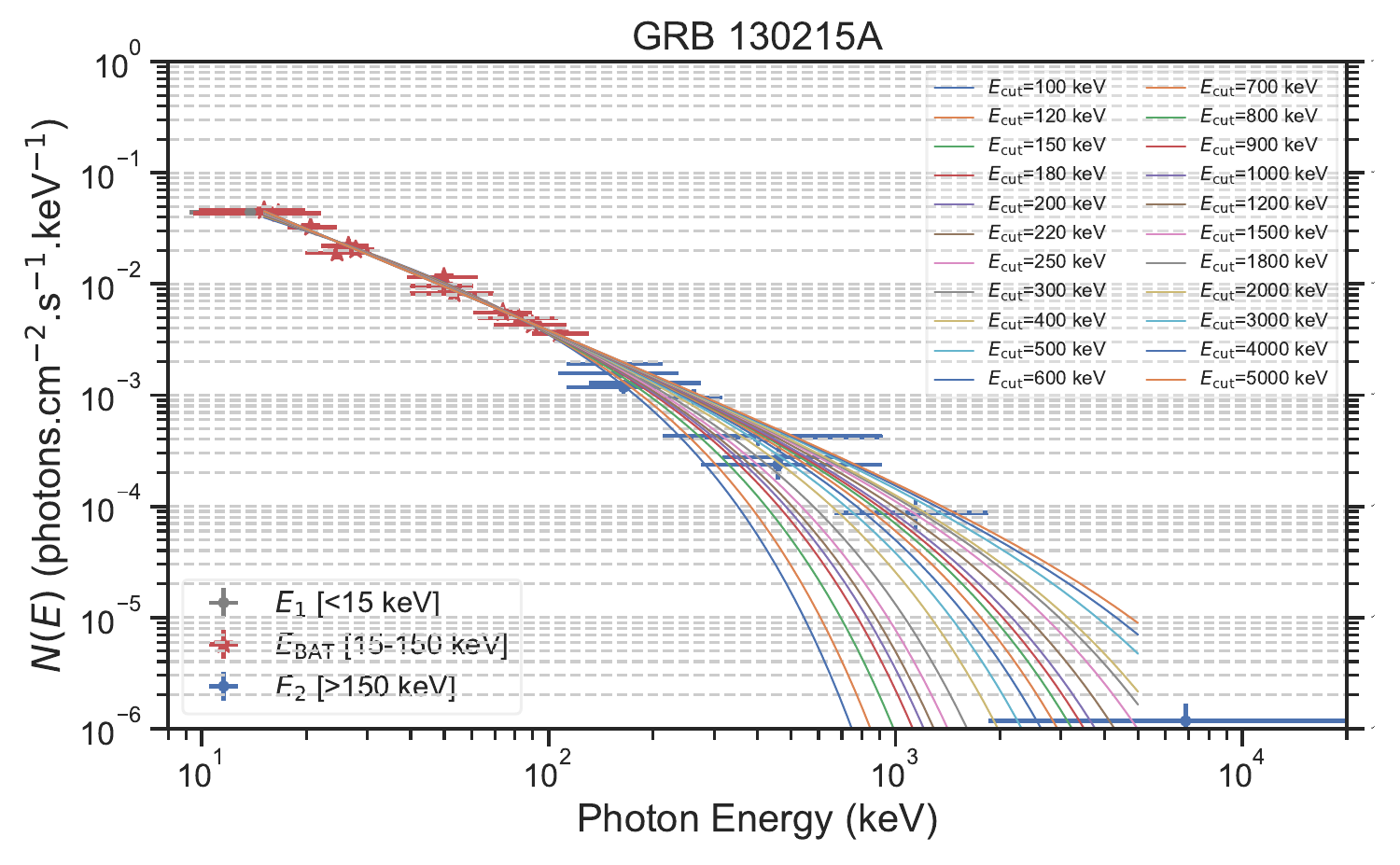}
\includegraphics[angle=0,scale=0.45]{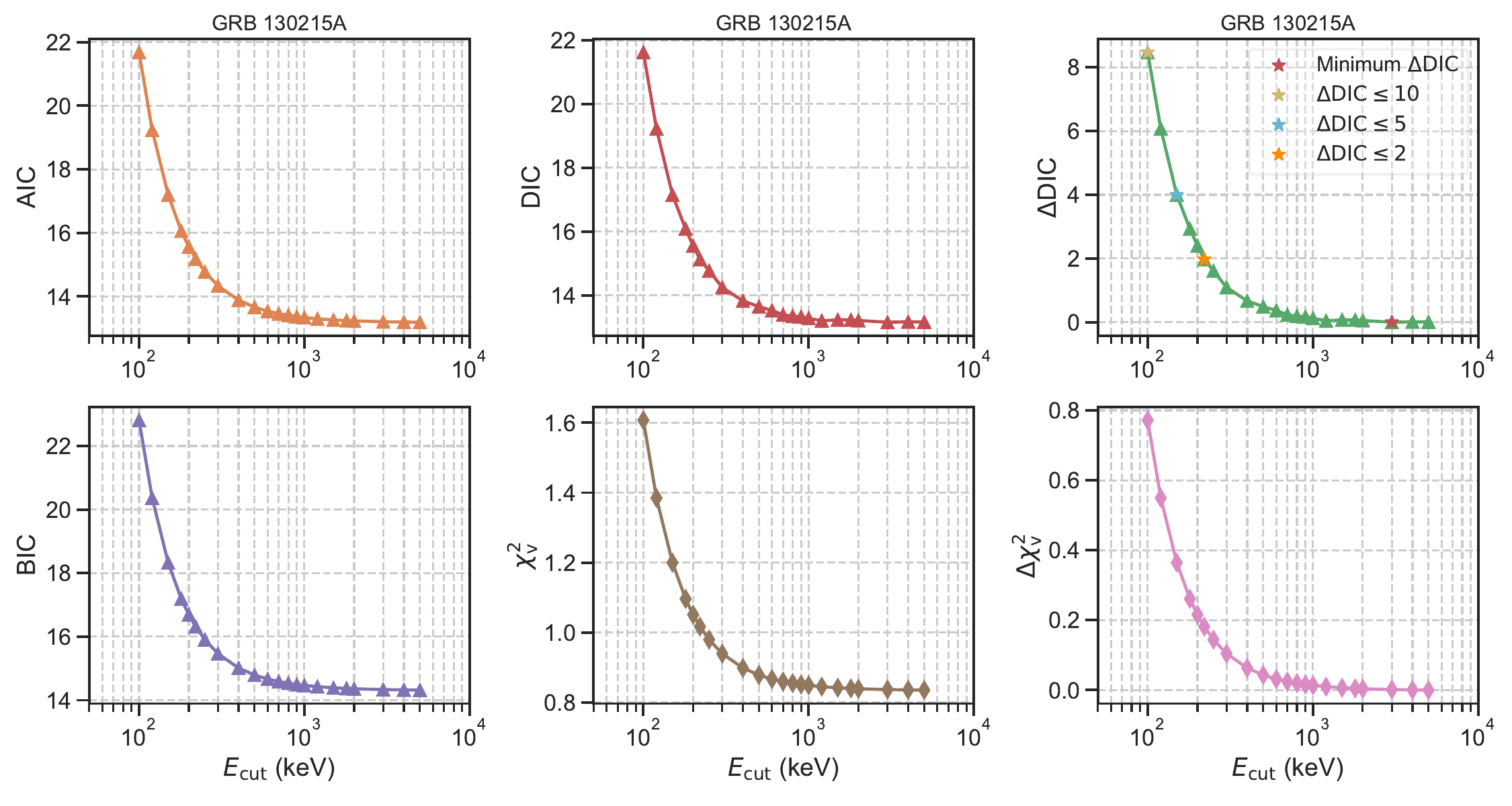}
\includegraphics[angle=0,scale=0.45]{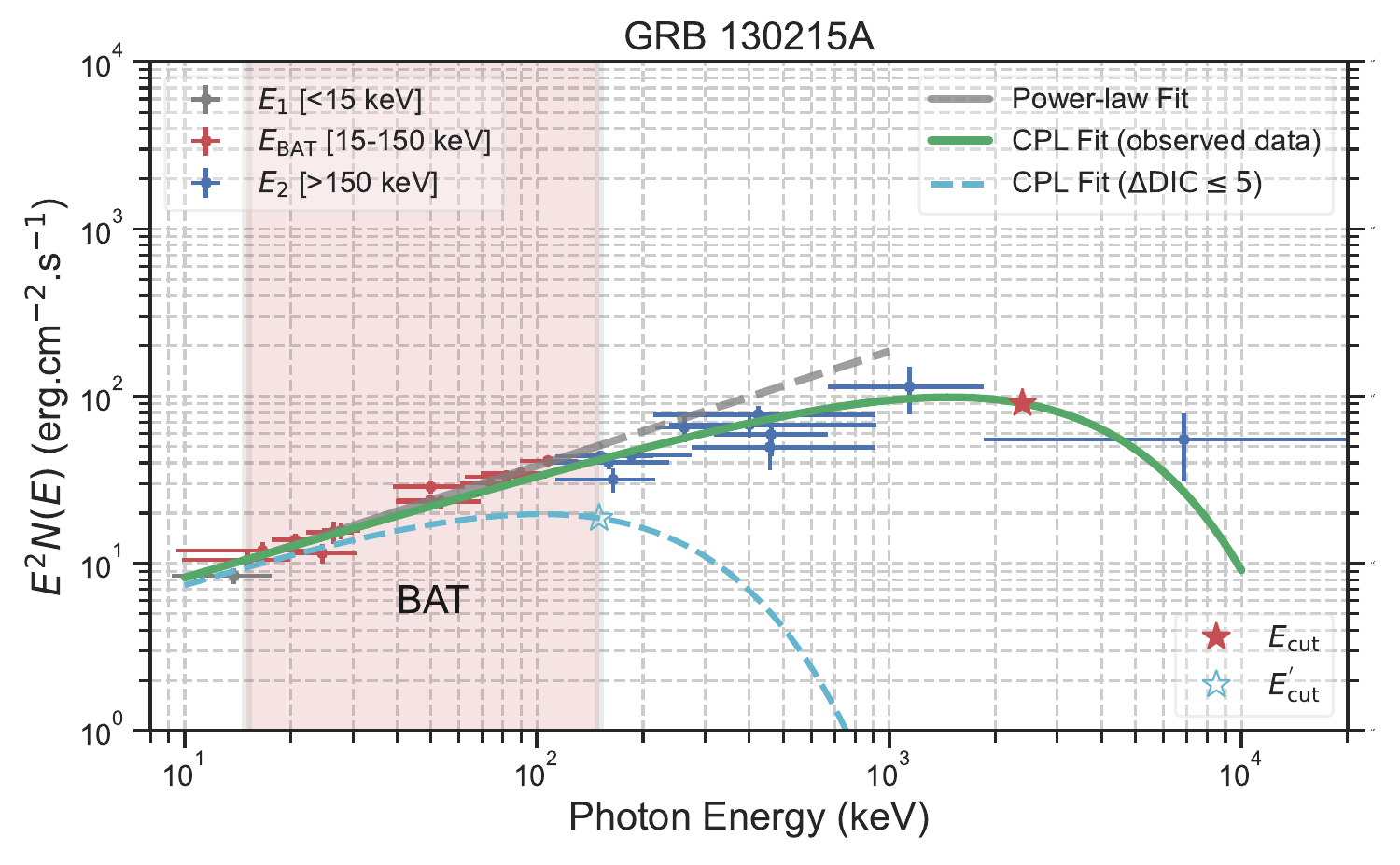}
\caption{Same as Figure \ref{fig:140703A} but for GRB 130215A. Note that the best CPL model fit to the original spectral data from GBM observation gives $E_{\rm c}$=2387$^{+1798}_{-970}$ keV, $A$=(2.0$^{+0.3}_{-0.3}$), $\alpha$=-1.38$^{+0.04}_{-0.04}$.}\label{fig:130215A}
\end{figure*}

To eliminate the potential influence of variations in $\alpha$ on the results, we selected 10 bursts from the GBM catalog, where $\alpha$ remained relatively constant within a typical and narrow range ($\sim$-0.80 to $\sim$-1.20), and the $E_{\rm c}$ gradually increased from 150 keV up to several MeV with a large change. Figures \ref{fig:140703A}-\ref{fig:130215A} present the preliminary estimation results of the spectral cutoff energy ($E^{'}_{\rm c}$) for 10 bursts, whose intrinsic spectra follow a CPL-like model. The intrinsic $E_{\rm c}$ values of these GRBs range from 173 keV to 2837 keV, corresponding to GRB 140703A, GRB 150301B, GRB 160629A, GRB 090529, GRB 091003, GRB 140606B, GRB 120624B, GRB 090328, GRB 120711A, and GRB 130215A, respectively, in the order shown in the Figures.

In the top panel of each figure, we show the raw spectral data for each analyzed burst alongside the CPL function fitted using the extrapolation method with fixed $E^{'}_{\rm c}$ values. The vertical axis represents photon flux density (photons cm$^{-2}$.s$^{-1}$.keV$^{-1}$), as defined in the CPL model. The gray, red, and blue data points represent spectral data from three distinct energy ranges: $E_{\rm below}$ (below the BAT energy band, E$<$15 keV), $E_{\rm BAT}$ (within the BAT band, 15 keV$\leq$E$\leq$150 keV), and $E_{\rm above}$ (above the BAT band, E$>$150 keV), respectively. The colored curves correspond to the best-fit CPL models for fixed $E^{'}_{\rm c}$ values of 100 keV, 120 keV, 150 keV, 180 keV, 200 keV, 220 keV, 250 keV, 300 keV, 400 keV, 500 keV, 600 keV, 700 keV, 800 keV, 900 keV, 1000 keV, 1200 keV, 1500 keV, 1800 keV, 2000 keV, 3000 keV, 4000 keV, and 5000 keV. It is important to note that, to simulate the scenario where only \emph{Swift}/BAT data is available, we used only the raw spectral data from the GBM energy range overlapping with BAT. Detailed data analysis and fitting methods are described in Section \S 2.3. For each fixed $E^{'}_{\rm c}$, we derived the CPL model parameters $A$ and $\alpha$, along with statistical criteria to evaluate goodness-of-fit, such as $\chi^{2}$, degrees of freedom (DOF), reduced $\chi^{2}$, AIC, BIC, DIC, and $p_{\rm DIC}$. These results are summarized in Tables \ref{tab:140703A}-\ref{tab:130215A}.

In the middle panel of each figure, we plot the statistical criteria (AIC, BIC, DIC, and $\chi^{2}$) and their deviations from the minimum value ($\Delta$DIC, and $\Delta$$\chi^{2}$) as a function of $E^{'}_{\rm c}$. These curves generally fall into two categories: (1) ``U"-shaped curves with a distinct minimum (e.g., GRB 140703A, GRB 160629A, and GRB 091003), where the optimal fit function and the corresponding $E^{'}_{\rm c}$ can be easily identified by identifying the minimum value of the curve; (2) monotonic decreasing curves resembling a ``slide" (GRB 150301B, GRB 090529, GRB 140606B, GRB 120624B, GRB 090328, GRB 120711A, and GRB 130215A), where the statistical criteria decrease as $E^{'}_{\rm c}$ increases, eventually reaching a bending point or plateau. The $E^{'}_{\rm c}$ at this bending point provides a robust lower limit for the true $E_{\rm c}$. To quantitatively identify the bending point, we define thresholds of $\Delta$DIC$\leq$5 and $\chi_{\rm r}^{2}$$\leq$0.5. The value of the first $E_{\rm c}$ value below these threshold is the value we used. Four horizontal red dashed lines in the figures mark these thresholds ($\Delta$DIC$\leq$5 and $\chi_{\rm r}^{2}$$\leq$0.5).

In the bottom panel of each figure, we present the preliminary estimation of $E^{'}_{\rm c}$ in the $\nu F_{\nu}$ space. The solid green curve represents the best-fit model derived using the full GBM energy range ($E_{\rm below}$+$E_{\rm BAT}$+$E_{\rm above}$), from which the true observed $E_{\rm c}$ is extracted. The four dashed lines of different colors represent the $E^{'}_{\rm c}$ values determined using various statistical criteria, with vertical dashed lines marking these corresponding $E^{'}_{\rm c}$. Comparing different thresholds, we find that the criterion DIC=2 yields $E^{'}_{\rm c}$ values that are closest to the true $E^{'}_{\rm c}$.

\begin{figure*}
\centering
\includegraphics[angle=0,scale=0.45]{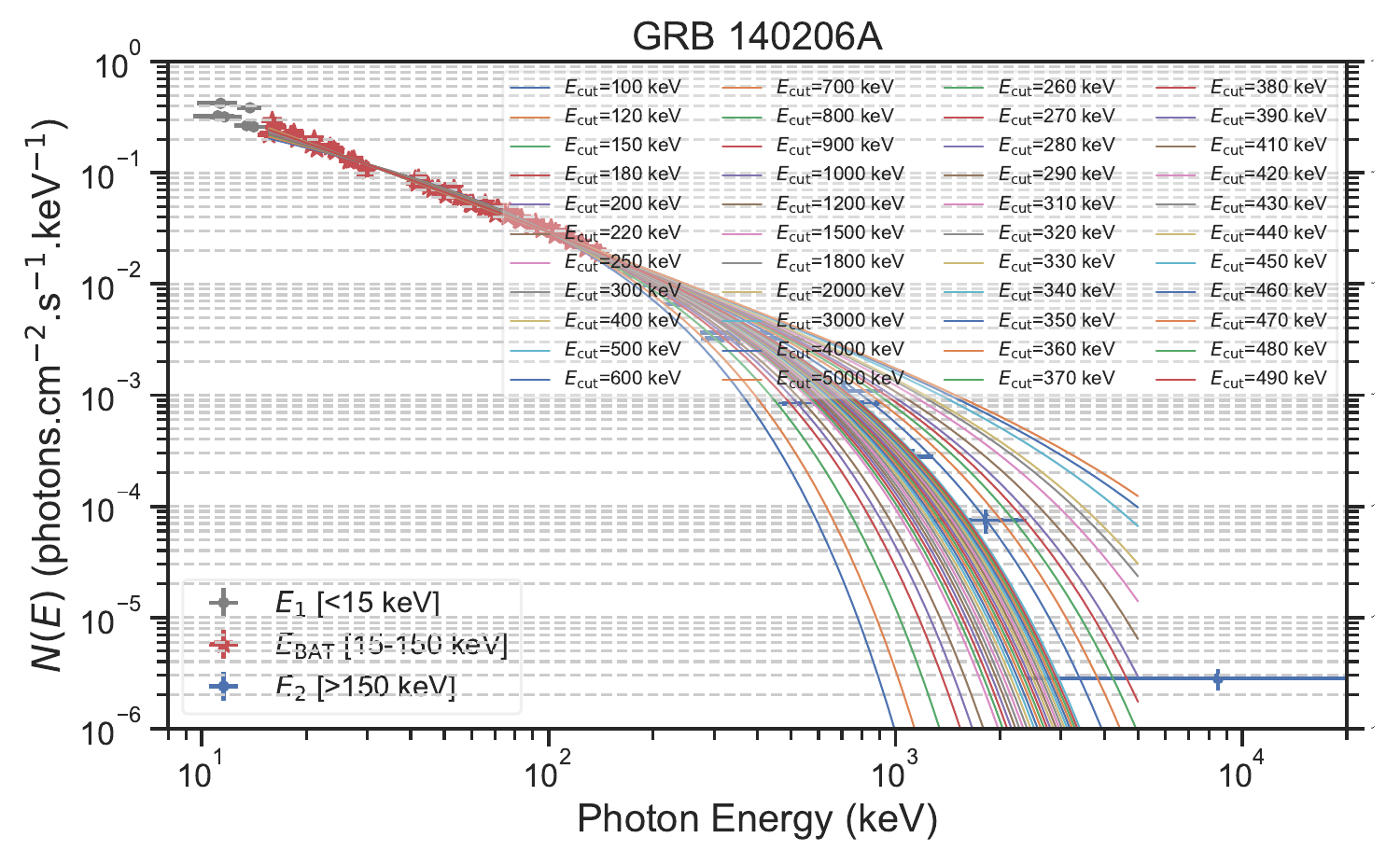}
\includegraphics[angle=0,scale=0.45]{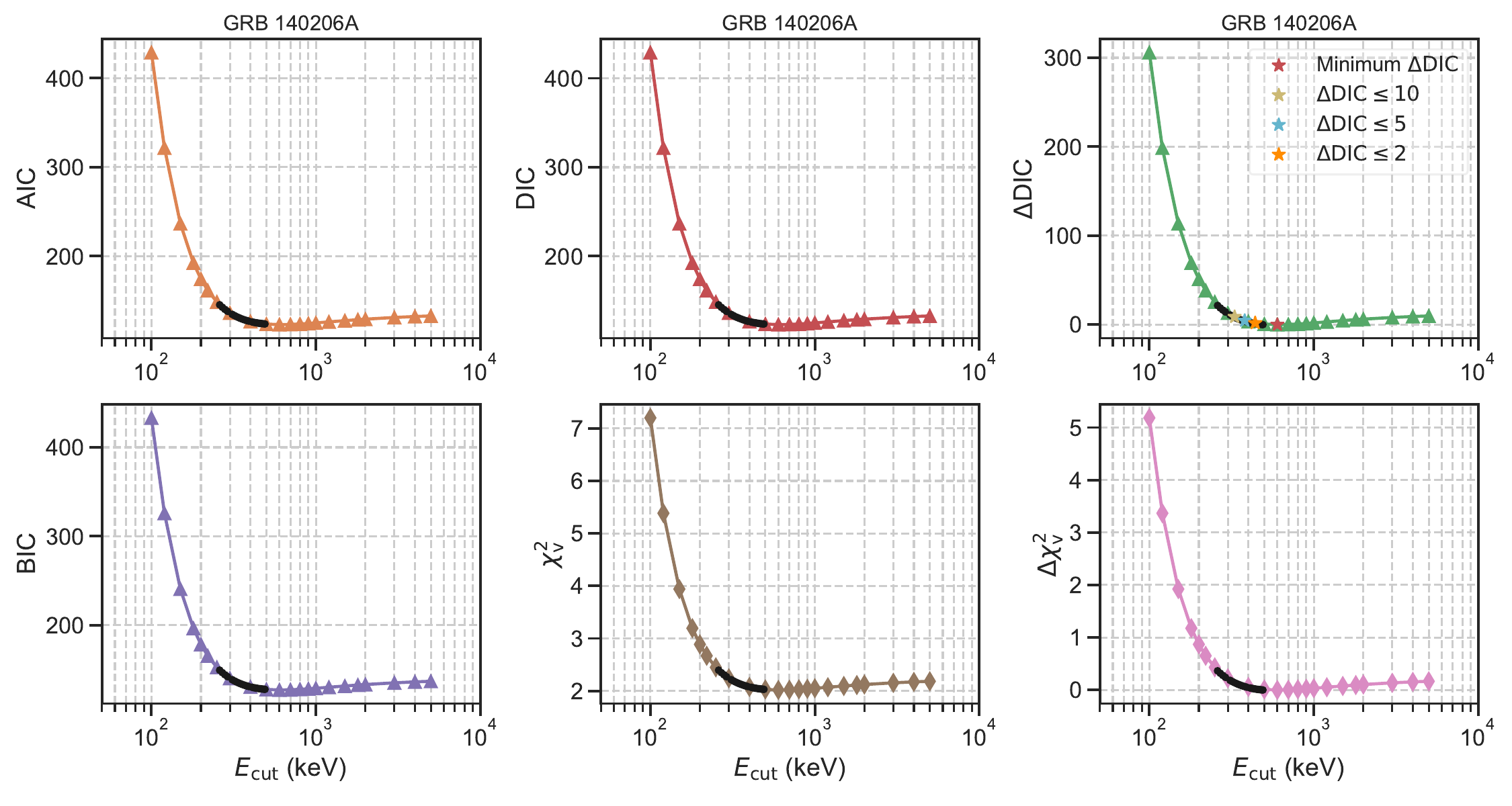}
\includegraphics[angle=0,scale=0.45]{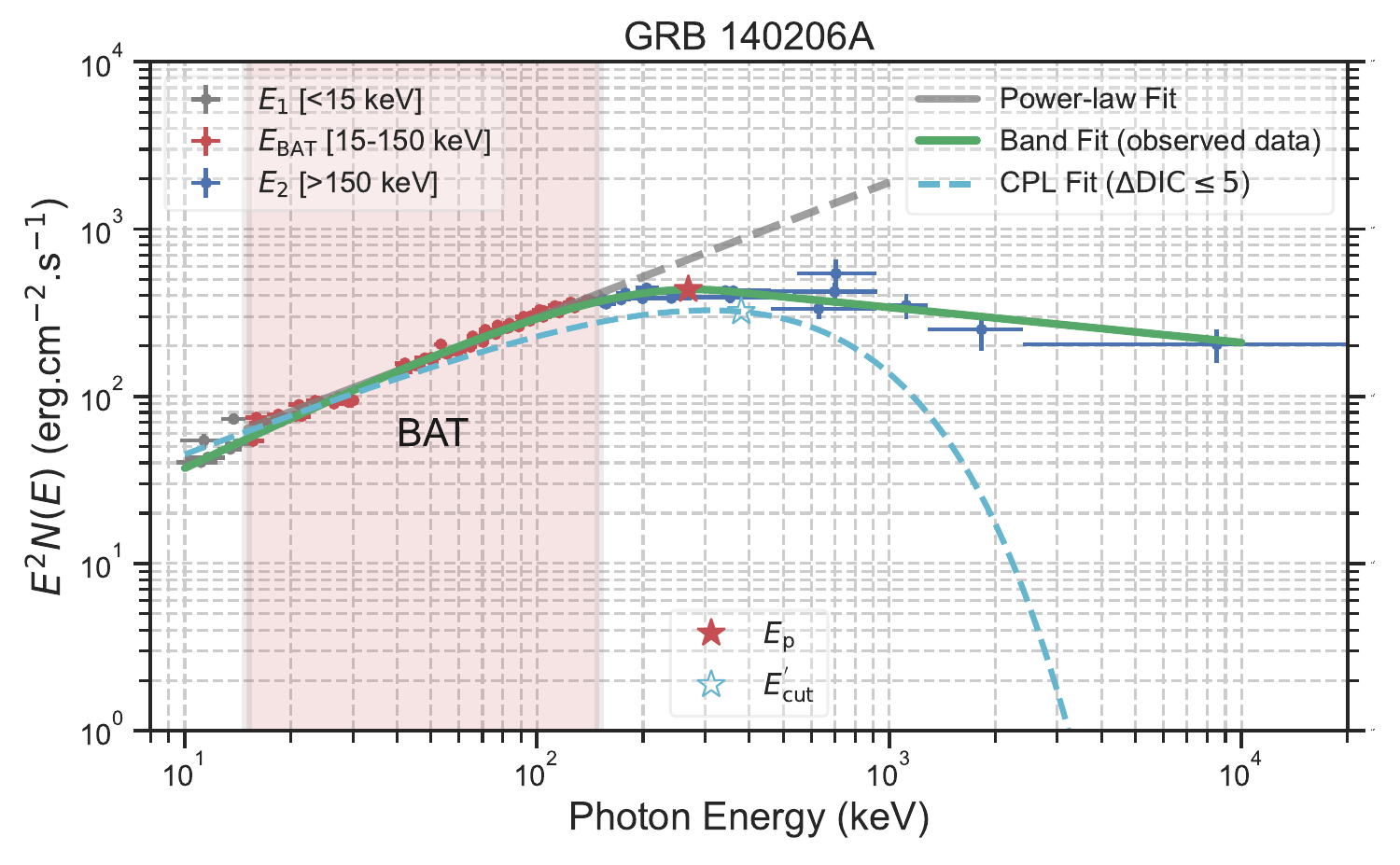}
\caption{Same as Figure \ref{fig:140703A} but for GRB 140206A. Note that the best-fit model in this case is the Band function, with fitted model parameters: $E_{\rm p}$=280$\pm$10 keV, $A$=(4.2$\pm$0.1)$\times$10$^{-2}$, $\alpha$=-0.96$\pm$0.16, $\beta$=-2.21$\pm$0.05. Color code and description as in Figure \ref{fig:140703A}.}\label{fig:140206A}
\end{figure*}

\begin{figure*}
\centering
\includegraphics[angle=0,scale=0.45]{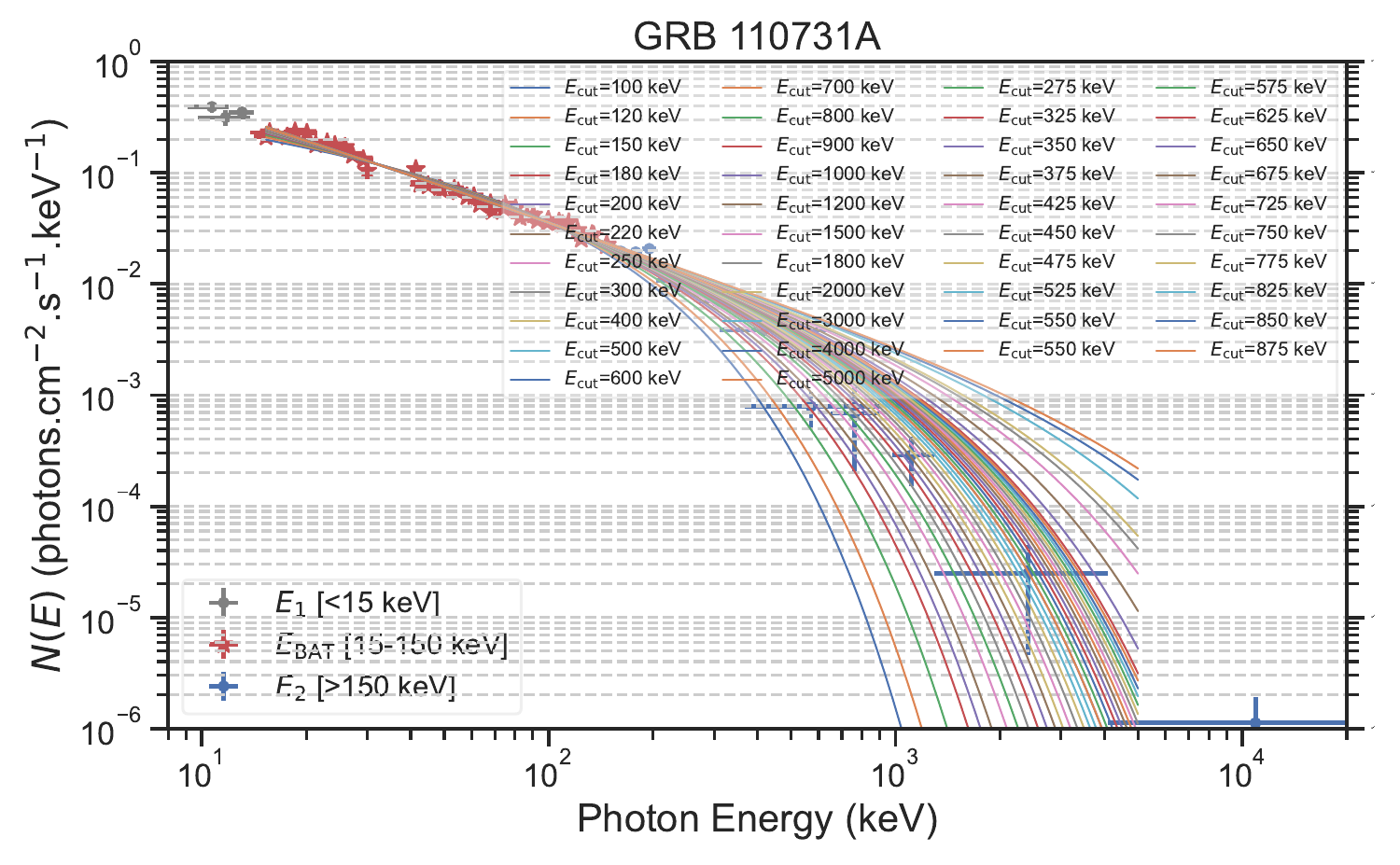}
\includegraphics[angle=0,scale=0.45]{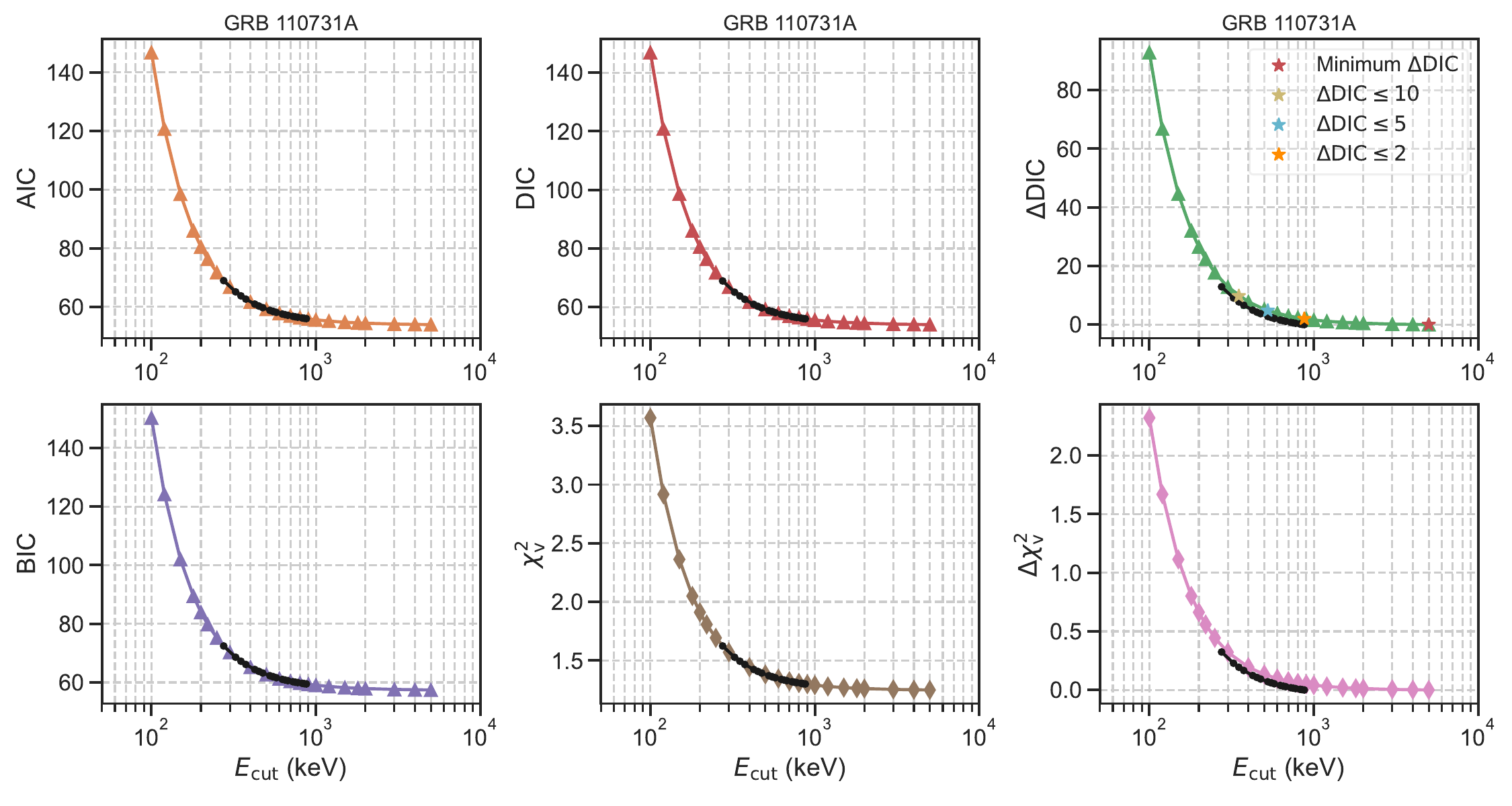}
\includegraphics[angle=0,scale=0.45]{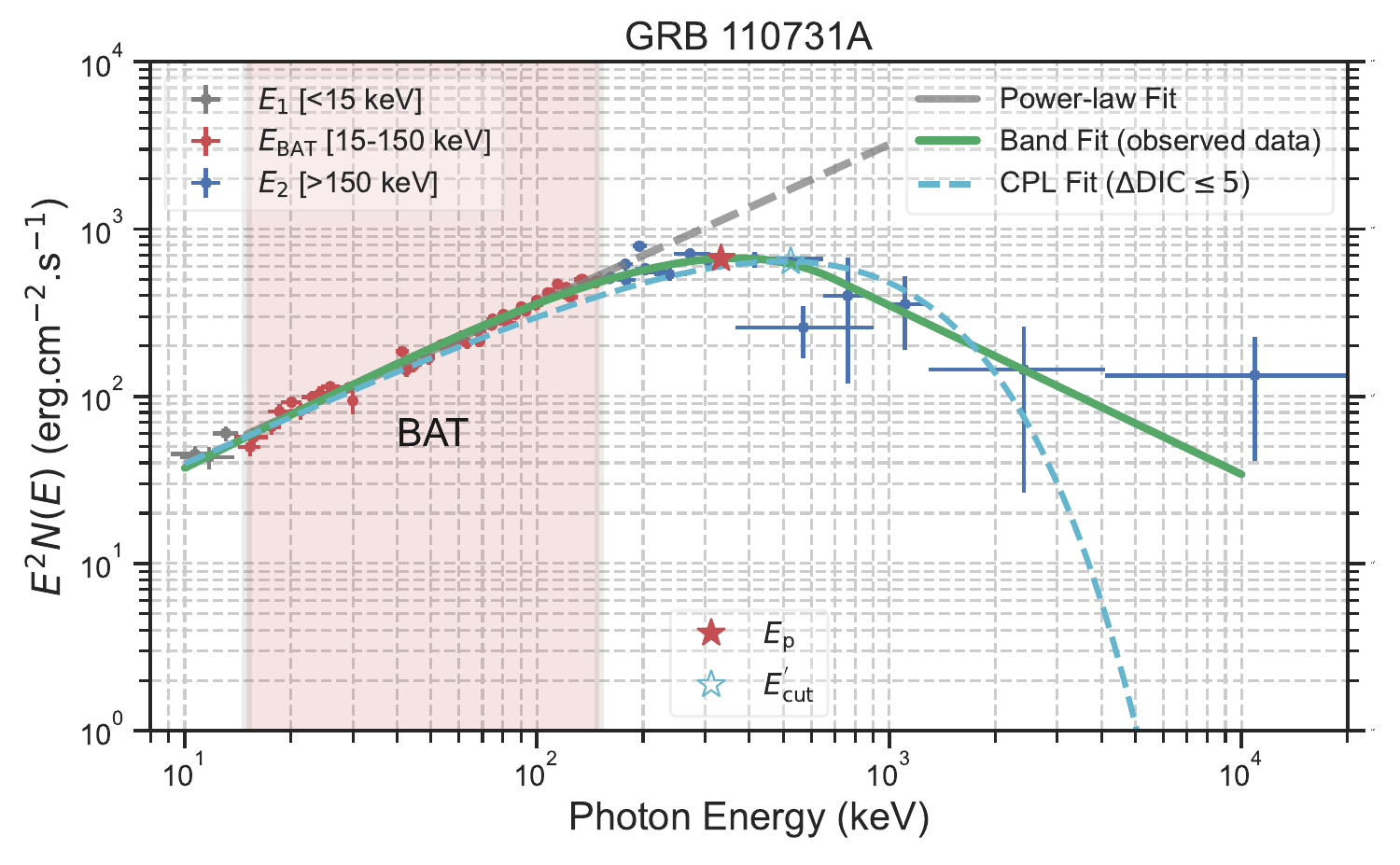}
\caption{Same as Figure \ref{fig:140206A} but for GRB 110731A. Note that the best Band model fit to the original spectral data from GBM observation gives $E_{\rm p}$=366$\pm$27 keV, $A$=(4.8$\pm$0.2)$\times$10$^{-2}$, $\alpha$=-0.90$\pm$0.03, $\beta$=-3.0$\pm$0.7.}
\label{fig:110731A}
\end{figure*}

\begin{figure*}
\centering
\includegraphics[angle=0,scale=0.45]{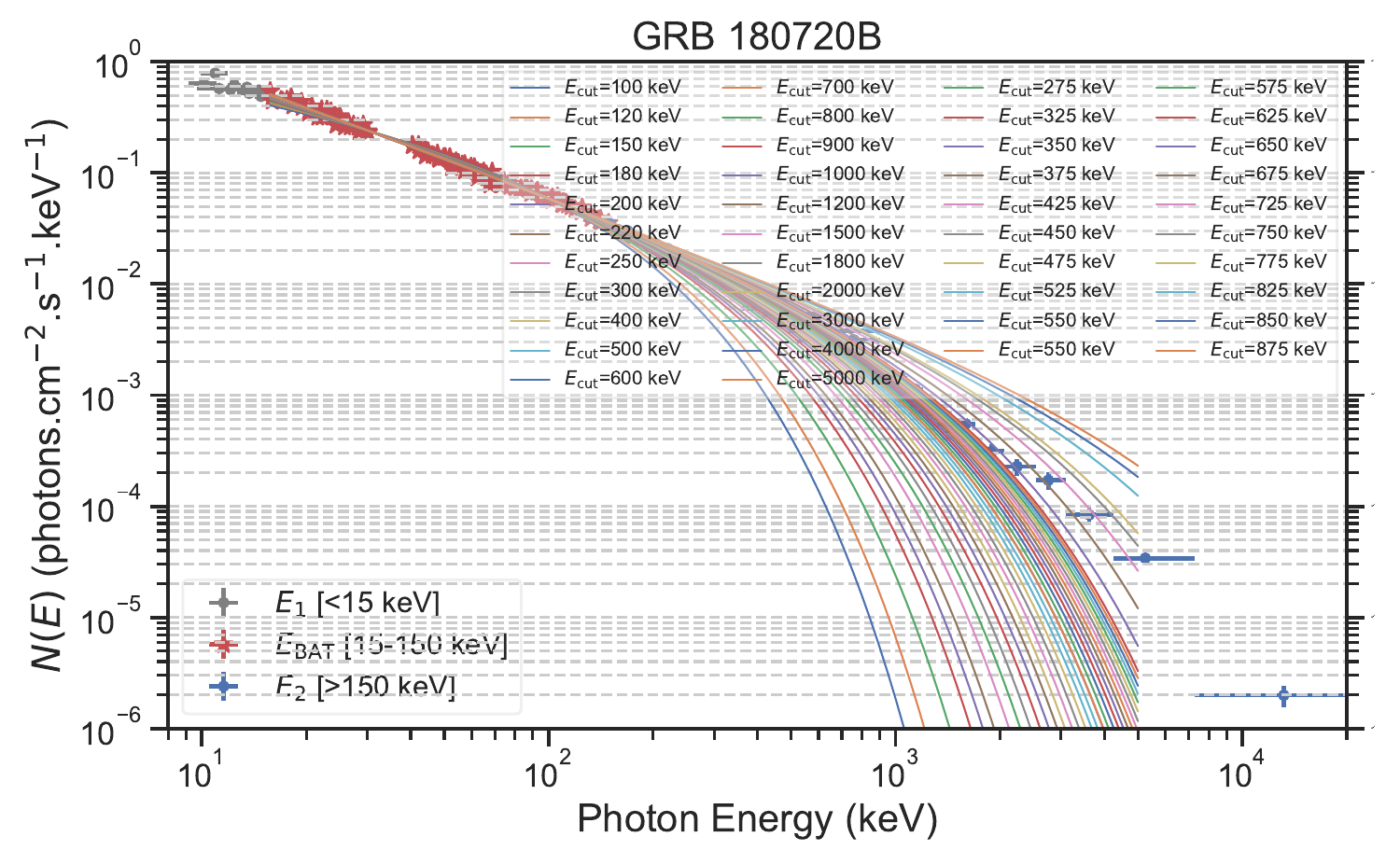}
\includegraphics[angle=0,scale=0.45]{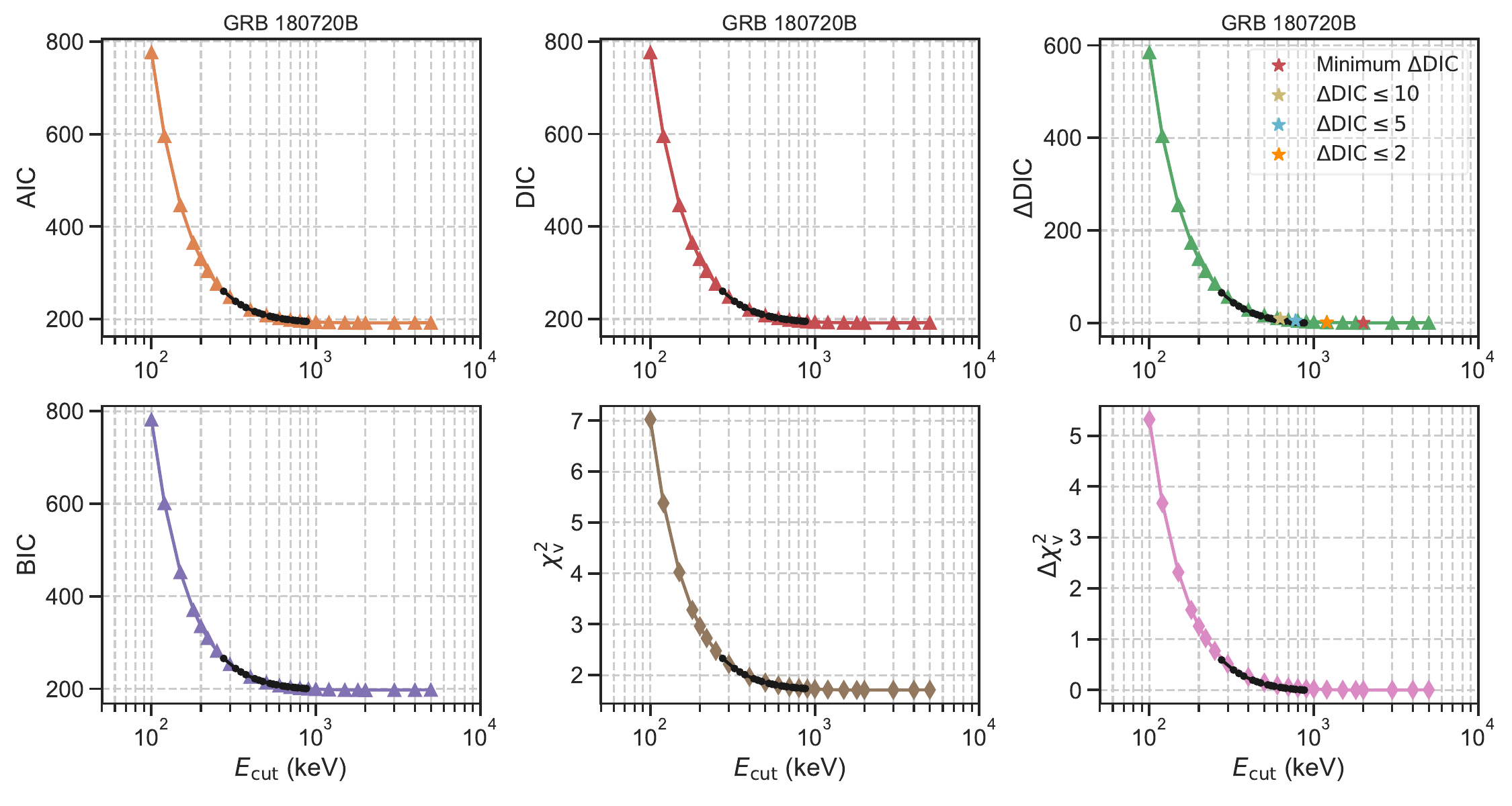}
\includegraphics[angle=0,scale=0.45]{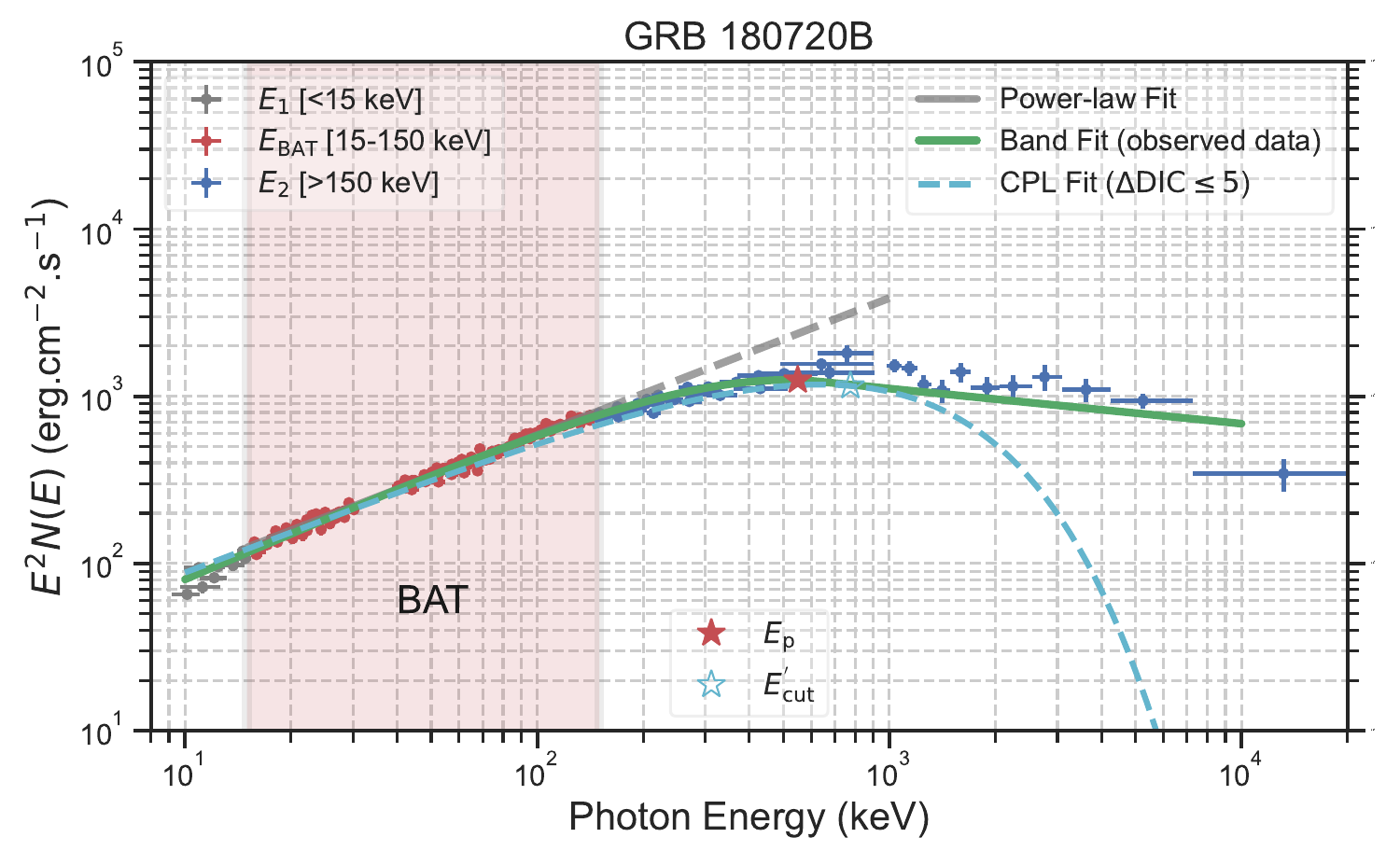}
\caption{Same as Figure \ref{fig:140206A} but for GRB 180720B. Note that the best Band model fit to the original spectral data from GBM observation gives $E_{\rm p}$=511$\pm$26 keV, $A$=(7.0$\pm$0.1)$\times$10$^{-2}$, $\alpha$=-1.07$\pm$0.01, $\beta$=-2.21$\pm$0.03.}
\label{fig:180720B}
\end{figure*}

\begin{figure*}
\centering
\includegraphics[angle=0,scale=0.45]{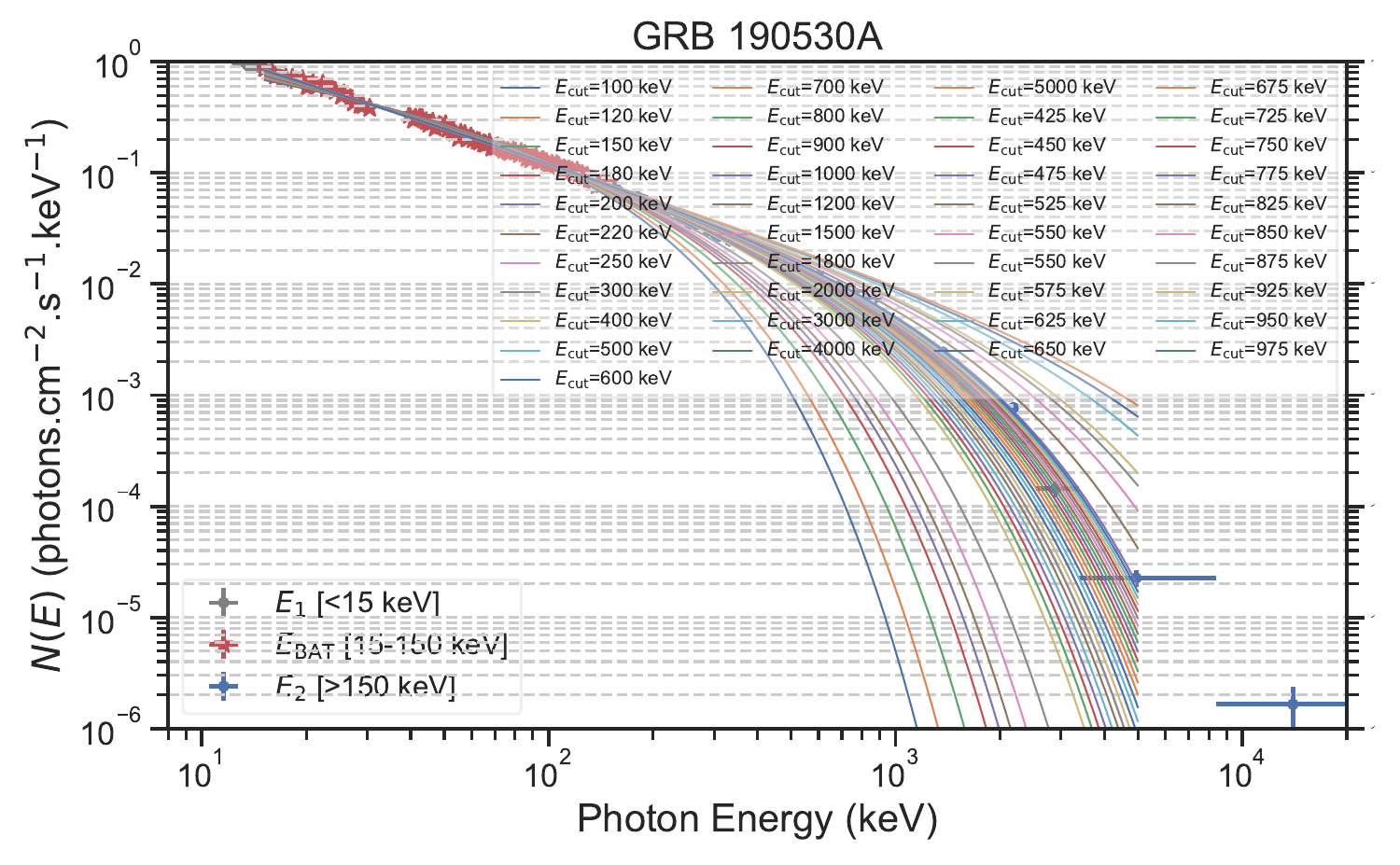}
\includegraphics[angle=0,scale=0.45]{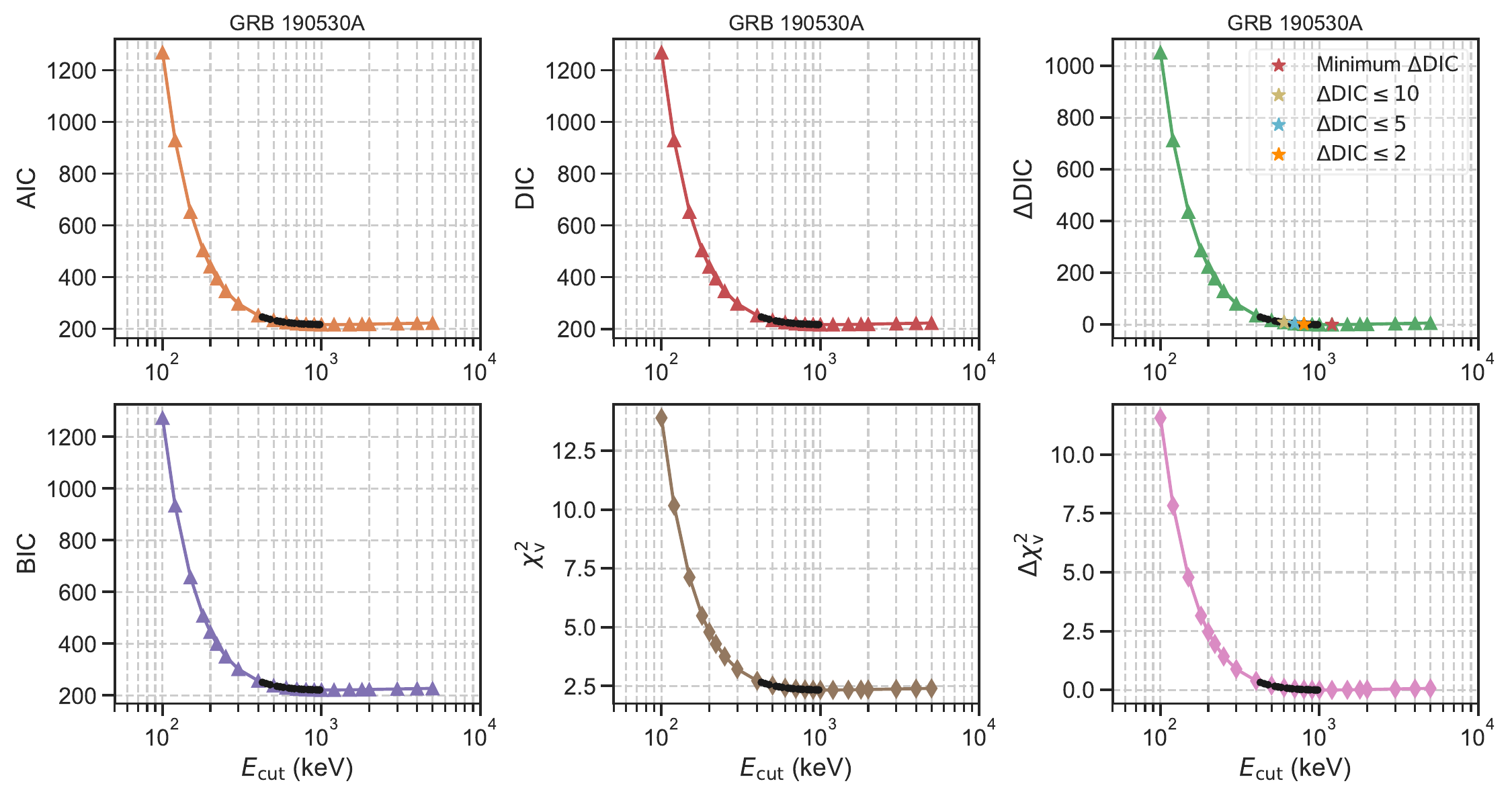}
\includegraphics[angle=0,scale=0.45]{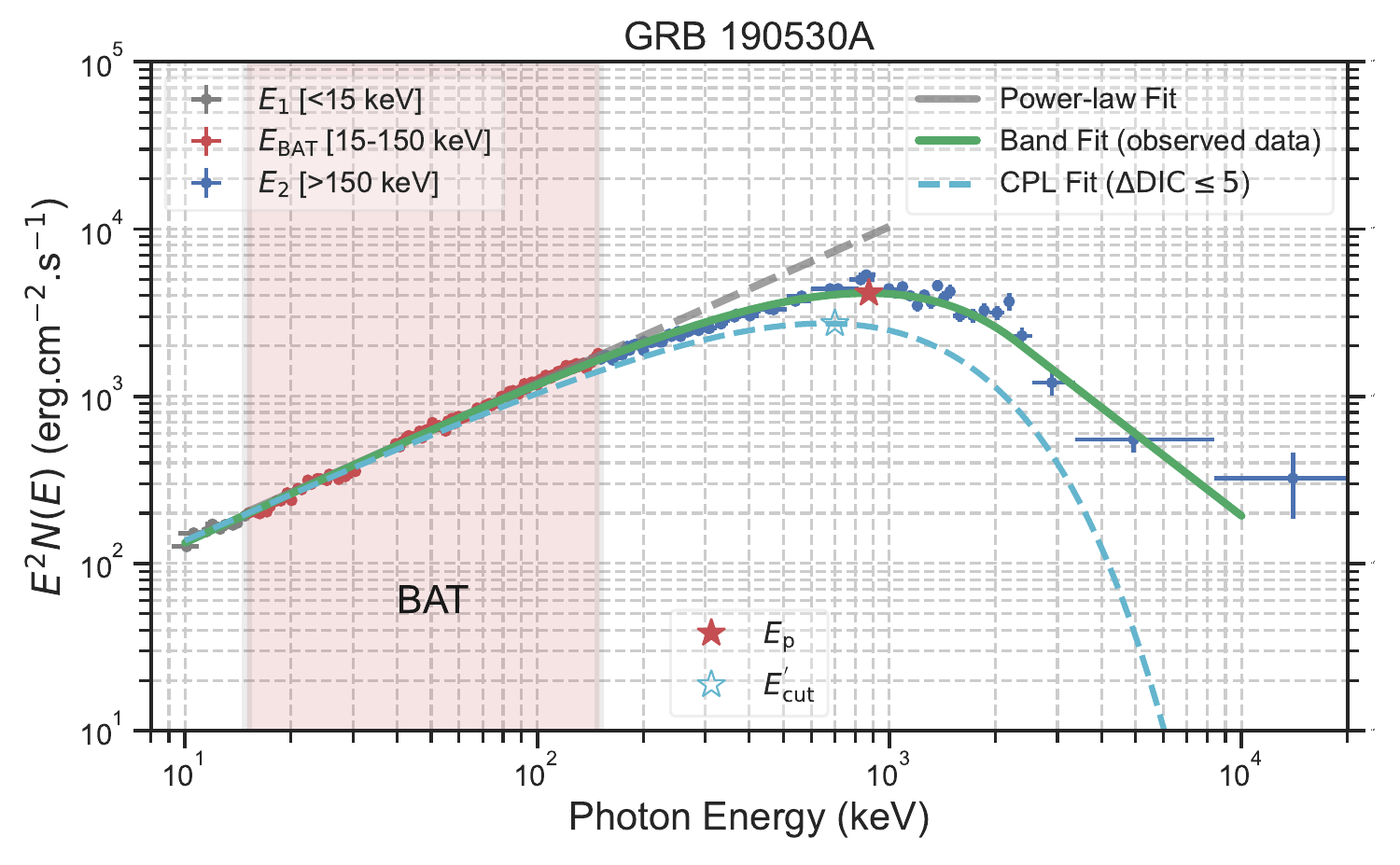}
\caption{Same as Figure \ref{fig:140206A} but for GRB GRB 190530A. Note that the best Band model fit to the original spectral data from GBM observation gives $E_{\rm p}$=866$\pm$13 keV, $A$=(1.3$\pm$0.01)$\times$10$^{-1}$, $\alpha$=-1.01$\pm$0.01, $\beta$=-3.63$\pm$0.20.}\label{fig:190530A}
\end{figure*}

\begin{figure*}
\centering
\includegraphics[angle=0,scale=0.45]{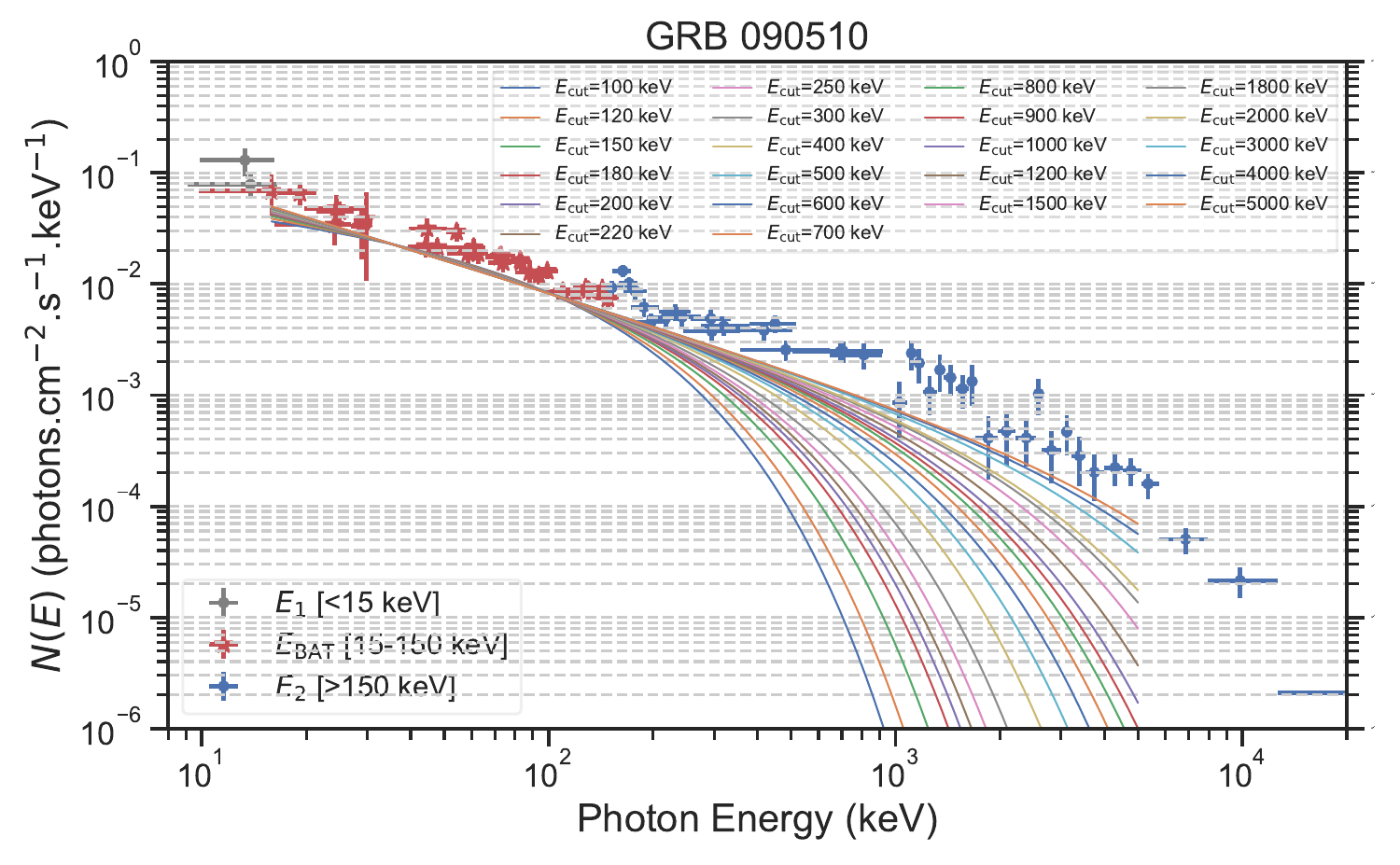}
\includegraphics[angle=0,scale=0.45]{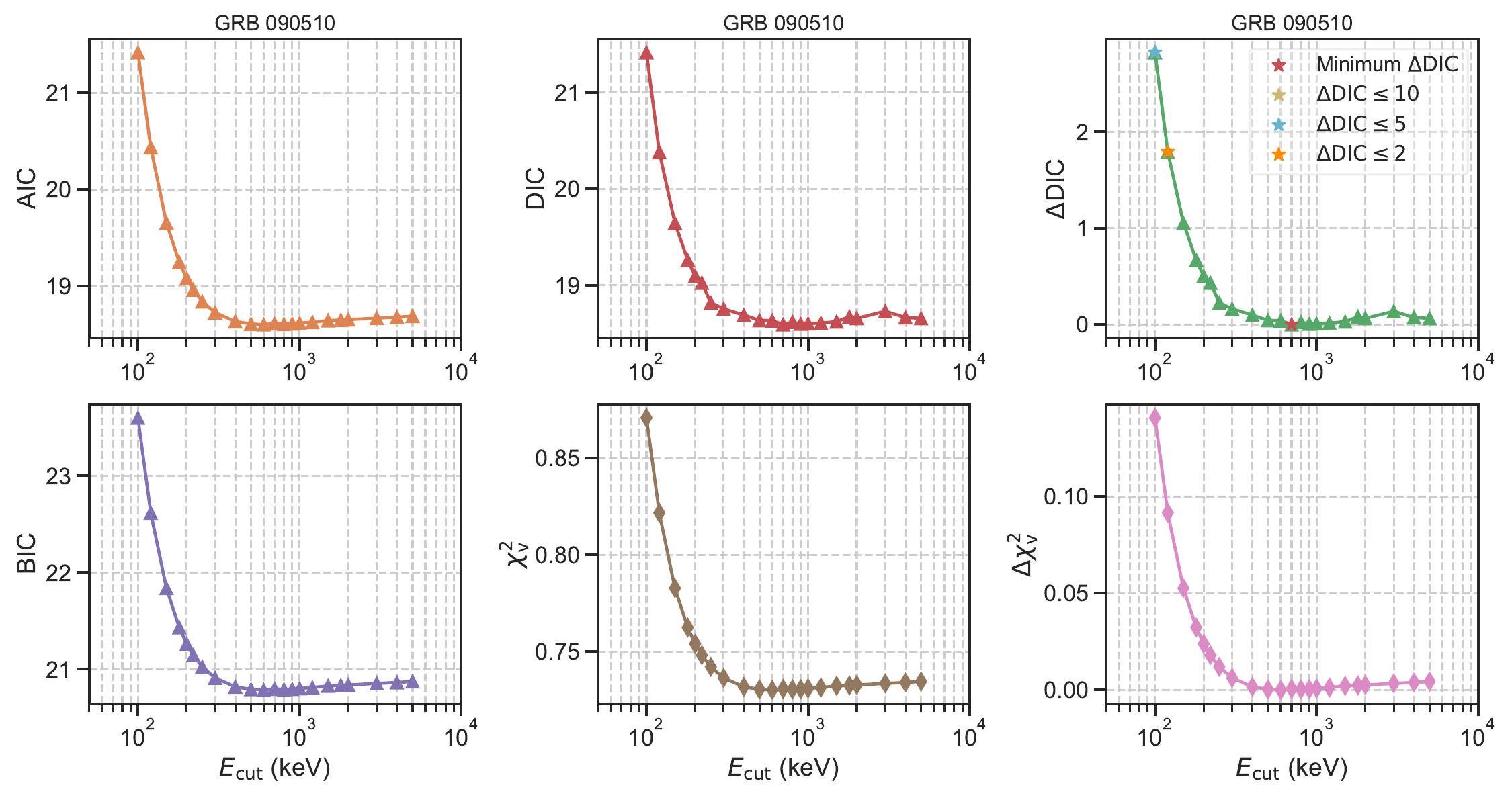}
\includegraphics[angle=0,scale=0.45]{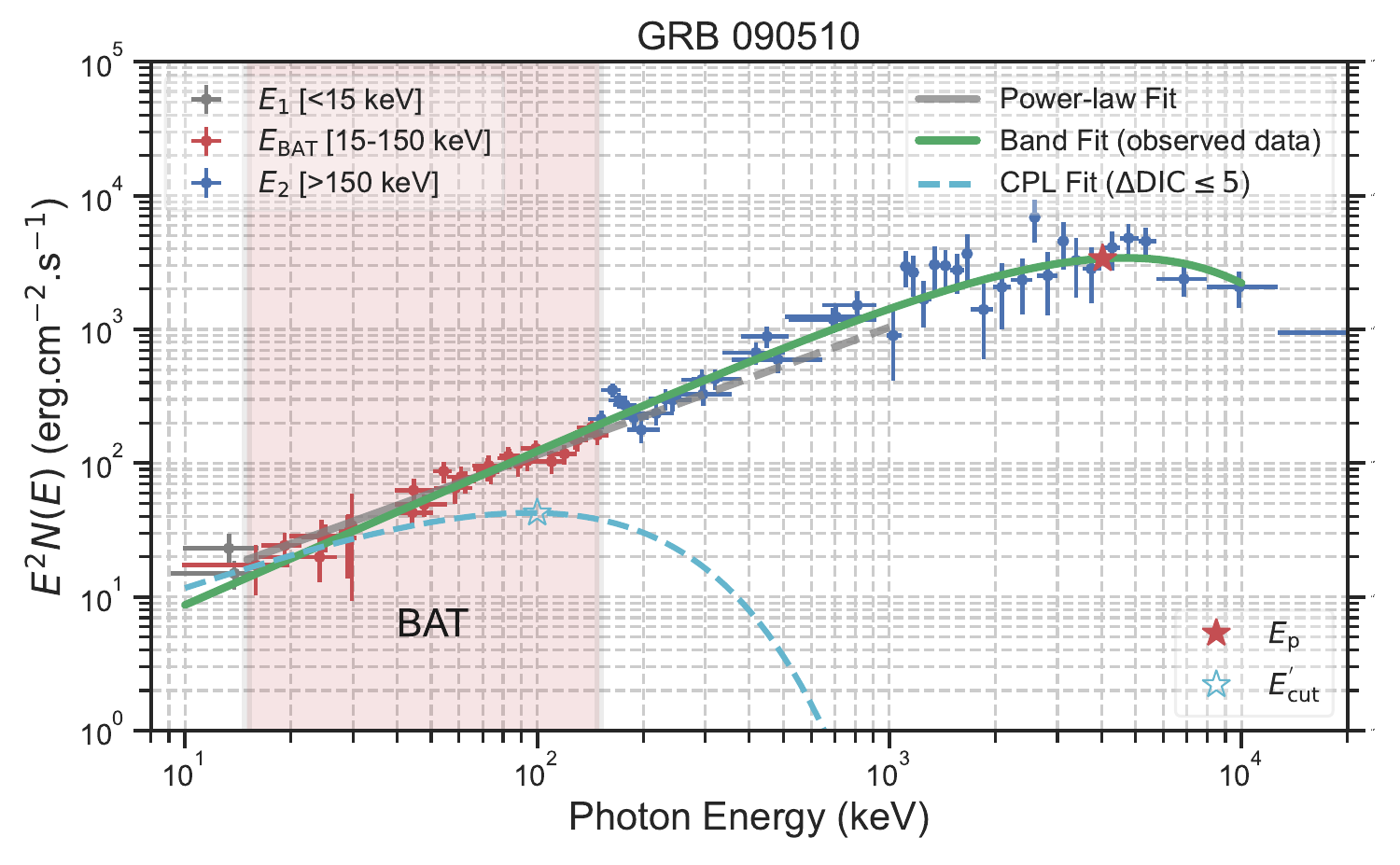}
\caption{Same as Figure \ref{fig:140206A} but for GRB GRB GRB 090510. Note that the best Band model fit to the original spectral data from GBM observation gives $E_{\rm p}$=5200$^{+574}_{-584}$ keV, $A$=(9.0$\pm$0.4)$\times$10$^{-3}$, $\alpha$=-0.89$\pm$0.04, $\beta$=-3.4$\pm$1.3.}
\label{fig:090510}
\end{figure*}

To test the applicability of our extrapolation method for estimating the peak energy ($E_{\rm p}$) of GRBs with Band-like spectra, we selected five GRBs with intrinsic Band spectral models (Figures \ref{fig:140206A}-\ref{fig:090510} and Tables \ref{tab:140206A}-\ref{tab:090510}). Their true $E_{\rm p}$ values range from 280 keV to 5200 keV. Figures \ref{fig:140206A}-\ref{fig:090510} present the preliminary $E_{\rm p}$ estimation results for these bursts. The data analysis procedure is identical to that used for CPL-like GRBs, with the primary distinction appearing in the bottom panels of the figures. In the bottom panels, the green solid lines represent the best-fit Band models derived from the GBM energy range, rather than CPL models. Additionally, the estimated $E^{'}_{\rm c}$ obtained via the extrapolation method is converted to an $E_{\rm p}$ estimate using the relation, here we define $E^{'}_{\rm p}$=(2+$\alpha$)$E^{'}_{\rm c}$.

Another important question to address is whether the variation in $\alpha$ values, as measured within the BAT energy range, has a significant impact on the resulting cutoff energy, $E^{'}_{\rm c}$.

\begin{figure*}
\centering
\includegraphics[angle=0,scale=0.45]{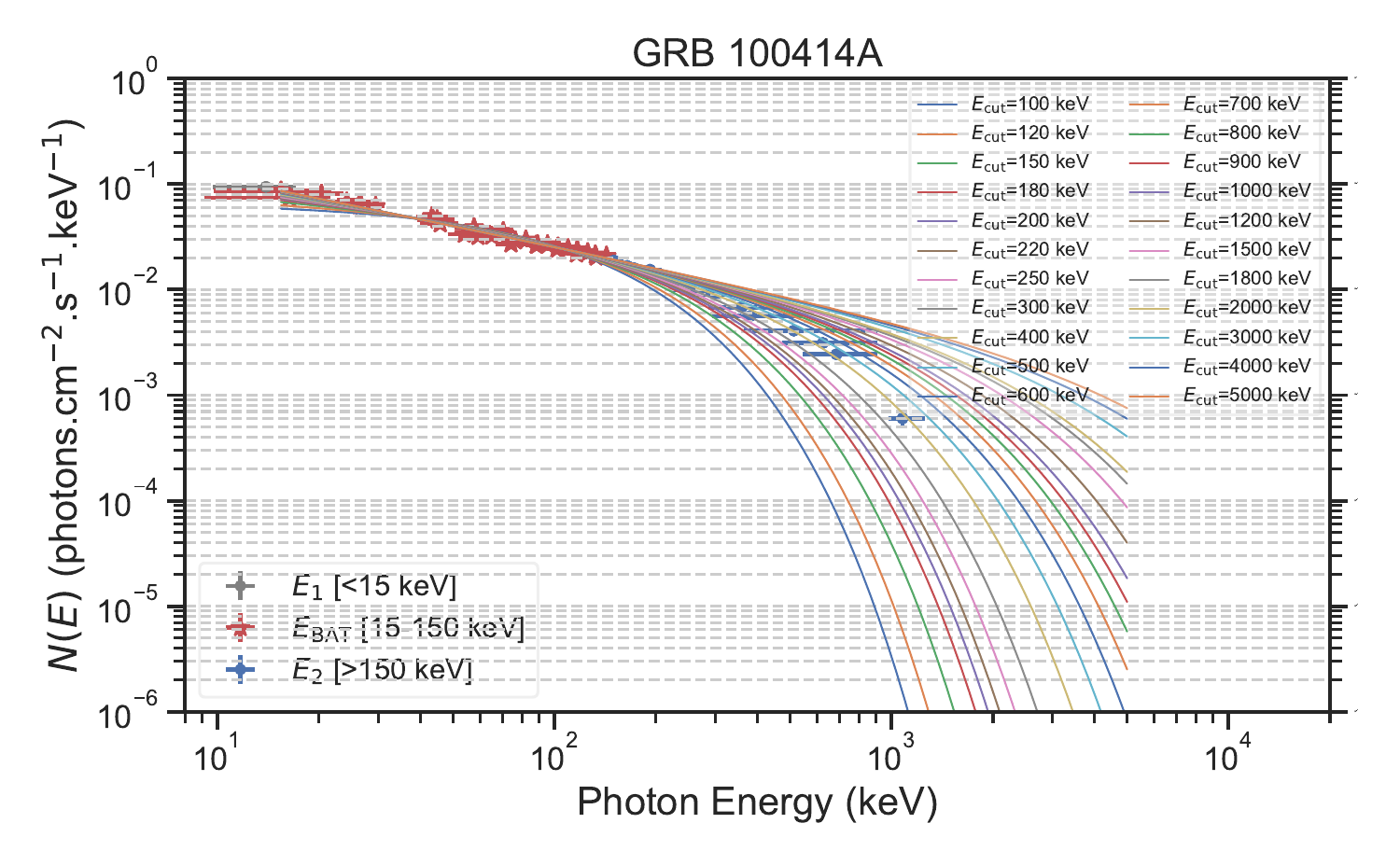}
\includegraphics[angle=0,scale=0.45]{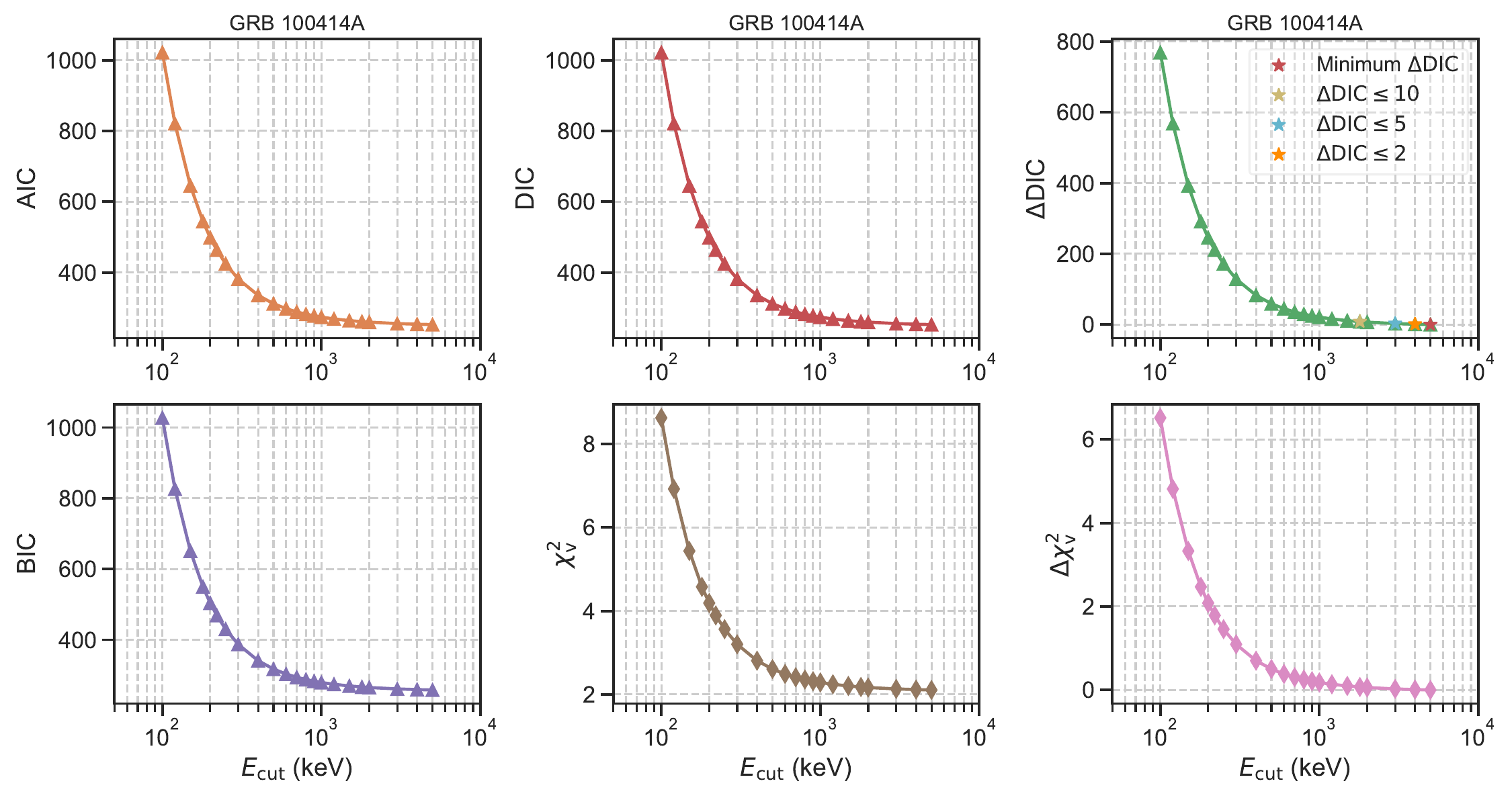}
\includegraphics[angle=0,scale=0.45]{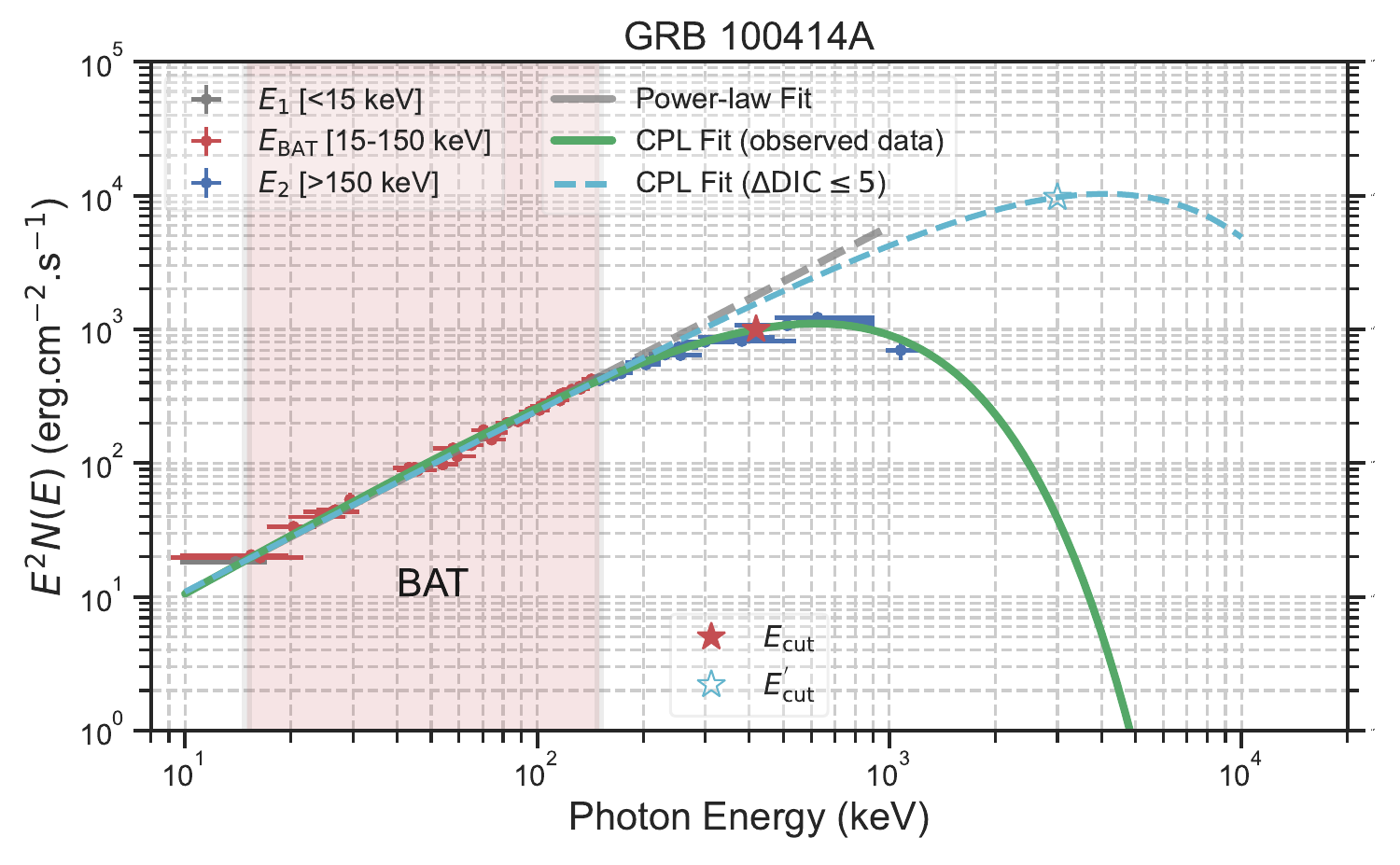}
\caption{Same as Figure \ref{fig:140703A} but for GRB 100414A. Note that the best CPL model fit to the original spectral data from GBM observation gives $E_{\rm c}$=418$^{+21}_{-20}$ keV, $A$=(3.59$^{+0.40}_{-0.34}$)$\times$10$^{-1}$, $\alpha$=-0.52$^{+0.02}_{-0.02}$.}\label{fig:100414A}
\end{figure*}

\begin{figure*}
\centering
\includegraphics[angle=0,scale=0.45]{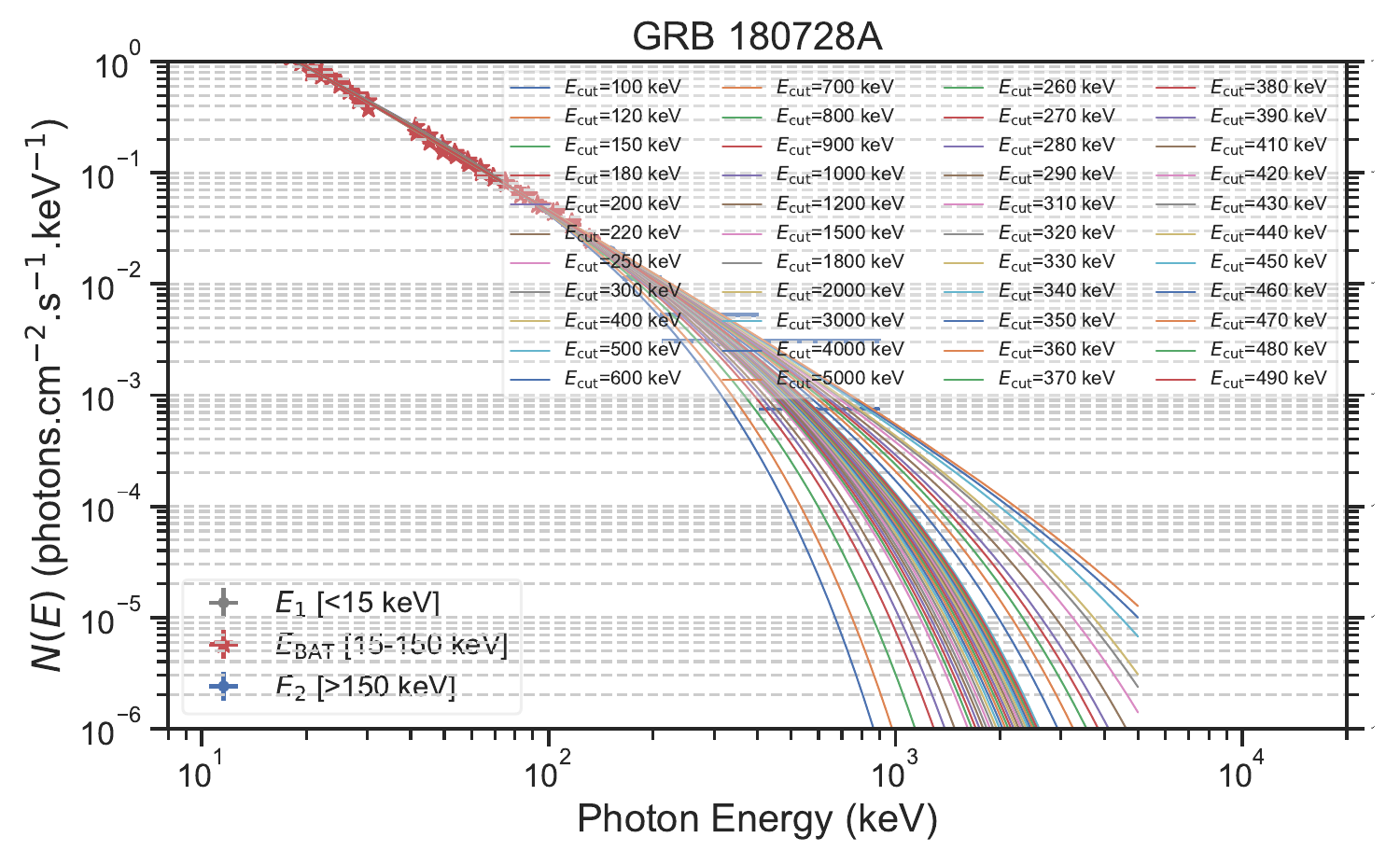}
\includegraphics[angle=0,scale=0.45]{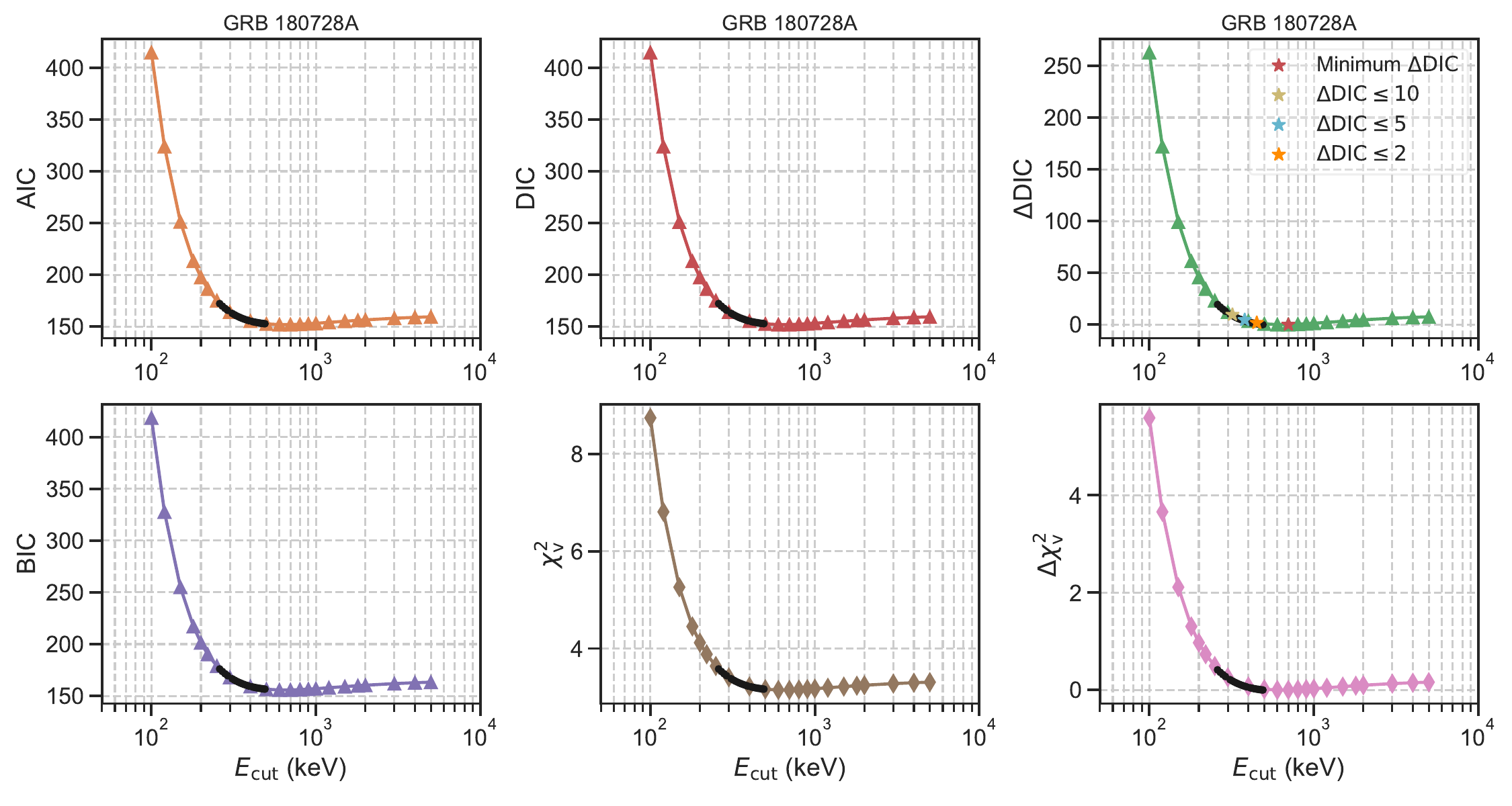}
\includegraphics[angle=0,scale=0.45]{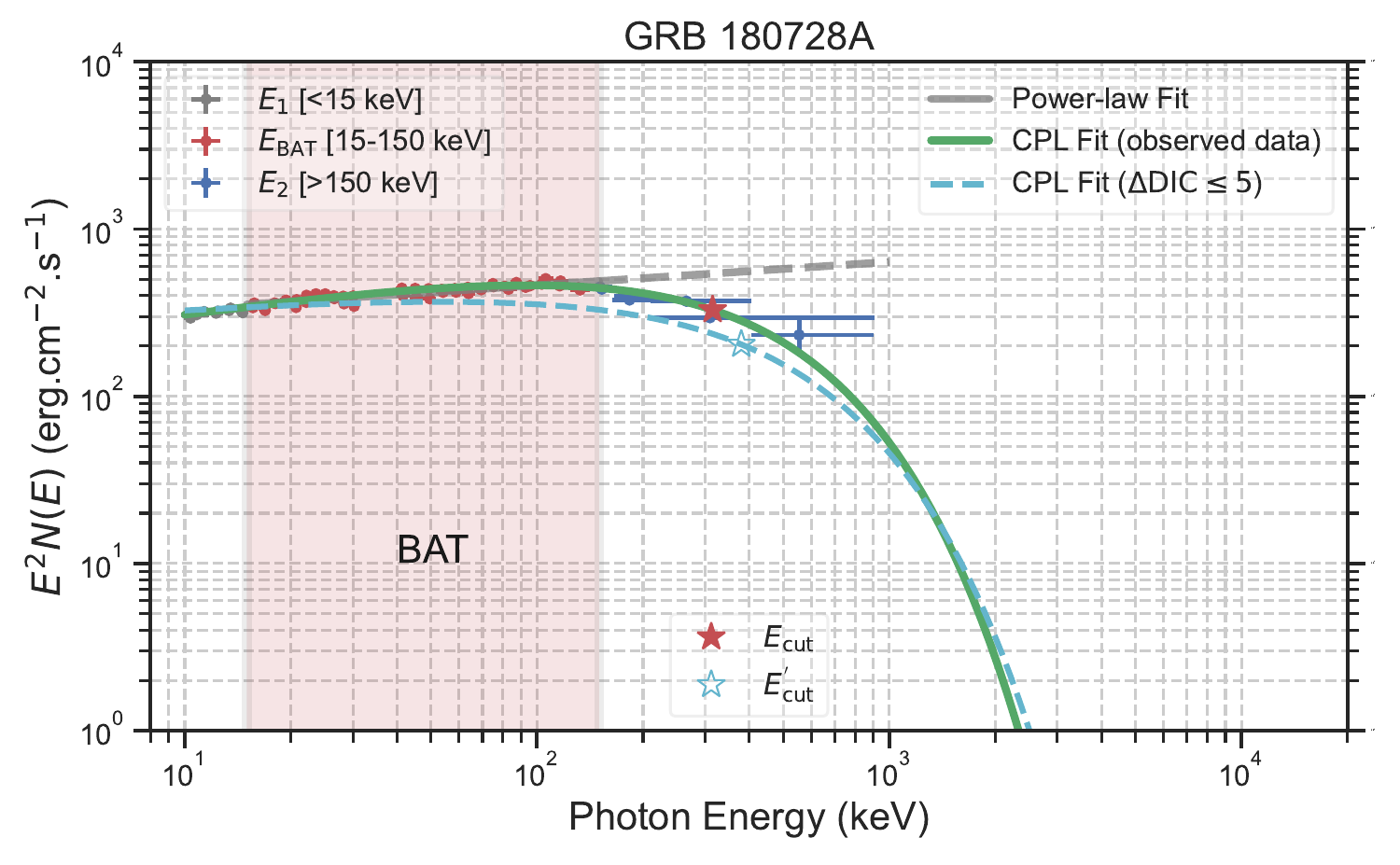}
\caption{Same as Figure \ref{fig:140703A} but for GRB 180728A. Note that the best CPL model fit to the original spectral data from GBM observation gives $E_{\rm c}$=315$^{+19}_{-18}$ keV, $A$=(1.59$^{+0.06}_{-0.05}$)$\times$10$^{2}$, $\alpha$=-1.70$^{+0.01}_{-0.01}$.}\label{fig:180728A}
\end{figure*}

Similarly, to exclude the potential influence of variations in $E_{\rm c}$ on the results, we selected two GRBs from the GBM catalog, where $E_{\rm c}$ remained relatively stable within a typical and narrow range (200 keV to 400 keV), while $\alpha$ gradually increased from -1.9 to -0.4 with a large change (see the lower-panel of Table \ref{tab:CPLresults}). The data analysis procedure is identical to that used for CPL-like GRBs. Figures \ref{fig:100414A}-\ref{fig:180728A} present the preliminary estimation results of the spectral cutoff energy ($E_{\rm c}$) for the two bursts, whose intrinsic spectra follow a CPL-like model. The intrinsic $E_{\rm c}$ values of these GRBs are $E_{\rm c}=418^{+21}_{-20}$ and $E_{\rm c}=315^{+19}_{-18}$, and the intrinsic $\alpha$ values of these GRBs range from $-0.51^{+0.02}_{-0.02}$ to $-1.70^{+0.01}_{-0.01}$, corresponding to GRB 100414A, and GRB 180728A, respectively.

We observed that when the low-energy spectral index $\alpha$ is either very hard (e.g., GRB 100414A with $-0.51^{+0.02}_{-0.02}$) or very soft (e.g., GRB 180728A with $-1.70^{+0.01}_{-0.01}$), the $E^{'}_{\rm c}$ values obtained through the extrapolation method tend to be higher than the true observed $E^{'}_{\rm c}$, showing a noticeable deviation. This deviation is particularly pronounced in cases with a harder $\alpha$ (e.g., GRB 100414A). To quantitatively assess these deviations, we define the ratio $k_\xi\equiv (E^{'}_{\rm c}/E_{\rm c})$ as the value of $E^{'}_{\rm c}$ obtained through the extrapolation method divided by the true observed $E^{'}_{\rm c}$. For GRB 100414A, we find $k_\xi=12.7$ (the statistical criteria $\Delta \text{DIC} \leq 2$ are used), while for GRB 180728A, $k_\xi=1.6$  ($\Delta \text{DIC} \leq 2$). The results indicate that both GRBs have $k_\xi$ values significantly greater than 1, with $k_\xi$ for GRB 100414A being much higher than that for GRB 180728A. This demonstrates that the deviation for GRB 100414A is considerably larger than that for GRB 180728A.

\subsubsection{Refined Estimation of $E^{'}_{\rm c}$}\label{sec:RefEstimation}

We obtained a relatively coarse value of $E^{'}_{\rm c}$ from the preliminary estimation. These $E^{'}_{\rm c}$ values, derived through the extrapolation method, are consistent with the true observed values for most GRBs. If the precision requirement for $E_{\rm c}$ estimation is low, the preliminary estimation suffices. However, for higher precision demands, a refined estimation can be conducted based on the initial $E_{\rm c}$ results, aiming to confine the uncertainty to a smaller range. The process of refined estimation involves the following steps:

From the preliminary estimation, we found that the statistical criterion $\Delta \text{DIC}\leq 5$ effectively identifies $E_{\rm c}$. Thus, we take the $E_{\rm c}$ value obtained under $\Delta \text{DIC}\leq 5$ as the starting point for refined estimation. For each GRB with $E_{\rm c}\leq500$ keV, we select 10 additional $E_{\rm c}$ values at uniform intervals before the starting point and 5 additional $E_{\rm c}$ values after it. The interval i is determined as follows: i=10 for $E_{\rm c} \leq 500$ keV, i=25 for $500\ \text{keV} \leq E_{\rm c} \leq 1000\ \text{keV}$, and i=50 for $E_{\rm c} \geq 1000$ keV.

We repeat all data analysis steps from the preliminary estimation to obtain a finer set of $E_{\rm c}$ results with smaller intervals and reduced uncertainties. The fitting results (e.g., $A$, $\alpha$, $\chi^{2}$, DOF, $\chi_{\rm r}^{2}$, AIC, BIC, DIC, and $p_{\rm DIC}$) are summarized in the lower-panel of Tables \ref{tab:140703A}-\ref{tab:190530A}. Three GRBs were excluded from this refined analysis because the $E_{\rm c}$ values obtained from the preliminary estimation differed significantly from their observed values, rendering further detailed estimation meaningless.

Combining the preliminary and refined estimations, we find that among the 10 GRBs in Group I, 8 GRBs have their estimated $E^{'}_{\rm c}$ values closely matching the observed $E_{\rm c}$. Only two GRBs (GRB 120711A and GRB 130215A) show significant deviation between the estimated $E^{'}_{\rm c}$ and their observed $E_{\rm c}$ values, both of which have $E_{\rm c}> 1000$ keV, far exceeding the upper energy limit of the BAT instrument (150 keV). For the two GRBs in Group II, one (GRB 180728A) exhibits an estimated $E^{'}_{\rm c}$ that matches the observed value closely, albeit slightly higher. This GRB is characterized by a very soft intrinsic $\alpha$ index (-1.70$\pm$0.01). In contrast, the other burst (GRB 100414A) shows a large deviation between its estimated $E^{'}_{\rm c}$ and observed $E_{\rm c}$, with the intrinsic $\alpha$ index being notably hard (-0.52$\pm$0.02).

\section{Discussions}\label{sec:Dis}

\subsection{Curvature in the CPL function within the Swift/BAT Energy Range}

The curvature of a curve is defined as the measure of its deviation from a straight line, providing a quantitative description of the degree of bending. In the context of gamma-ray burst spectral analysis, curvature can be utilized to examine the bending behavior of the CPL function within the energy range observed by \emph{Swift}-BAT (15-150 keV). This analysis is particularly relevant for evaluating how well the CPL function captures spectral features in this energy band, especially when the intrinsic spectral shape exhibits significant deviation from linearity. We, therefore, use {\tt curvature} to describe the degree of bending of a curve at a point, reflecting how rapidly its geometric shape changes. The formula for curvature in two-dimensional space is given by:
\begin{equation}
\kappa = \frac{|f''(E)|}{\left(1 + \left(f'(E)\right)^2\right)^{3/2}}
\end{equation}
where $f(E)$ is the function value corresponding to energy $E$ (e.g., the photon spectrum density in the CPL or Band model), $f'(E)$ is the first derivative of $f(E)$ with respect to $E$. $f''(E)$ is the second derivative of $f(E)$ with respect to $E$.

The total curvature is computed using the following integral expression:
\begin{equation}
C = \int_{E_{\text{min}}}^{E_{\text{max}}} \kappa(E) \, dE
\end{equation}
where $\kappa(E) = \frac{|f''(E)|}{\left(1 + \left(f'(E)\right)^2\right)^{3/2}}$ is the curvature at energy $E$, $E_{\text{min}}$ and $E_{\text{max}}$ are the lower and upper bounds of the energy range (e.g., 15 keV and 150 keV).

This expression is numerically evaluated using the trapezoidal rule, implemented in Python with the \texttt{np.trapz} function, which approximates the integral by summing the areas of trapezoids formed under the curve.

\begin{figure*}
\includegraphics[angle=0,scale=0.37]{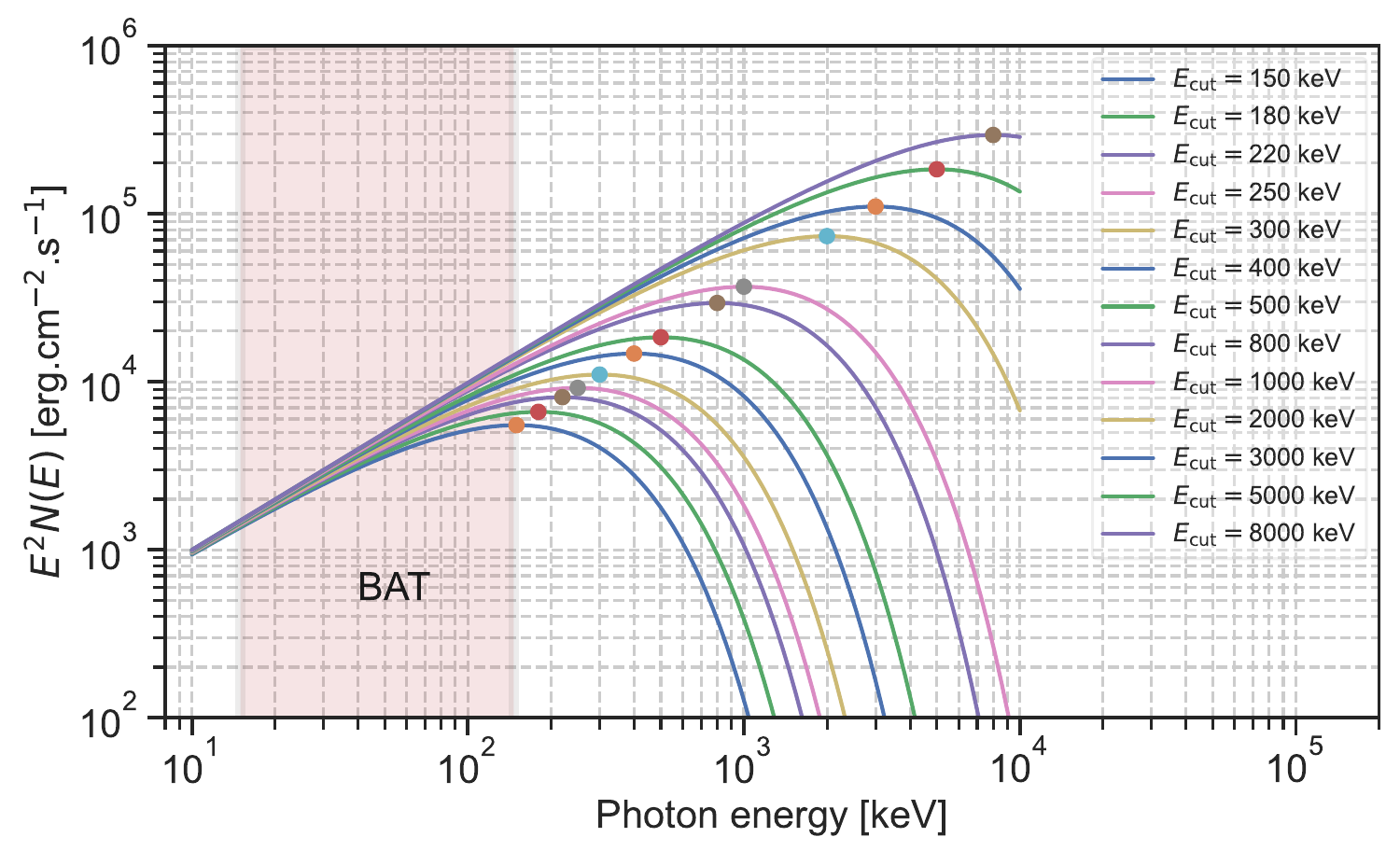}
\includegraphics[angle=0,scale=0.39]{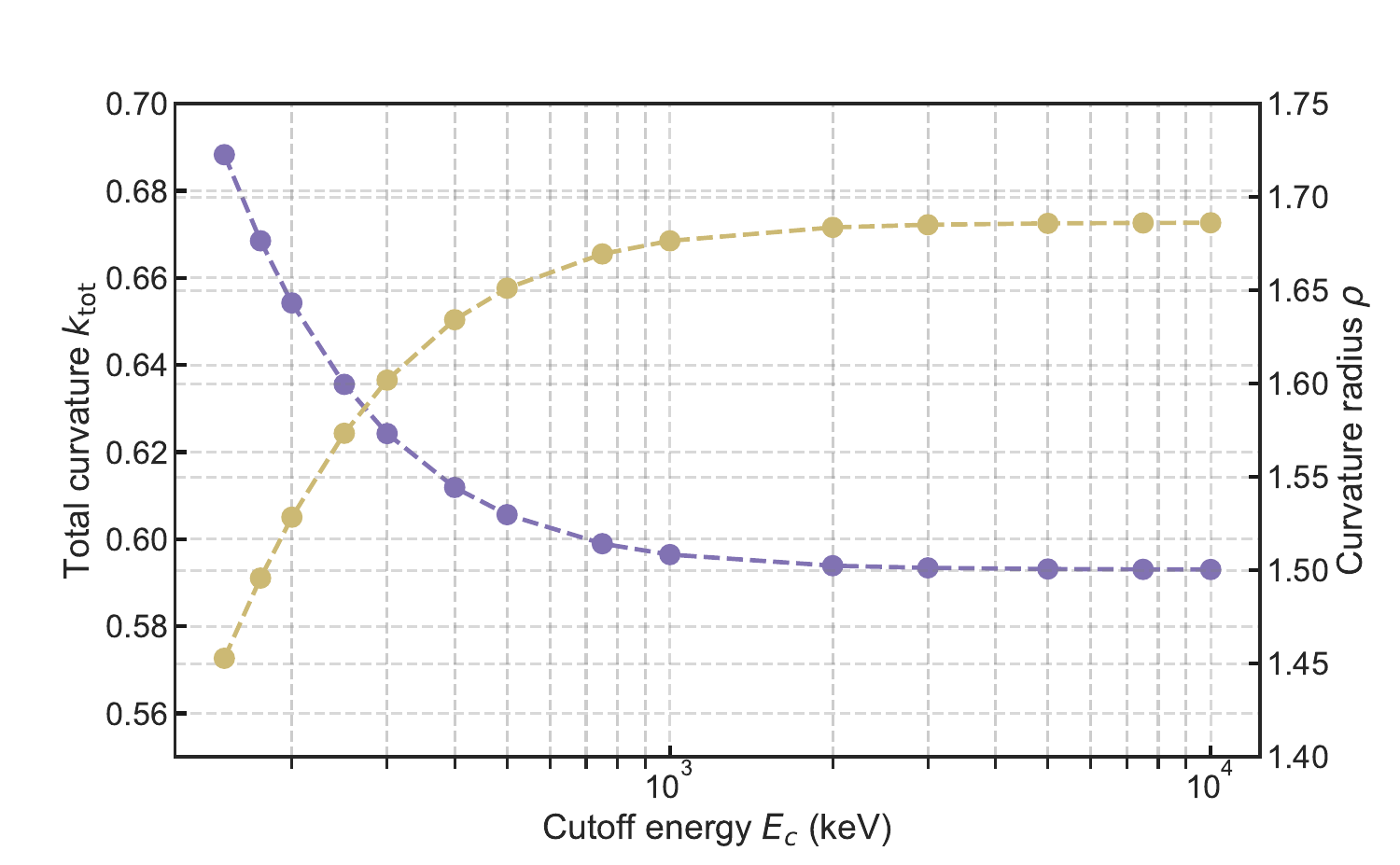}
\includegraphics[angle=0,scale=0.37]{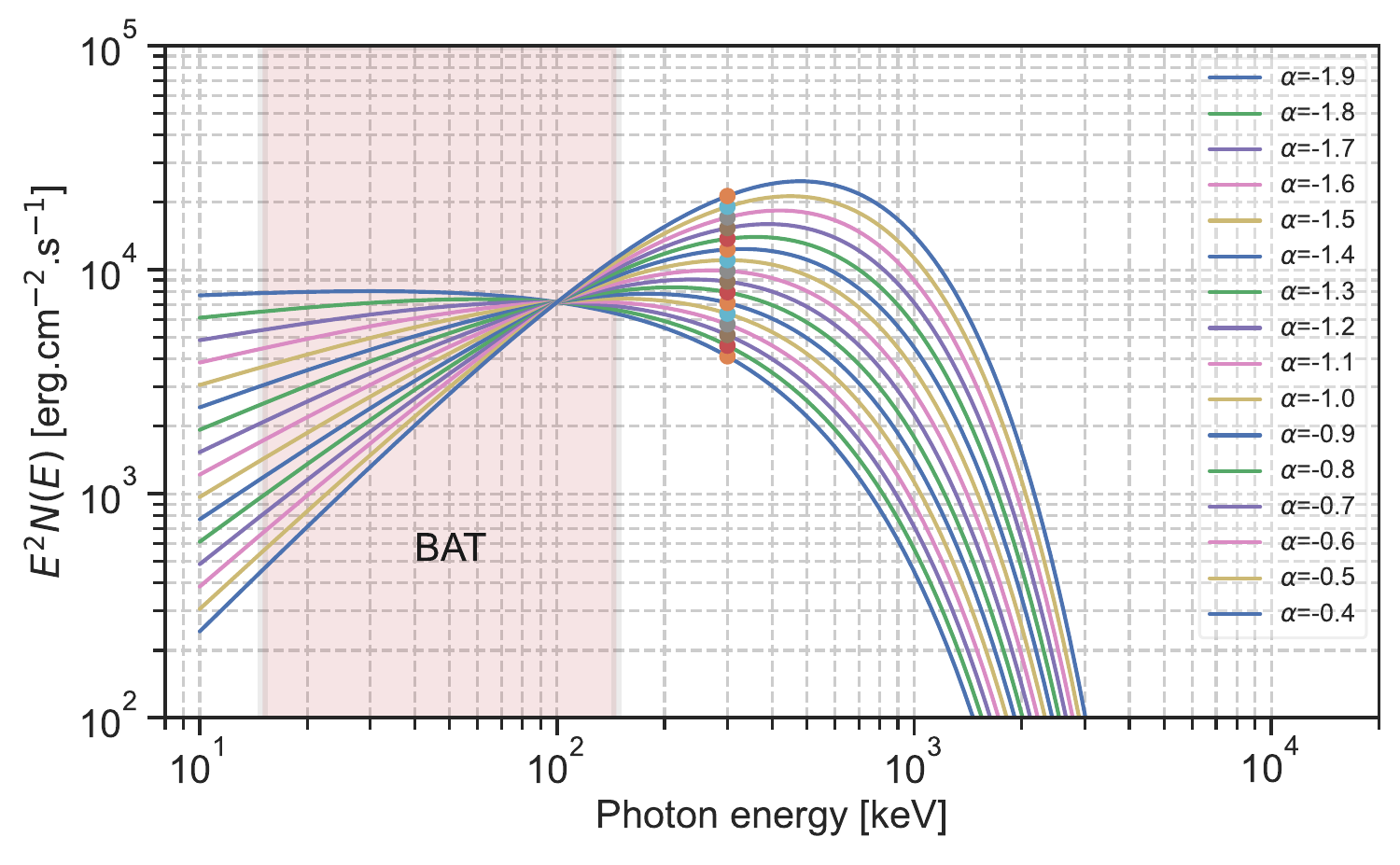}
\includegraphics[angle=0,scale=0.39]{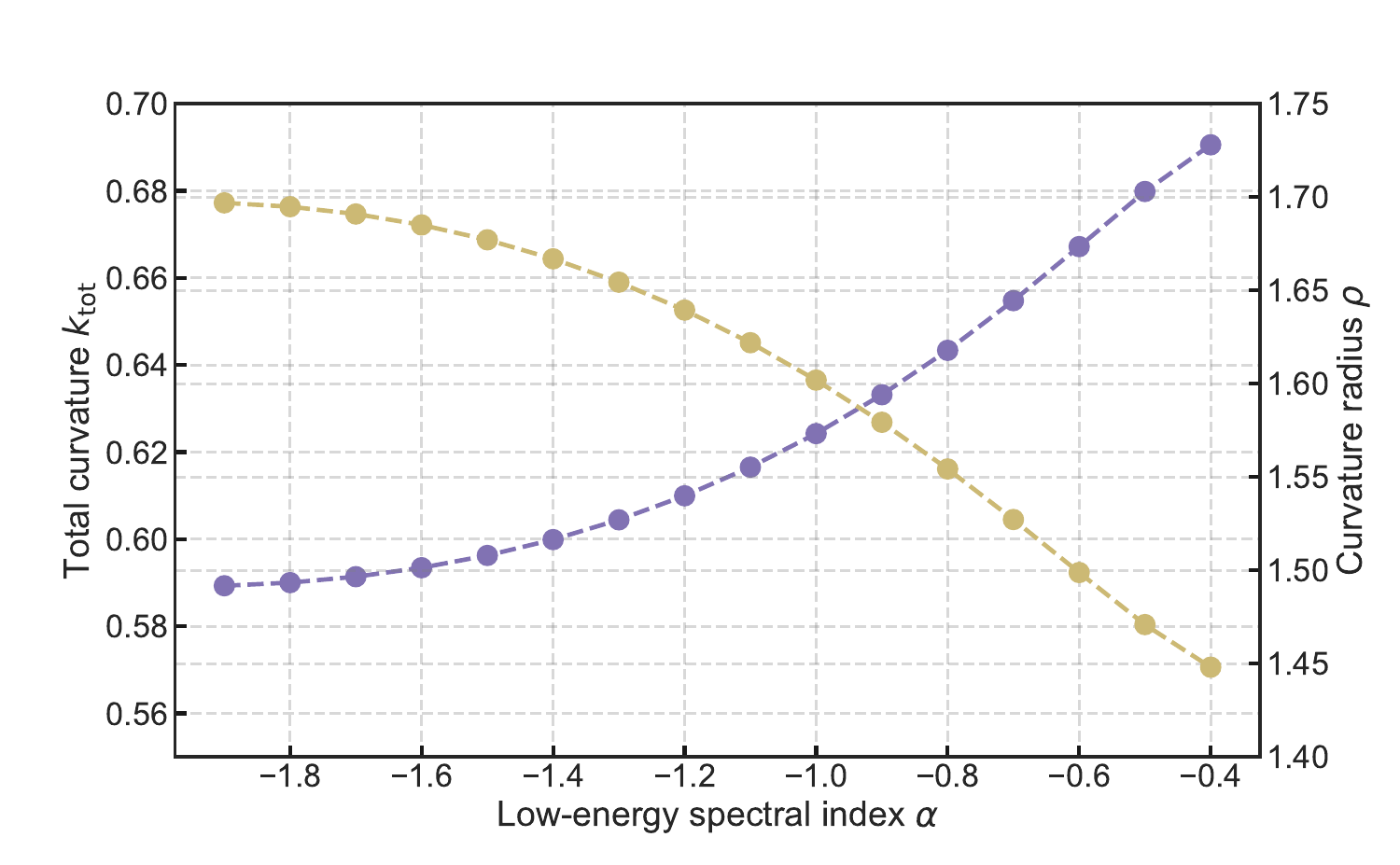}
\includegraphics[angle=0,scale=0.37]{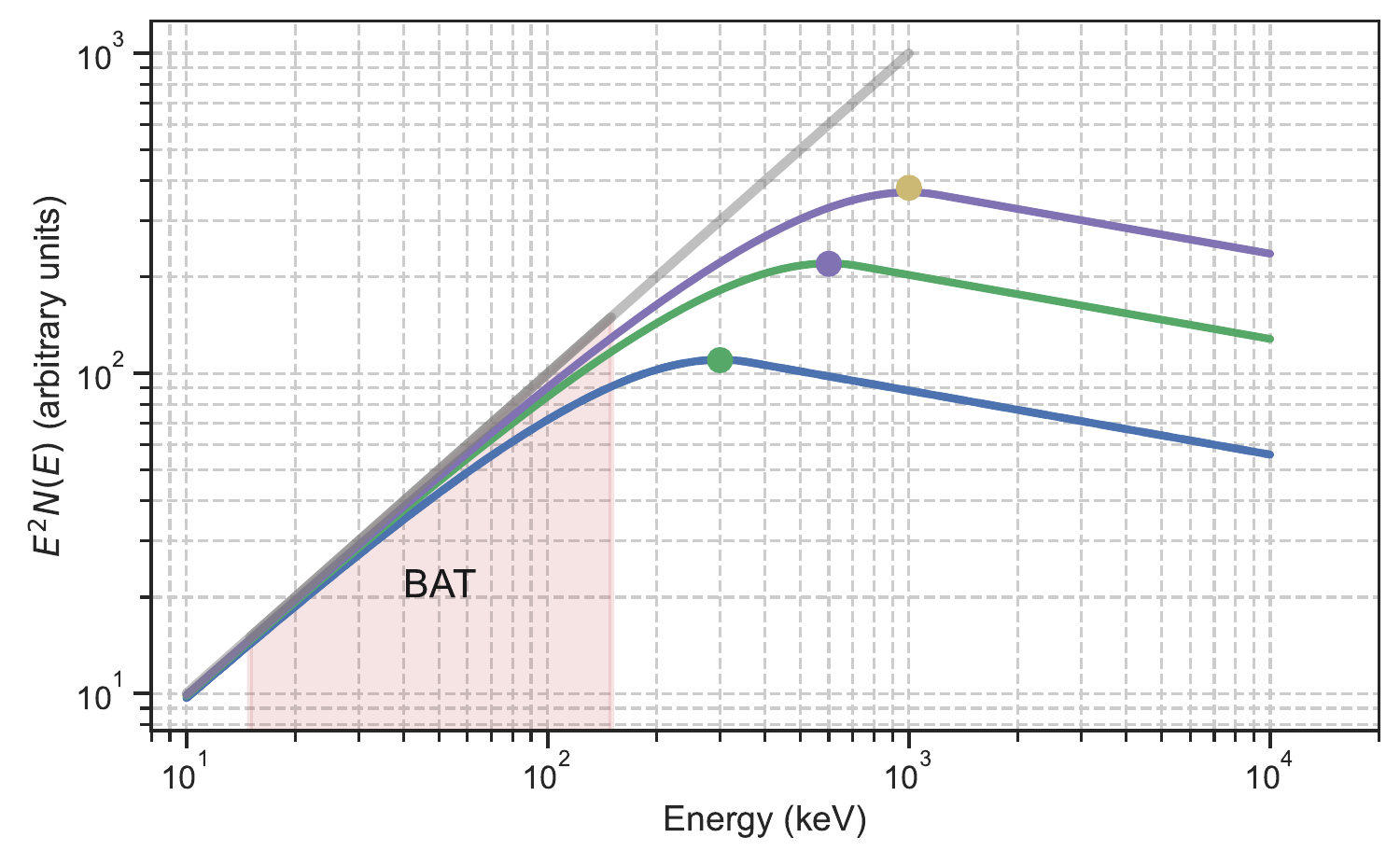}
\includegraphics[angle=0,scale=0.37]{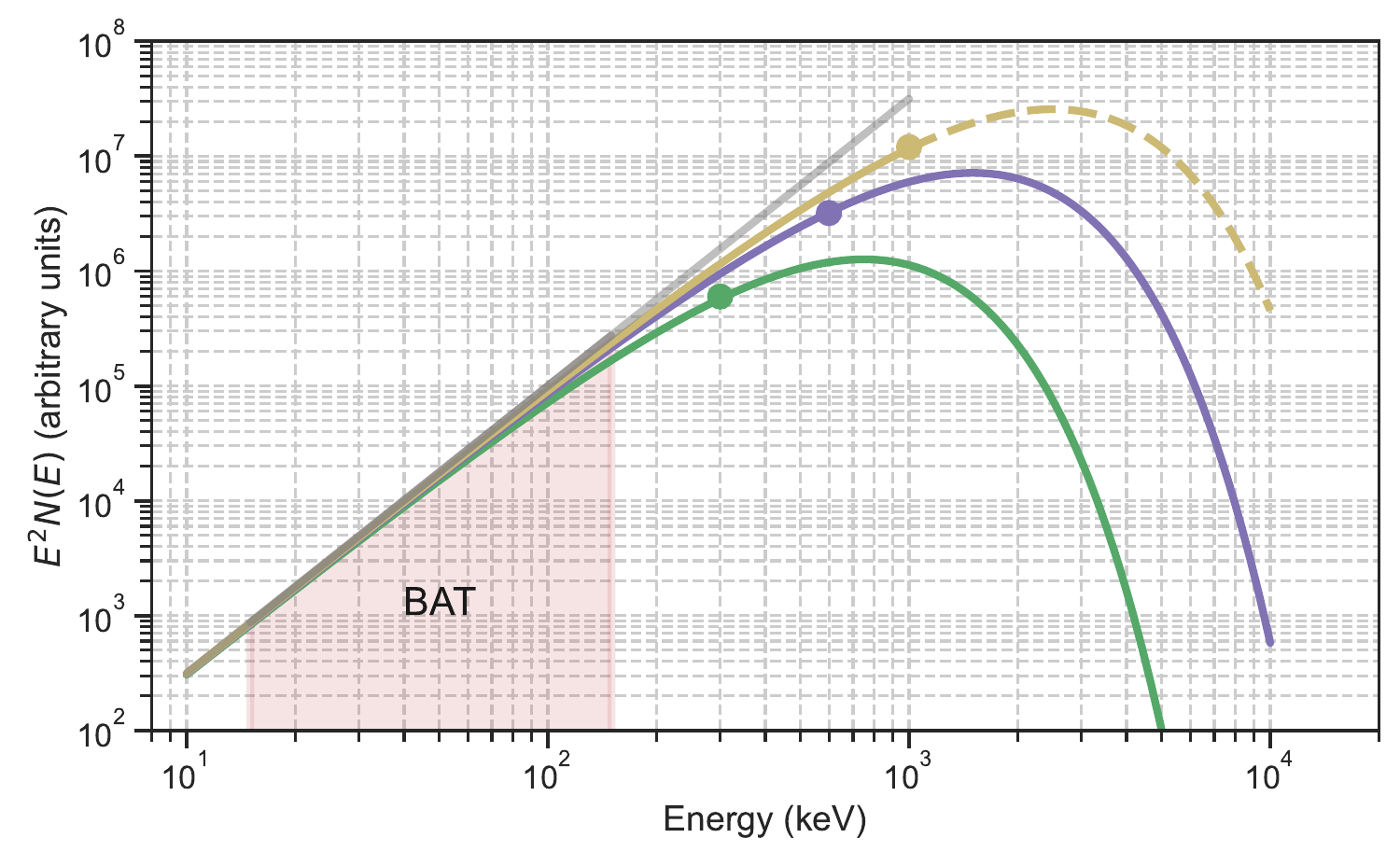}
\caption{{\bf a,} Spectra of CPL models with the same low-energy power-law index $\alpha$ but different fixed $E^{'}_{\rm c}$ values, plotted in the $\nu F_\nu$ space. The solid points in different colors represent the corresponding cut-off energies $E^{'}_{\rm c}$ values.
{\bf b,} Total curvature and the associated curvature radius as a function of $E^{'}_{\rm c}$.
{\bf c,} Same as {\bf (a)} but for variations in the $\alpha$ index.
{\bf d,} Same as {\bf (b)} but for variations in the $\alpha$ index.
{\bf e,} Spectra of Band models with different peak energies ($E_{\rm p}$=300 keV, 600 keV, and 1000 keV), as well as the baseline power-law model (the same $A$ fixed 1 and $\alpha$ fixed -1). 
{\bf f,} Same as {\bf (e),} but for the CPL model.}
\label{fig:Curvature}
\end{figure*}

To quantitatively evaluate the impact of changes in the cut-off energy $E_{\rm c}$ on the curvature of the CPL function within the BAT energy range (15-150 keV), we conducted the following analysis. First, we fixed two parameters, the normalization ($A=1$) and the low-energy spectral index ($\alpha=-1$), at typical values. We then allowed $E_{\rm c}$ to vary, selecting several representative observational values ranging from 150 keV to 10,000 keV (see the upper-left panel of Figure \ref{fig:Curvature}). For each $E_{\rm c}$, we calculated the total curvature $C$ (see the upper-right panel of Figure \ref{fig:Curvature}) and the corresponding curvature radius $\rho$ (see the lower-right panel of Figure \ref{fig:Curvature}). Our results indicate that the total curvature decreases as $E_{\rm c}$ increases. This suggests that when the intrinsical $E_{\rm c}$ of a CPL spectrum approaches the upper limit of the BAT energy range (150 keV), the extrapolated curvature provides a more reliable estimate of the $E_{\rm c}$, even though the decrease in curvature with increasing $E_{\rm c}$ is not particularly steep.
  
Similarly, we analyzed the effect of changes in the low-energy spectral index $\alpha$ on the curvature of the CPL function in the BAT energy band. For this analysis, we fixed $A=1$ and $E_{\rm c}$ keV as typical values and allowed $\alpha$ to vary. We selected several representative observational values for $\alpha$ ranging from -1.9 to -0.4 (see the lower-left panel of Figure \ref{fig:Curvature}). For each $\alpha$, we calculated the total curvature $C$ (see the left-panel of Figure \ref{fig:Curvature}) and the corresponding curvature radius $\rho$ (see the lower-right panel of Figure \ref{fig:Curvature}). We found that the total curvature decreases as $\alpha$ becomes harder, indicating that when the intrinsically $\alpha$ of a CPL spectrum is harder, the extrapolated curvature method yields a more reliable estimate of the true peak energy $E_{\rm p}$.

The general effects of each parameter on the curvature of the CPL function based on theory: 
(1) Normalization $A$: Typically, the curvature remains nearly unaffected by changes in $A$ because it scales the amplitude of the function without altering the overall shape.
(2) Low-energy spectral index $\alpha$: Changing $\alpha$ often has a noticeable effect on curvature, especially at lower energies, as it alters the slope of the function in this region.
(3) Peak energy $E_{\rm p}$: Modifying $E_{\rm p}$ shifts the spectrum's peak, which can significantly affect the curvature, particularly around the peak's location in the energy band.
(4)  High-energy spectral index $\beta$: Variations in $\beta$ influence the curvature at higher energies, though their impact on curvature in the 15-150 keV range might be less prominent unless the spectrum has significant high-energy contributions.

\subsection{Comparison with Other Methods}

\cite{Sakamoto2009} proposed a correlation between the peak energy ($E_{\rm p}$) in the $\nu F_\nu$ spectrum and the photon index $\Gamma$ derived from a simple power-law (SPL) model, expressed as: log $E_{\rm p}$=3.258-0.829$\Gamma$, (1.3$\leqslant\Gamma\leqslant$2.3). To evaluate this method in comparison to ours, we applied the SPL model to fit the spectral data in the BAT energy range. The fitted photon index $\Gamma$ and the $E_{\rm p1}$ values derived using the above relation are summarized in Columns 3 and 11 of Table \ref{tab:PLresults}. Moreover, we identified 9 GRBs observed simultaneously by both GBM and BAT. For these bursts, we collected fitting results provided by the \emph{Swift} Science Team. It is worth noting that almost all these GRBs were poorly fitted by the SPL model, yet the SPL model was still reported as the best-fitting model on the \emph{Swift} Science Data Centre website\footnote{\url{https://swift.gsfc.nasa.gov/results/batgrbcat/}}.

\begin{figure*}
\centering
\includegraphics[angle=0,scale=0.6]{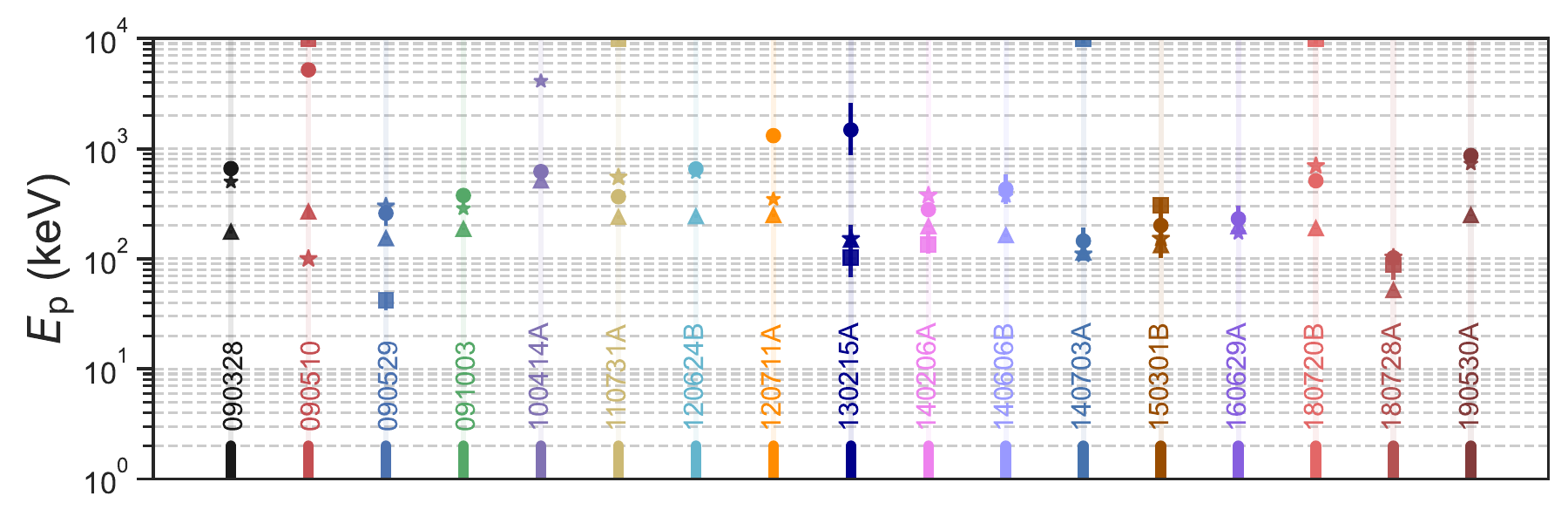}
\caption{Comparison of several methods for estimating the $E_{\rm p}$ of \emph{Swift} GRBs. The solid circles represent the true observed $E_{\rm p}$ values, the stars denote the $E_{\rm p}$ values estimated using our spectral curvature extrapolation method, the triangles indicate the $E_{\rm p}$ values derived from the empirical relation proposed by \cite{Sakamoto2009}, and the squares show the $E_{\rm p}$ values provided by the Science Data Centre website based on BAT data fitting. Different colors correspond to different bursts.}\label{fig:Ep_GRBs}
\end{figure*}

We compared several methods for estimating the $E_{\rm p}$ of \emph{Swift} GRBs, and the results are presented in Figure \ref{fig:Ep_GRBs}. In the figure, the solid circles represent the true observed $E_{\rm p}$ values (obtained from GBM data), the stars indicate the $E_{\rm p}$ values estimated using our spectral curvature extrapolation method, the triangles represent the $E_{\rm p}$ values derived from the empirical relation proposed by \cite{Sakamoto2009}, and the squares denote the $E_{\rm p}$ values provided by the Science Data Centre website\footnote{\url{https://swift.gsfc.nasa.gov/sdc/}} based on BAT data fitting.

The comparison reveals that the $E_{\rm p}$ values provided by the \emph{Swift} Science Team show the poorest agreement with the true observed values (GBM data). When compared to the method of \cite{Sakamoto2009}, our extrapolation method yields significantly better $E_{\rm p}$ estimates for nine GRBs. However, for three bursts (GRB 090510, GRB 120711A, and GRB 130215A) neither method performs well. Notably, the true $E_{\rm p}$ values of these three bursts are exceptionally high ($E_{\rm p}>1000$ keV), likely exceeding the applicable range of both methods.

Interestingly, only one burst (GRB 100414A) shows better $E_{\rm p}$ estimation using the method describe in  \cite{Sakamoto2009} compared to ours. This exception may be attributed to the harder spectral index ($\alpha$) of this GRB. Based on this result, we suggest that the method in \cite{Sakamoto2009} might be more reliable for GRBs with harder $\alpha$ values, whereas our curvature extrapolation method is generally more effective for the majority of GRBs.

\section{Conclusions}\label{sec:Con}

The \emph{Swift} Burst Alert Telescope (BAT), operating in the 15--150 keV energy band, has significantly advanced our understanding of nature of gamma-ray bursts, in both theory and observation aspects \citep{Gehrels2009}. However, its limited energy coverage poses challenges when analyzing bursts whose spectral peak energies exceed 150 keV, as most GRBs have $E_{\rm p}$ values typically distributed between 200-300 keV, exceeding BAT's upper limit. Given that \emph{Swift}/BAT operates within the 15-150 keV energy band, most GRBs with peak energies exceeding 150 keV present a significant challenge for precise $E_{\rm p}$ estimation. In this paper, we developed an innovative method to reliably estimate the cut-off energy ($E^{'}_{\rm c}$) and peak energy ($E^{'}_{\rm p}$) of GRB spectra observed by the \emph{Swift}/BAT satellite, particularly when the $E_{\rm c}$ (or $E_{\rm p}$) lies above the instrument's observational energy range ($>$150 keV). 

We selected 17 bright GRBs for performing such a test. In order to perform a self-consistency check, all the bursts are selected from the GBM observations, which provides a broader energy observation window covering both the BAT energy range and higher energy bands beyond BAT's upper limit. Using the intrinsic curvature of the CPL spectrum within the BAT range, we extrapolated the $E^{'}_{\rm c}$ values for energies beyond BAT's range. To examine whether the estimated $E^{'}_{\rm c}$ values are influenced by the true observed $E_{\rm c}$ and the low-energy spectral index $\alpha$, we divided the sample into three Groups. Group I: Consists of 10 GRBs with $\alpha$ values consistently within the typical range, while $E_{\rm c}$ varies significantly from 150 keV to several MeV. Group II: Includes 2 GRBs with $E_{\rm c}$ approximately fixed within the typical distribution range but with $\alpha$ values varying significantly. Group III: Includes 5 GRBs where the optimal spectral model is the Band model, providing an additional test for consistency.

We then performed a two-step estimation process for these samples: preliminary estimation and refined estimation. We selected 22 reasonably spaced and gradually increasing $E_{\rm c}$ values ranging from 150 keV to several MeV. For each GRB in the sample, we fixed $E_{\rm c}$ and used the CPL model to fit the spectral data within the BAT energy range. Using a statistical criterion of $\Delta \text{DIC} \leq 5$, we preliminarily selected a rough $E_{\rm c}$ estimate. Based on the results of the preliminary estimation, we then selected a narrower range of $E^{'}_{\rm c}$ values for each GRB and conducted refined estimation. The number of $E^{'}_{\rm c}$ values chosen for refinement varied depending on the preliminary results. In Group I, 8 GRBs showed excellent agreement between the extrapolated $E^{'}_{\rm c}$ and the true observed values. These GRBs all had observed $E_{\rm c}$ values within a moderate range (150--1000 keV). However, two bursts (GRB 120711A and GRB 130215A) with significantly mismatched $E_{\rm c}$ estimates had observed $E_{\rm c}$ values exceeding 1000 keV, far beyond BAT's upper energy limit. In Group II, one burst (GRB 180728A) showed good agreement between the estimated $E^{'}_{\rm c}$ and observed $E_{\rm c}$, with its $\alpha$ index being relatively soft. In contrast, the other burst (GRB 100414A), characterized by a very hard $\alpha$ index, exhibited a significant mismatch between the estimated $E^{'}_{\rm c}$ and observed $E_{\rm c}$. In Group III, which included GRBs with intrinsic Band-like spectra, the results were consistent with the findings from Groups I and II. These findings align with our simulation results discussed earlier. High $E_{\rm c}$ values and hard $\alpha$ indices correspond to less pronounced curvature within the BAT energy range, leading to greater uncertainty in $E_{\rm c}$ estimates derived from this method. Conversely, GRBs with moderate $E_{\rm c}$ values (150-1000 keV) and $\alpha$ indices near typical values yield reliable extrapolated $E_{\rm c}$ estimates.

Overall, our analysis indicates that when $E_{\rm c}$ lies within a moderate range (150 keV $<E_{\rm c}\lesssim$ 1000 keV) and the intrinsic $\alpha$ index is not too hard ($\alpha \gtrsim-2/3$), the extrapolated $E^{'}_{\rm c}$ values obtained through this curvature method are consistent with their observed $E_{\rm c}$ values. Conversely, when $E_{\rm c}$ is excessively high ($E_{\rm c}> 1000$ keV) or $\alpha$ is too hard (e.g., beyond the synchrotron radiation death line, \citealt{Preece1998}, $\alpha=-2/3$), significant deviation arise between the estimated $E^{'}_{\rm c}$ and observed $E_{\rm c}$ values. In general, our method achieves higher accuracy compared to directly fitting the spectrum or estimating $E_{\rm c}$ (or $E_{\rm p}$) through empirical relations.

\acknowledgments

This work is supported by the Natural Science Foundation of China (grant No. 11874033), the KC Wong Magna Foundation at Ningbo University, and made use of the High Energy Astrophysics Science Archive Research Center (HEASARC) Online Service at the NASA/Goddard Space Flight Center (GSFC). The computations were supported by the high performance computing center at Ningbo University.

\vspace{5mm}
\facilities{\emph{Fermi}/GBM}
\software{
{\tt 3ML} \citep{Vianello2015}, 
{\tt matplotlib} \citep{Hunter2007}, 
{\tt NumPy} \citep{Harris2020,Walt2011}, 
{\tt SciPy} \citep{Virtanen2020}, 
{\tt $lmfit$} \citep{Newville2016}, 
{\tt astropy} \citep{AstropyCollaboration2013},
{\tt pandas} \citep{Reback2022},
{\tt emcee} \citep{Foreman-Mackey2013},
{\tt seaborn} \citep{Waskom2017}}  
\bibliography{Myreferences.bib}

\clearpage
\setlength{\tabcolsep}{0.25em}
\begin{deluxetable*}{ccccccccccc}
\tabletypesize{\scriptsize}
\tablecaption{Results of GRB 140703A}
\tablehead{
\colhead{$E_{\rm c}$}
&\colhead{Level}
&\colhead{$A$}
&\colhead{$\alpha$}
&\colhead{$\chi^{2}$}
&\colhead{DOF}
&\colhead{$\chi_{\rm r}^{2}$}
&\colhead{AIC}
&\colhead{BIC}
&\colhead{DIC}
&\colhead{pDIC}\\
(Fixed)&(Estimation)&(ph.s$^{-1}$.cm$^{-2}$keV$^{-1}$)
}
\colnumbers
\startdata
\hline\noalign{\smallskip}
\multicolumn{10}{c}{{\bf Preliminary estimation}}\tabularnewline
\hline\noalign{\smallskip}
100&Preliminary&1.08$\times$10$^{-2}$&0.90&36.2&26&1.39&40.2&42.8&40.2&4.05\\
120&Preliminary&9.71$\times$10$^{-3}$&0.99&33.1&26&1.27&37.1&39.7&37.1&4.02\\
150&Preliminary&8.70$\times$10$^{-3}$&1.09&31.7&26&1.22&35.7&38.3&35.6&3.98\\
180&Preliminary&8.09$\times$10$^{-3}$&1.14&31.7&26&1.22&35.7&38.3&35.7&4.01\\
200&Preliminary&7.80$\times$10$^{-3}$&1.18&31.9&26&1.23&35.9&38.6&35.9&4.00\\
220&Preliminary&7.57$\times$10$^{-3}$&1.20&32.3&26&1.24&36.3&39.0&36.3&4.04\\
250&Preliminary&7.29$\times$10$^{-3}$&1.23&32.9&26&1.27&36.9&39.6&36.9&4.04\\
\enddata 
\vspace{3mm}
\label{tab:140703A}
\end{deluxetable*}

\setlength{\tabcolsep}{0.35em}
\begin{deluxetable*}{ccccccccccc}
\tabletypesize{\scriptsize}
\tablecaption{Results of GRB 160629A}
\tablehead{
\colhead{$E_{\rm c}$}
&\colhead{Level}
&\colhead{$A$}
&\colhead{$\alpha$}
&\colhead{$\chi^{2}$}
&\colhead{DOF}
&\colhead{$\chi_{\rm r}^{2}$}
&\colhead{AIC}
&\colhead{BIC}
&\colhead{DIC}
&\colhead{pDIC}\\
(Fixed)&(Estimation)&(ph.s$^{-1}$.cm$^{-2}$keV$^{-1}$)
}
\colnumbers
\startdata
\hline\noalign{\smallskip}
\multicolumn{10}{c}{{\bf Preliminary estimation}}\tabularnewline
\hline\noalign{\smallskip}
100&Preliminary&1.76$\times$10$^{-2}$&0.63&19.9&13&1.53&23.9&25.3&23.8&3.96\\
120&Preliminary&1.59$\times$10$^{-2}$&0.72&15.4&13&1.18&19.4&20.8&19.3&3.93\\
150&Preliminary&1.44$\times$10$^{-2}$&0.80&12.1&13&0.93&16.1&17.5&16.0&3.90\\
180&Preliminary&1.34$\times$10$^{-2}$&0.86&10.5&13&0.81&14.5&16.0&14.5&3.96\\
200&Preliminary&1.30$\times$10$^{-2}$&0.89&10.0&13&0.77&14.0&15.4&14.0&3.98\\
220&Preliminary&1.26$\times$10$^{-2}$&0.92&9.6&13&0.74&13.6&15.0&13.6&4.00\\
250&Preliminary&1.22$\times$10$^{-2}$&0.94&9.3&13&0.71&13.3&14.7&13.2&3.92\\
\enddata 
\vspace{3mm}
\label{tab:160629A}
\end{deluxetable*}

\setlength{\tabcolsep}{0.35em}
\begin{deluxetable*}{ccccccccccc}
\tabletypesize{\scriptsize}
\tablecaption{Results of GRB 090529}
\tablehead{
\colhead{$E_{\rm c}$}
&\colhead{Level}
&\colhead{$A$}
&\colhead{$\alpha$}
&\colhead{$\chi^{2}$}
&\colhead{DOF}
&\colhead{$\chi_{\rm r}^{2}$}
&\colhead{AIC}
&\colhead{BIC}
&\colhead{DIC}
&\colhead{pDIC}\\
(Fixed)&(Estimation)&(ph.s$^{-1}$.cm$^{-2}$keV$^{-1}$)
}
\colnumbers
\startdata
\hline\noalign{\smallskip}
\multicolumn{10}{c}{{\bf Preliminary estimation}}\tabularnewline
\hline\noalign{\smallskip}
100&Preliminary&5.83$\times$10$^{-2}$&0.75&195.7&62&3.16&199.7&204.0&199.6&3.95\\
120&Preliminary&5.28$\times$10$^{-2}$&0.84&175.3&62&2.83&179.3&183.6&179.3&3.96\\
150&Preliminary&4.78$\times$10$^{-2}$&0.93&158.8&62&2.56&162.8&167.1&162.7&3.92\\
180&Preliminary&4.47$\times$10$^{-2}$&0.99&149.9&62&2.42&153.9&158.2&153.9&3.98\\
200&Preliminary&4.33$\times$10$^{-2}$&1.02&146.2&62&2.36&150.2&154.5&150.1&3.95\\
220&Preliminary&4.20$\times$10$^{-2}$&1.04&143.4&62&2.31&147.4&151.8&147.3&3.88\\
250&Preliminary&4.07$\times$10$^{-2}$&1.07&140.6&62&2.27&144.6&148.9&144.5&3.90\\
\enddata 
\vspace{3mm}
\label{tab:090529}
\end{deluxetable*}

\setlength{\tabcolsep}{0.35em}
\begin{deluxetable*}{ccccccccccc}
\tabletypesize{\scriptsize}
\tablecaption{Results of GRB 091003}
\tablehead{
\colhead{$E_{\rm c}$}
&\colhead{Level}
&\colhead{$A$}
&\colhead{$\alpha$}
&\colhead{$\chi^{2}$}
&\colhead{DOF}
&\colhead{$\chi_{\rm r}^{2}$}
&\colhead{AIC}
&\colhead{BIC}
&\colhead{DIC}
&\colhead{pDIC}\\
(Fixed)&(Estimation)&(ph.s$^{-1}$.cm$^{-2}$keV$^{-1}$)
}
\colnumbers
\startdata
\hline\noalign{\smallskip}
\multicolumn{10}{c}{{\bf Preliminary estimation}}\tabularnewline
\hline\noalign{\smallskip}
100&Preliminary&7.90$\times$10$^{-2}$&0.60&211.3&60&3.52&215.3&219.5&215.3&4.03\\
120&Preliminary&7.17$\times$10$^{-2}$&0.70&156.8&60&2.61&160.8&165.1&160.8&4.01\\
150&Preliminary&6.50$\times$10$^{-2}$&0.80&114.6&60&1.91&118.6&122.9&118.6&3.98\\
180&Preliminary&6.09$\times$10$^{-2}$&0.86&93.6&60&1.56&97.6&101.9&97.6&4.02\\
200&Preliminary&5.89$\times$10$^{-2}$&0.90&85.3&60&1.42&89.3&93.6&89.3&4.00\\
220&Preliminary&5.73$\times$10$^{-2}$&0.92&79.7&60&1.33&83.7&87.9&83.7&4.01\\
250&Preliminary&5.55$\times$10$^{-2}$&0.96&74.2&60&1.24&78.2&82.5&78.2&4.01\\
\enddata 
\vspace{3mm}
\label{tab:091003}
\end{deluxetable*}

\setlength{\tabcolsep}{0.35em}
\begin{deluxetable*}{ccccccccccc}
\tabletypesize{\scriptsize}
\tablecaption{Results of GRB 140606B}
\tablehead{
\colhead{$E_{\rm c}$}
&\colhead{Level}
&\colhead{$A$}
&\colhead{$\alpha$}
&\colhead{$\chi^{2}$}
&\colhead{DOF}
&\colhead{$\chi_{\rm r}^{2}$}
&\colhead{AIC}
&\colhead{BIC}
&\colhead{DIC}
&\colhead{pDIC}\\
(Fixed)&(Estimation)&(ph.s$^{-1}$.cm$^{-2}$keV$^{-1}$)
}
\colnumbers
\startdata
\hline\noalign{\smallskip}
\multicolumn{10}{c}{{\bf Preliminary estimation}}\tabularnewline
\hline\noalign{\smallskip}
100&Preliminary&4.07$\times$10$^{-2}$&0.70&102.6&47&2.18&106.6&110.3&106.6&4.07\\
120&Preliminary&3.70$\times$10$^{-2}$&0.80&89.0&47&1.89&93.0&96.7&93.0&4.02\\
150&Preliminary&3.36$\times$10$^{-2}$&0.89&77.2&47&1.64&81.2&85.0&81.3&4.03\\
180&Preliminary&3.15$\times$10$^{-2}$&0.95&70.5&47&1.50&74.5&78.3&74.5&3.97\\
200&Preliminary&3.05$\times$10$^{-2}$&0.98&67.5&47&1.44&71.5&75.3&71.6&4.04\\
220&Preliminary&2.97$\times$10$^{-2}$&1.01&65.2&47&1.39&69.2&73.0&69.3&4.02\\
250&Preliminary&2.87$\times$10$^{-2}$&1.04&62.7&47&1.33&66.7&70.5&66.8&4.06\\
\enddata 
\vspace{3mm}
\label{tab:140606B}
\end{deluxetable*}

\setlength{\tabcolsep}{0.35em}
\begin{deluxetable*}{ccccccccccc}
\tabletypesize{\scriptsize}
\tablecaption{Results of GRB 120624B}
\tablehead{
\colhead{$E_{\rm c}$}
&\colhead{Level}
&\colhead{$A$}
&\colhead{$\alpha$}
&\colhead{$\chi^{2}$}
&\colhead{DOF}
&\colhead{$\chi_{\rm r}^{2}$}
&\colhead{AIC}
&\colhead{BIC}
&\colhead{DIC}
&\colhead{pDIC}\\
(Fixed)&(Estimation)&(ph.s$^{-1}$.cm$^{-2}$keV$^{-1}$)
}
\colnumbers
\startdata
\hline\noalign{\smallskip}
\multicolumn{10}{c}{{\bf Preliminary estimation}}\tabularnewline
\hline\noalign{\smallskip}
100&Preliminary&4.54$\times$10$^{-2}$&0.40&181.2&34&5.33&185.2&188.3&185.2&4.00\\
120&Preliminary&4.14$\times$10$^{-2}$&0.51&144.5&34&4.25&148.5&151.7&148.5&3.98\\
150&Preliminary&3.78$\times$10$^{-2}$&0.62&113.5&34&3.34&117.5&120.7&117.5&4.01\\
180&Preliminary&3.55$\times$10$^{-2}$&0.69&96.1&34&2.83&100.1&103.3&100.1&4.02\\
200&Preliminary&3.44$\times$10$^{-2}$&0.73&88.5&34&2.60&92.5&95.6&92.5&4.00\\
220&Preliminary&3.35$\times$10$^{-2}$&0.76&82.7&34&2.43&86.7&89.9&86.8&4.03\\
250&Preliminary&3.25$\times$10$^{-2}$&0.79&76.5&34&2.25&80.5&83.6&80.4&3.95\\
\enddata 
\vspace{3mm}
\label{tab:120624B}
\end{deluxetable*}

\setlength{\tabcolsep}{0.25em}
\begin{deluxetable*}{ccccccccccc}
\tabletypesize{\scriptsize}
\tablecaption{Results of GRB 090328}
\tablehead{
\colhead{$E_{\rm c}$}
&\colhead{Level}
&\colhead{$A$}
&\colhead{$\alpha$}
&\colhead{$\chi^{2}$}
&\colhead{DOF}
&\colhead{$\chi_{\rm r}^{2}$}
&\colhead{AIC}
&\colhead{BIC}
&\colhead{DIC}
&\colhead{pDIC}\\
(Fixed)&(Estimation)&(ph.s$^{-1}$.cm$^{-2}$keV$^{-1}$)
}
\colnumbers
\startdata
\hline\noalign{\smallskip}
\multicolumn{10}{c}{{\bf Preliminary estimation}}\tabularnewline
\hline\noalign{\smallskip}
100&Preliminary&3.32$\times$10$^{-2}$&0.65&150.3&29&5.18&154.3&157.1&154.3&4.00\\
120&Preliminary&3.01$\times$10$^{-2}$&0.74&118.1&29&4.07&122.1&124.9&122.1&4.03\\
150&Preliminary&2.73$\times$10$^{-2}$&0.84&91.1&29&3.14&95.1&98.0&95.1&3.98\\
180&Preliminary&2.55$\times$10$^{-2}$&0.90&76.2&29&2.63&80.2&83.0&80.2&4.00\\
200&Preliminary&2.47$\times$10$^{-2}$&0.93&69.6&29&2.40&73.6&76.5&73.6&3.97\\
220&Preliminary&2.40$\times$10$^{-2}$&0.96&64.7&29&2.23&68.7&71.6&68.8&4.02\\
250&Preliminary&2.33$\times$10$^{-2}$&0.99&59.5&29&2.05&63.5&66.3&63.4&3.98\\
\enddata 
\vspace{3mm}
\label{tab:090328}
\end{deluxetable*}

\setlength{\tabcolsep}{0.35em}
\begin{deluxetable*}{ccccccccccc}
\tabletypesize{\scriptsize}
\tablecaption{Results of GRB 150301B}
\tablehead{
\colhead{$E_{\rm c}$}
&\colhead{Level}
&\colhead{$A$}
&\colhead{$\alpha$}
&\colhead{$\chi^{2}$}
&\colhead{DOF}
&\colhead{$\chi_{\rm r}^{2}$}
&\colhead{AIC}
&\colhead{BIC}
&\colhead{DIC}
&\colhead{pDIC}\\
(Fixed)&(Estimation)&(ph.s$^{-1}$.cm$^{-2}$keV$^{-1}$)
}
\colnumbers
\startdata
\hline\noalign{\smallskip}
\multicolumn{10}{c}{{\bf Preliminary estimation}}\tabularnewline
\hline\noalign{\smallskip}
100&Preliminary&2.25$\times$10$^{-2}$&0.78&30.1&18&1.67&34.1&36.1&34.1&3.96\\
120&Preliminary&2.04$\times$10$^{-2}$&0.88&27.0&18&1.50&31.0&33.0&31.0&4.04\\
150&Preliminary&1.85$\times$10$^{-2}$&0.98&24.3&18&1.35&28.3&30.3&28.3&3.98\\
180&Preliminary&1.73$\times$10$^{-2}$&1.04&22.9&18&1.27&26.9&28.9&26.9&3.98\\
200&Preliminary&1.67$\times$10$^{-2}$&1.07&22.3&18&1.24&26.3&28.3&26.2&3.99\\
220&Preliminary&1.63$\times$10$^{-2}$&1.10&21.8&18&1.21&25.8&27.8&25.8&3.98\\
250&Preliminary&1.58$\times$10$^{-2}$&1.13&21.3&18&1.18&25.3&27.3&25.3&3.98\\
\enddata 
\vspace{3mm}
\label{tab:???}
\end{deluxetable*}

\setlength{\tabcolsep}{0.35em}
\begin{deluxetable*}{ccccccccccc}
\tabletypesize{\scriptsize}
\tablecaption{Results of GRB 180728A}
\tablehead{
\colhead{$E_{\rm c}$}
&\colhead{Level}
&\colhead{$A$}
&\colhead{$\alpha$}
&\colhead{$\chi^{2}$}
&\colhead{DOF}
&\colhead{$\chi_{\rm r}^{2}$}
&\colhead{AIC}
&\colhead{BIC}
&\colhead{DIC}
&\colhead{pDIC}\\
(Fixed)&(Estimation)&(ph.s$^{-1}$.cm$^{-2}$keV$^{-1}$)
}
\colnumbers
\startdata
\hline\noalign{\smallskip}
\multicolumn{10}{c}{{\bf Preliminary estimation}}\tabularnewline
\hline\noalign{\smallskip}
100&Preliminary&3.02$\times$10$^{-1}$&1.40&410.7&47&8.74&414.7&418.5&414.7&3.99\\
120&Preliminary&2.74$\times$10$^{-1}$&1.47&319.9&47&6.81&323.9&327.7&323.9&3.94\\
150&Preliminary&2.49$\times$10$^{-1}$&1.55&247.2&47&5.26&251.2&255.0&251.1&3.91\\
180&Preliminary&2.33$\times$10$^{-1}$&1.60&209.3&47&4.45&213.3&217.1&213.2&3.90\\
200&Preliminary&2.26$\times$10$^{-1}$&1.62&193.6&47&4.12&197.6&201.4&197.5&3.93\\
220&Preliminary&2.20$\times$10$^{-1}$&1.64&182.4&47&3.88&186.4&190.2&186.4&3.93\\
250&Preliminary&2.13$\times$10$^{-1}$&1.67&171.1&47&3.64&175.1&178.8&175.0&3.96\\
\enddata 
\vspace{3mm}
\label{tab:180728A}
\end{deluxetable*}

\setlength{\tabcolsep}{0.35em}
\begin{deluxetable*}{ccccccccccc}
\tabletypesize{\scriptsize}
\tablecaption{Results of GRB 140206A}
\tablehead{
\colhead{$E_{\rm c}$}
&\colhead{Level}
&\colhead{$A$}
&\colhead{$\alpha$}
&\colhead{$\chi^{2}$}
&\colhead{DOF}
&\colhead{$\chi_{\rm r}^{2}$}
&\colhead{AIC}
&\colhead{BIC}
&\colhead{DIC}
&\colhead{pDIC}\\
(Fixed)&(Estimation)&(ph.s$^{-1}$.cm$^{-2}$keV$^{-1}$)
}
\colnumbers
\startdata
\hline\noalign{\smallskip}
\multicolumn{10}{c}{{\bf Preliminary estimation}}\tabularnewline
\hline\noalign{\smallskip}
100&Preliminary&1.20$\times$10$^{-1}$&0.59&424.9&59&7.20&428.9&433.1&429.0&4.10\\
120&Preliminary&1.09$\times$10$^{-1}$&0.69&317.6&59&5.38&321.6&325.9&321.6&3.98\\
150&Preliminary&9.86$\times$10$^{-2}$&0.78&232.4&59&3.94&236.4&240.7&236.5&4.08\\
180&Preliminary&9.22$\times$10$^{-2}$&0.84&188.4&59&3.19&192.4&196.6&192.4&4.02\\
200&Preliminary&8.92$\times$10$^{-2}$&0.88&170.3&59&2.89&174.3&178.5&174.3&4.03\\
220&Preliminary&8.68$\times$10$^{-2}$&0.90&157.5&59&2.67&161.5&165.7&161.5&4.02\\
250&Preliminary&8.40$\times$10$^{-2}$&0.93&144.5&59&2.45&148.5&152.7&148.6&4.10\\
\enddata 
\vspace{3mm}
\label{tab:140206A}
\end{deluxetable*}

\setlength{\tabcolsep}{0.35em}
\begin{deluxetable*}{ccccccccccc}
\tabletypesize{\scriptsize}
\tablecaption{Results of GRB 110731A}
\tablehead{
\colhead{$E_{\rm c}$}
&\colhead{Level}
&\colhead{$A$}
&\colhead{$\alpha$}
&\colhead{$\chi^{2}$}
&\colhead{DOF}
&\colhead{$\chi_{\rm r}^{2}$}
&\colhead{AIC}
&\colhead{BIC}
&\colhead{DIC}
&\colhead{pDIC}\\
(Fixed)&(Estimation)&(ph.s$^{-1}$.cm$^{-2}$keV$^{-1}$)
}
\colnumbers
\startdata
\hline\noalign{\smallskip}
\multicolumn{10}{c}{{\bf Preliminary estimation}}\tabularnewline
\hline\noalign{\smallskip}
100&Preliminary&1.34$\times$10$^{-1}$&0.46&142.8&40&3.57&146.8&150.3&146.9&4.08\\
120&Preliminary&1.22$\times$10$^{-1}$&0.56&116.7&40&2.92&120.7&124.2&120.8&4.14\\
150&Preliminary&1.11$\times$10$^{-1}$&0.66&94.5&40&2.36&98.5&102.0&98.6&4.16\\
180&Preliminary&1.04$\times$10$^{-1}$&0.73&82.0&40&2.05&86.0&89.5&86.1&4.13\\
200&Preliminary&1.00$\times$10$^{-1}$&0.76&76.4&40&1.91&80.4&83.9&80.6&4.12\\
220&Preliminary&9.78$\times$10$^{-2}$&0.79&72.3&40&1.81&76.3&79.8&76.4&4.13\\
250&Preliminary&9.47$\times$10$^{-2}$&0.82&67.7&40&1.69&71.7&75.2&71.8&4.07\\
\enddata 
\vspace{3mm}
\label{tab:110731A}
\end{deluxetable*}

\setlength{\tabcolsep}{0.35em}
\begin{deluxetable*}{ccccccccccc}
\tabletypesize{\scriptsize}
\tablecaption{Results of GRB 180720B}
\tablehead{
\colhead{$E_{\rm c}$}
&\colhead{Level}
&\colhead{$A$}
&\colhead{$\alpha$}
&\colhead{$\chi^{2}$}
&\colhead{DOF}
&\colhead{$\chi_{\rm r}^{2}$}
&\colhead{AIC}
&\colhead{BIC}
&\colhead{DIC}
&\colhead{pDIC}\\
(Fixed)&(Estimation)&(ph.s$^{-1}$.cm$^{-2}$keV$^{-1}$)
}
\colnumbers
\startdata
\hline\noalign{\smallskip}
\multicolumn{10}{c}{{\bf Preliminary estimation}}\tabularnewline
\hline\noalign{\smallskip}
100&Preliminary&2.40$\times$10$^{-1}$&0.59&772.5&110&7.02&776.5&781.9&776.4&3.91\\
120&Preliminary&2.18$\times$10$^{-1}$&0.69&591.8&110&5.38&595.8&601.3&595.8&3.92\\
150&Preliminary&1.98$\times$10$^{-1}$&0.79&442.4&110&4.02&446.4&451.9&446.4&3.96\\
180&Preliminary&1.85$\times$10$^{-1}$&0.85&361.0&110&3.28&365.0&370.4&364.9&3.95\\
200&Preliminary&1.79$\times$10$^{-1}$&0.88&325.9&110&2.96&329.9&335.3&329.8&3.94\\
220&Preliminary&1.74$\times$10$^{-1}$&0.91&300.0&110&2.73&304.0&309.5&304.0&3.96\\
250&Preliminary&1.69$\times$10$^{-1}$&0.94&272.5&110&2.48&276.5&281.9&276.4&3.96\\
\enddata 
\vspace{3mm}
\label{tab:180720B}
\end{deluxetable*}

\setlength{\tabcolsep}{0.35em}
\begin{deluxetable*}{ccccccccccc}
\tabletypesize{\scriptsize}
\tablecaption{Results of GRB 190530A}
\tablehead{
\colhead{$E_{\rm c}$}
&\colhead{Level}
&\colhead{$A$}
&\colhead{$\alpha$}
&\colhead{$\chi^{2}$}
&\colhead{DOF}
&\colhead{$\chi_{\rm r}^{2}$}
&\colhead{AIC}
&\colhead{BIC}
&\colhead{DIC}
&\colhead{pDIC}\\
(Fixed)&(Estimation)&(ph.s$^{-1}$.cm$^{-2}$keV$^{-1}$)
}
\colnumbers
\startdata
\hline\noalign{\smallskip}
\multicolumn{10}{c}{{\bf Preliminary estimation}}\tabularnewline
\hline\noalign{\smallskip}
100&Preliminary&4.54$\times$10$^{-1}$&0.47&1264.4&91&13.89&1268.4&1273.4&1268.4&4.00\\
120&Preliminary&4.11$\times$10$^{-1}$&0.57&925.0&91&10.16&929.0&934.0&929.0&4.03\\
150&Preliminary&3.72$\times$10$^{-1}$&0.66&647.8&91&7.12&651.8&656.9&651.8&3.99\\
180&Preliminary&3.48$\times$10$^{-1}$&0.72&499.2&91&5.49&503.2&508.3&503.3&4.02\\
200&Preliminary&3.36$\times$10$^{-1}$&0.76&436.1&91&4.79&440.1&445.2&440.1&3.98\\
220&Preliminary&3.27$\times$10$^{-1}$&0.78&390.2&91&4.29&394.2&399.2&394.1&3.98\\
250&Preliminary&3.17$\times$10$^{-1}$&0.81&341.8&91&3.76&345.8&350.9&345.8&3.99\\
\enddata 
\vspace{3mm}
\label{tab:190530A}
\end{deluxetable*}

\setlength{\tabcolsep}{0.35em}
\begin{deluxetable*}{ccccccccccc}
\tablewidth{900pt}
\tabletypesize{\scriptsize}
\tablecaption{Results of GRB 100414A}
\tablehead{
\colhead{$E_{\rm c}$}
&\colhead{Level}
&\colhead{$A$}
&\colhead{$\alpha$}
&\colhead{$\chi^{2}$}
&\colhead{DOF}
&\colhead{$\chi_{\rm r}^{2}$}
&\colhead{AIC}
&\colhead{BIC}
&\colhead{DIC}
&\colhead{pDIC}\\
(Fixed)&(Estimation)&(ph.s$^{-1}$.cm$^{-2}$keV$^{-1}$)
}
\colnumbers
\startdata
100&Preliminary&6.95$\times$10$^{-2}$&-0.02&1017.6&118&8.62&1021.6&1027.2&1021.6&4.00\\
120&Preliminary&6.36$\times$10$^{-2}$&0.10&816.6&118&6.92&820.6&826.2&820.6&3.95\\
150&Preliminary&5.81$\times$10$^{-2}$&0.22&641.3&118&5.44&645.3&650.9&645.3&4.00\\
180&Preliminary&5.47$\times$10$^{-2}$&0.29&540.0&118&4.58&544.0&549.5&544.0&4.02\\
200&Preliminary&5.30$\times$10$^{-2}$&0.33&494.2&118&4.19&498.2&503.8&498.2&4.01\\
220&Preliminary&5.17$\times$10$^{-2}$&0.36&459.2&118&3.89&463.2&468.8&463.3&4.01\\
250&Preliminary&5.02$\times$10$^{-2}$&0.40&420.4&118&3.56&424.4&430.0&424.4&4.01\\
300&Preliminary&4.83$\times$10$^{-2}$&0.44&377.5&118&3.20&381.5&387.1&381.5&3.98\\
\enddata 
\vspace{3mm}
\label{tab:100414A}
\end{deluxetable*}

\setlength{\tabcolsep}{0.35em}
\begin{deluxetable*}{ccccccccccc}
\tabletypesize{\scriptsize}
\tablecaption{Results of GRB 120711A}
\tablehead{
\colhead{$E_{\rm c}$}
&\colhead{Level}
&\colhead{$A$}
&\colhead{$\alpha$}
&\colhead{$\chi^{2}$}
&\colhead{DOF}
&\colhead{$\chi_{\rm r}^{2}$}
&\colhead{AIC}
&\colhead{BIC}
&\colhead{DIC}
&\colhead{pDIC}\\
(Fixed)&(Estimation)&(ph.s$^{-1}$.cm$^{-2}$keV$^{-1}$)
}
\colnumbers
\startdata
100&Preliminary&5.28$\times$10$^{-2}$&0.42&91.9&9&10.22&95.9&96.7&95.9&4.01\\
120&Preliminary&4.79$\times$10$^{-2}$&0.53&67.4&9&7.49&71.4&72.2&71.3&3.94\\
150&Preliminary&4.35$\times$10$^{-2}$&0.63&47.5&9&5.28&51.5&52.3&51.5&3.99\\
180&Preliminary&4.08$\times$10$^{-2}$&0.70&37.0&9&4.12&41.0&41.8&41.0&3.98\\
200&Preliminary&3.95$\times$10$^{-2}$&0.74&32.6&9&3.63&36.6&37.4&36.6&3.96\\
220&Preliminary&3.84$\times$10$^{-2}$&0.77&29.5&9&3.28&33.5&34.3&33.4&3.96\\
250&Preliminary&3.72$\times$10$^{-2}$&0.80&26.2&9&2.91&30.2&31.0&30.2&3.96\\
300&Preliminary&3.58$\times$10$^{-2}$&0.84&23.0&9&2.55&27.0&27.8&27.0&3.98\\
\enddata 
\vspace{3mm}
\label{tab:120711A}
\end{deluxetable*}

\setlength{\tabcolsep}{0.35em}
\begin{deluxetable*}{ccccccccccc}
\tablewidth{1000pt}
\tabletypesize{\scriptsize}
\tablecaption{Results of GRB 130215A}
\tablehead{
\colhead{$E_{\rm c}$}
&\colhead{Level}
&\colhead{$A$}
&\colhead{$\alpha$}
&\colhead{$\chi^{2}$}
&\colhead{DOF}
&\colhead{$\chi_{\rm r}^{2}$}
&\colhead{AIC}
&\colhead{BIC}
&\colhead{DIC}
&\colhead{pDIC}\\
(Fixed)&(Estimation)&(ph.s$^{-1}$.cm$^{-2}$keV$^{-1}$)
}
\colnumbers
\startdata
100&Preliminary&1.70$\times$10$^{-2}$&0.84&17.7&11&1.61&21.7&22.8&21.6&3.94\\
120&Preliminary&1.55$\times$10$^{-2}$&0.92&15.2&11&1.38&19.2&20.4&19.2&3.99\\
150&Preliminary&1.40$\times$10$^{-2}$&0.99&13.2&11&1.20&17.2&18.3&17.1&3.95\\
180&Preliminary&1.32$\times$10$^{-2}$&1.05&12.1&11&1.10&16.1&17.2&16.1&4.03\\
200&Preliminary&1.27$\times$10$^{-2}$&1.07&11.6&11&1.05&15.6&16.7&15.6&3.99\\
220&Preliminary&1.24$\times$10$^{-2}$&1.10&11.2&11&1.02&15.2&16.3&15.1&3.94\\
250&Preliminary&1.20$\times$10$^{-2}$&1.12&10.8&11&0.98&14.8&15.9&14.8&3.99\\
300&Preliminary&1.16$\times$10$^{-2}$&1.15&10.3&11&0.94&14.3&15.5&14.2&3.91\\
\enddata 
\vspace{3mm}
\label{tab:130215A}
\end{deluxetable*}

\setlength{\tabcolsep}{0.35em}
\begin{deluxetable*}{ccccccccccc}
\tablewidth{700pt}
\tabletypesize{\scriptsize}
\tablecaption{Results of GRB 090510}
\tablehead{
\colhead{$E_{\rm c}$}
&\colhead{Level}
&\colhead{$A$}
&\colhead{$\alpha$}
&\colhead{$\chi^{2}$}
&\colhead{DOF}
&\colhead{$\chi_{\rm r}^{2}$}
&\colhead{AIC}
&\colhead{BIC}
&\colhead{DIC}
&\colhead{pDIC}\\
(Fixed)&(Estimation)&(ph.s$^{-1}$.cm$^{-2}$keV$^{-1}$)
}
\colnumbers
\startdata
100&Preliminary&2.88$\times$10$^{-2}$&0.35&17.4&20&0.87&21.4&23.6&21.4&4.00\\
120&Preliminary&2.62$\times$10$^{-2}$&0.46&16.4&20&0.82&20.4&22.6&20.4&3.95\\
150&Preliminary&2.38$\times$10$^{-2}$&0.56&15.7&20&0.78&19.7&21.8&19.6&3.99\\
180&Preliminary&2.24$\times$10$^{-2}$&0.63&15.3&20&0.76&19.3&21.4&19.3&4.01\\
200&Preliminary&2.17$\times$10$^{-2}$&0.67&15.1&20&0.75&19.1&21.3&19.1&4.01\\
220&Preliminary&2.12$\times$10$^{-2}$&0.69&15.0&20&0.75&19.0&21.1&19.0&4.06\\
250&Preliminary&2.05$\times$10$^{-2}$&0.73&14.8&20&0.74&18.8&21.0&18.8&3.97\\
300&Preliminary&1.97$\times$10$^{-2}$&0.77&14.7&20&0.74&18.7&20.9&18.8&4.03\\
\enddata 
\vspace{3mm}
\label{tab:090510}
\end{deluxetable*}

\clearpage
\startlongtable
\setlength{\tabcolsep}{0.15em}
\begin{deluxetable*}{cccccccccccc}
\tabletypesize{\scriptsize}
\tablecaption{Results of $E^{'}_{\rm c}$ Estimation of the CPL model}
\tablehead{
\colhead{GRB}
&\colhead{Group}
&\colhead{Model}
&\colhead{$E_{\rm c}$}
&\colhead{$A$}
&\colhead{$\alpha$}
&\colhead{$E_{\rm p}$}
&\colhead{$E^{'}_{\rm c}$}
&\colhead{$\alpha^{'}$}
&\colhead{$E^{'}_{\rm p}$}
&\colhead{$E_{\rm c}$ vs. $E^{'}_{\rm c}$}
&\colhead{Interpretation}\\
&&&(Observed)&(Observed)&(Observed)&(Observed)&($\Delta$DIC$\leq$5)&($\Delta$DIC$\leq$5)&($\Delta$DIC$\leq$5)&(Match)&(For $E_{\rm c}$ vs. $E^{'}_{\rm c}$)
}
\colnumbers
\startdata
140703A(026)&I&CPL&173$^{+51}_{-38}$&0.8$^{+0.3}_{-0.2}$&-1.16$^{+0.11}_{-0.12}$&145$^{+47}_{-38}$&100&-0.90&110&Nicely&Moderated $E_{\rm c}$ and $\alpha$\\
150301B(818)&I&CPL&239$^{+178}_{-100}$&1.5$^{+1.5}_{-0.7}$&-1.16$^{+0.21}_{-0.22}$&201$^{+158}_{-99}$&150&-0.98&153&Nicely&Moderated $E_{\rm c}$ and $\alpha$\\
160629A(930)&I&CPL&212$^{+64}_{-49}$&0.5$^{+0.2}_{-0.1}$&-0.91$^{+0.12}_{-0.11}$&231$^{+74}_{-58}$&140&-0.78&171&Nicely&Moderated $E_{\rm c}$ and $\alpha$\\
090529(564)&I&CPL&305$^{+66}_{-55}$&3.6$^{+0.8}_{-0.6}$&-1.15$^{+0.06}_{-0.06}$&259$^{+59}_{-50}$&300&-1.11&267&Nicely&Moderated $E_{\rm c}$ and $\alpha$\\
091003(191)&I&CPL&388$^{+41}_{-38}$&2.9$\pm$0.3&-1.03$\pm$0.03&376$^{+41}_{-39}$&280&-0.98&286&Nicely&Moderated $E_{\rm c}$ and $\alpha$\\
140606B(133)&I&CPL&531$^{+188}_{-137}$&2.7$^{+0.7}_{-0.5}$&-1.19$\pm$0.07&430$^{+157}_{-117}$&425&-1.13&370&Nicely&Moderated $E_{\rm c}$ and $\alpha$\\
120624B(933)&I&CPL&642$^{+90}_{-79}$&1.3$^{+0.2}_{-0.2}$&-0.98$\pm$0.04&655$^{+95}_{-85}$&575&-0.94&610&Nicely&Moderated $E_{\rm c}$ and $\alpha$\\
090328(401)&I&CPL&759$^{+104}_{-95}$&1.7$\pm$0.1&-1.13$^{+0.03}_{-0.02}$&660$^{+93}_{-84}$&500&-1.11&445&Nicely&Moderated $E_{\rm c}$ and $\alpha$\\
\enddata 
\vspace{3mm}
\label{tab:CPLresults}
\end{deluxetable*}

\startlongtable
\setlength{\tabcolsep}{0.15em}
\begin{deluxetable*}{cccccccccccc}
\tabletypesize{\scriptsize}
\tablecaption{Results of $E^{'}_{\rm p}$ Estimation of the Band model}
\tablehead{
\colhead{GRB}
&\colhead{Group}
&\colhead{Model}
&\colhead{$E_{\rm p}$}
&\colhead{$A$}
&\colhead{$\alpha$}
&\colhead{$\beta$}
&\colhead{$E^{'}_{\rm c}$}
&\colhead{$\alpha^{'}$}
&\colhead{$E^{'}_{\rm p}$}
&\colhead{$E_{\rm p}$ vs. $E^{'}_{\rm p}$}
&\colhead{Interpretation}\\
&&&(Observed)&(Observed)&(Observed)&(Observed)&($\Delta$DIC$\leq$5)&($\Delta$DIC$\leq$5)&($\Delta$DIC$\leq$5)&(Match)&(For $E_{\rm p}$ vs. $E^{'}_{\rm p}$)
}
\colnumbers
\startdata
140206A(275)&III&Band&280$^{+10}_{-10}$&(4.2$\pm$0.1)$\times$10$^{-2}$&-0.96$\pm$0.16&-2.21$\pm$0.05&380&-1.01&376&Nicely&Moderated $E_{\rm p}$ and $\alpha$\\
110731A(465)&III&Band&366$^{+27}_{-27}$&(4.8$\pm$0.2)$\times$10$^{-2}$&-0.90$\pm$0.03&-3.00$\pm$0.70&525&-0.95&551&Nicely&Moderated $E_{\rm p}$ and $\alpha$\\
180720B(598)&III&Band&511$\pm$26&(7.0$\pm$0.1)$\times$10$^{-2}$&-1.07$\pm$0.01&-2.21$\pm$0.03&775&-1.10&697&Nicely&Moderated $E_{\rm p}$ and $\alpha$\\
190530A(430)&III&Band&866$^{+13}_{-13}$&(1.3$\pm$0.01)$\times$10$^{-1}$&-1.01$\pm$0.01&-3.63$\pm$0.20&700&-0.96&729&Nicely&Moderated $E_{\rm p}$ and $\alpha$\\
090510(016)&III&Band&5200$^{+574}_{-584}$&(9.0$\pm$0.4)$\times$10$^{-3}$&-0.89$\pm$0.04&-3.4$\pm$1.3&100&-0.34&165&{\bf Poorly}&$E_{\rm p}$ too high\\
\enddata 
\vspace{3mm}
\label{tab:Bandresults}
\end{deluxetable*}

\setlength{\tabcolsep}{0.25em}
\begin{deluxetable*}{cccccccccccc}
\tabletypesize{\scriptsize}
\tablecaption{Results of $E^{'}_{\rm p}$ Estimation Using the Empirical Relation.}
\tablehead{
\colhead{GRB}
&\colhead{$A$}
&\colhead{$\Gamma$}
&\colhead{$\chi^{2}$}
&\colhead{DOF}
&\colhead{$\chi_{\rm r}^{2}$}
&\colhead{AIC}
&\colhead{BIC}
&\colhead{DIC}
&\colhead{pDIC}
&\colhead{$E^{'}_{\rm p}$}
&\colhead{$E_{\rm p}$ vs. $E^{'}_{\rm p}$}\\
}
\colnumbers
\startdata
090328&(1.82$^{+0.02}_{-0.02}$)$\times$10$^{-2}$&1.22$^{+0.02}_{-0.01}$&40.48&29&1.40&44.48&47.35&44.43&3.95&176$\pm$7&{\bf Poorly}\\
090510&(2.27$^{+0.11}_{-0.11}$)$\times$10$^{-2}$&1.00$^{+0.07}_{-0.07}$&13.35&28&0.48&17.35&20.15&17.33&3.99&269$\pm$36&{\bf Poorly}\\
090529&(3.18$^{+0.04}_{-0.04}$)$\times$10$^{-2}$&1.29$^{+0.02}_{-0.02}$&133.88&62&2.16&137.88&142.20&137.88&4.00&154$\pm$6&{\bf Poorly}\\
091003&(4.35$^{+0.03}_{-0.03}$)$\times$10$^{-2}$&1.19$^{+0.01}_{-0.01}$&80.86&60&1.35&84.86&89.12&84.79&3.93&187$\pm$4&{\bf Poorly}\\
100414A&(3.99$^{+0.02}_{-0.02}$)$\times$10$^{-2}$&0.66$^{+0.01}_{-0.01}$&244.61&118&2.07&248.61&254.18&248.64&4.03&514$\pm$10&{\bf Poorly}\\
110731A&(7.45$^{+0.10}_{-0.10}$)$\times$10$^{-2}$&1.06$^{+0.02}_{-0.02}$&49.71&40&1.24&53.71&57.18&53.73&4.02&240$\pm$9&{\bf Poorly}\\
120624B&(2.58$^{+0.03}_{-0.03}$)$\times$10$^{-2}$&1.05$^{+0.02}_{-0.02}$&53.15&34&1.56&57.15&60.32&57.21&4.05&244$\pm$9&{\bf Poorly}\\
120711A&(2.92$^{+0.04}_{-0.04}$)$\times$10$^{-2}$&1.04$^{+0.02}_{-0.02}$&20.37&9&2.26&24.37&25.16&24.35&3.98&249$\pm$10&{\bf Poorly}\\
\enddata 
\vspace{3mm}
\label{tab:PLresults}
\end{deluxetable*}

\clearpage
\onecolumngrid

\clearpage

\clearpage

\clearpage

\clearpage

\clearpage

\clearpage

\clearpage

\clearpage

\clearpage

\clearpage

\clearpage

\clearpage

\clearpage

\clearpage

\clearpage

\clearpage

\clearpage

\clearpage

\clearpage

\clearpage

\clearpage
\appendix
\setcounter{figure}{0}    
\setcounter{section}{0}
\setcounter{table}{0}
\renewcommand{\thesection}{A\arabic{section}}
\renewcommand{\thefigure}{A\arabic{figure}}
\renewcommand{\thetable}{A\arabic{table}}
\renewcommand{\theequation}{A\arabic{equation}}

In this appendix, we present the amplitude normalization for different CPL models (Section \ref{sec: AmplitudeNorma}).

\section{Amplitude Normalization}\label{sec: AmplitudeNorma}

Details of the amplitude normalization procedure for different $E_\mathrm{c}$ values are provided here. According to the definition, we have:
\begin{equation}
N(E) =A \left(\frac{E}{E_{\rm piv}}\right)^{\alpha}\rm exp(-\frac{\it E}{\it E_{c}}),
\label{CPL}
\end{equation}

We consider two specific CPL spectra: one with the intrinsic cutoff energy ($E_{c,0}$), denoted as $N_0(A_0,\alpha_0,E_{\rm c,0})$, and another with a different cutoff energy, denoted as $N_i(A_i,\alpha_i,E_{\rm c,i})$. When $E \ll E_{\rm c}$, it follows that:
\begin{equation}
    \frac{E}{E_c} \approx 0, \quad e^{\left(-\frac{E}{E_c}\right)} \approx 1.
\end{equation}

and we require $N_0=N_i$, we can derive:
\begin{equation}
    A_0 \left(\frac{E}{E_{\rm piv,0}}\right)^{\alpha_0} = A_i \left(\frac{E}{E_{\rm piv,i}}\right)^{\alpha_i},
\end{equation}
since $\alpha$ indices are fixed, one has:
\begin{equation}
    \alpha_0 = \alpha_i=\alpha.
\end{equation}

Consequently, the final result is:
\begin{equation}
      A_i =  A_0 \left(\frac{E_{\rm piv,i}}{E_{\rm piv,0}}\right)^{\alpha}.
\end{equation}

\end{document}